\documentclass[trackchanges,twocolumn]{aastex631}

\shorttitle{StellarSpectraObservationFitting.jl}
\shortauthors{Gilbertson et al.}
\usepackage{hyperref}
\usepackage{graphicx}
\usepackage{amsmath}
\usepackage{gensymb}
\usepackage{afterpage}

\usepackage{xcolor}
\usepackage{rotating}
\usepackage{bbm}
\usepackage{longtable}
\usepackage[T1]{fontenc} 
\usepackage[utf8]{inputenc} 
\definecolor{cjg}{rgb}{0.0, 0.5, 0.5}
\definecolor{ebf}{rgb}{0.4, 0.0, 0.6}
\definecolor{sh}{rgb}{0.5, 0.5, 0.0}

\newcommand{\ssof}{{\tt SSOF} }
\newcommand{\ssofns}{{\tt SSOF}}
\newcommand{\ssofjl}{{\tt StellarSpectraObservationFitting.jl} }
\newcommand{\ssofjlns}{{\tt StellarSpectraObservationFitting.jl}}
\newcommand{\secref}[1]{\S\ref{#1}}
\newcommand{\nablajl}{{\tt Nabla.jl} }
\newcommand{\tgpjl}{{\tt TemporalGPs.jl} }
\newcommand{\tgpjlns}{{\tt TemporalGPs.jl}}
\newcommand{\optimjl}{{\tt Optim.jl} }

\newcommand{\julia}{{\tt Julia} }
\newcommand{\julians}{{\tt Julia}}
\newcommand{\wobble}{{\tt wobble} }
\newcommand{\wobblens}{{\tt wobble}}
\newcommand{\serval}{{\tt SERVAL} }
\newcommand{\servalns}{{\tt SERVAL}}

\newcommand\cms{$\mathrm{cm\ s^{-1}}$}
\newcommand\ms{$\mathrm{m\ s^{-1}}$}
\newcommand\kms{$\mathrm{km\ s^{-1}}$}
\newcommand\cahk{Ca II H\&K }

\newcommand{\PSUAA}{Department of Astronomy \& Astrophysics, Penn State University, 525 Davey Laboratory, 251 Pollock Road, University Park, PA 16802, USA}
\newcommand{\PSUCEHW}{Center for Exoplanets and Habitable Worlds, Penn State University, 525 Davey Laboratory, 251 Pollock Road, University Park, PA 16802, USA}
\newcommand{\ETH}{ETH Zurich, Institute for Particle Physics \& Astrophysics, Zurich, Switzerland}
\newcommand{\PSETI}{Penn State Extraterrestrial Intelligence Center, Penn State University, 525 Davey Laboratory, 251 Pollock Road, University Park, PA, 16802, USA}
\newcommand{\UCI}{Department of Physics \& Astronomy, The University of California, Irvine, Irvine, CA 92697, USA}

\begin{document}

\correspondingauthor{Christian Gilbertson}
\email{cjg66@psu.edu}

\title{Data-Driven Modeling of Telluric Features and Stellar Variability with StellarSpectraObservationFitting.jl}
\graphicspath{{./minfigs/}{}}

\author[0000-0002-1743-3684]{Christian Gilbertson}
\affiliation{Center for Computing Research, Sandia National Laboratories, 1450 Innovation Pkwy SE, Albuquerque, NM 87123 USA}
\affiliation{\PSUAA}
\affiliation{\PSUCEHW}
\affiliation{Penn State Astrobiology Research Center, University Park, PA 16802, USA}

\author[0000-0001-6545-639X]{Eric B. Ford}
\affiliation{\PSUAA}
\affiliation{\PSUCEHW}
\affiliation{Center for Astrostatistics, Penn State University, 525 Davey Laboratory, University, 251 Pollock Road Park, PA, 16802, USA}
\affiliation{Institute for Computational and Data Sciences, The Pennsylvania State University,  University Park, PA, 16802, USA}

\author[0000-0003-1312-9391]{Samuel Halverson}
\affiliation{Jet Propulsion Laboratory, 4800 Oak Grove Drive, Pasadena, CA 91109, USA}

\author[0000-0003-0199-9699]{Evan Fitzmaurice}
\affiliation{\PSUAA}
\affiliation{\PSUCEHW}
\affiliation{Institute for Computational and Data Sciences, 224B Computer Building, Penn State University,  University Park, PA, 16802, USA}

\author[0000-0002-6096-1749]{Cullen H. Blake}
\affiliation{Department of Physics \& Astronomy, University of Pennsylvania, 209 South 33rd Street, Philadelphia, PA 19104}

\author[0000-0001-7409-5688]{Guðmundur Stefánsson}
\affiliation{NASA Sagan Fellow}
\affiliation{Department of Astrophysical Sciences, Princeton University, 4 Ivy Lane, Princeton, NJ 08540, USA}

\author[0000-0001-9596-7983]{Suvrath Mahadevan}
\affil{\PSUAA}
\affil{\PSUCEHW}
\affil{\ETH}  

\author[0000-0001-6160-5888]{Jason T. Wright}
\affil{\PSUAA}
\affil{\PSUCEHW}
\affil{\PSETI}

\author[0000-0002-4927-9925]{Jacob K. Luhn}
\affil{\UCI}

\author[0000-0001-8720-5612]{Joe P.\ Ninan}
\affil{Department of Astronomy and Astrophysics, Tata Institute of Fundamental Research, Homi Bhabha Road, Colaba, Mumbai 400005, India}

\author[0000-0003-0149-9678]{Paul Robertson}
\affil{\UCI}

\author[0000-0001-8127-5775]{Arpita Roy}
\affiliation{Space Telescope Science Institute, 3700 San Martin Drive, Baltimore, MD 21218, USA}
\affiliation{Department of Physics and Astronomy, Johns Hopkins University, 3400 N Charles St, Baltimore, MD 21218, USA}

\author[0000-0002-4046-987X]{Christian Schwab}
\affiliation{School of Mathematical and Physical Sciences, Macquarie University, Balaclava Road, North Ryde, NSW 2109, Australia}

\author[0000-0002-4788-8858]{Ryan C. Terrien}
\affil{Carleton College, One North College St., Northfield, MN 55057, USA}


\date{\today}

\begin{abstract}

A significant barrier to achieving the radial velocity (RV) measurement accuracy and precision required to characterize terrestrial mass exoplanets is the existence of time-variable features in the measured spectra, from both telluric absorption and stellar variability, which affect measured line shapes and can cause apparent RV shifts. 
Reaching the desired accuracy using traditional techniques often requires avoiding lines contaminated by stellar variability and/or changing tellurics, and thus discarding a large fraction of the spectrum, lowering precision. 
New data-driven methods can help achieve extremely precise and accurate RVs by enabling the use of a larger fraction of the available data.
While there exist methods for modeling telluric features or the stellar variability individually, there is a need for additional tools that are capable of modeling them simultaneously at the spectral level.
Here we present \ssofjl (\ssofns), a \julia package for measuring Doppler shifts and creating data-driven models (with fast, physically-motivated Gaussian Process regularization) for the time-variable spectral features for both the telluric transmission and stellar spectrum, while accounting for the wavelength-dependent instrumental line-spread function. 
We demonstrate \ssofns's state-of-the-art performance on data from the NEID RV spectrograph on the WIYN 3.5m Telescope for multiple stars.
We show \ssofns's ,ability to accurately identify and characterize spectral variability and provide $\sim$2-6x smaller photon-limited errors over the NEID CCF-based pipeline and match the performance of \servalns, a leading template-based pipeline, using only observed EPRV spectra.
\end{abstract}


\section{Introduction}\label{sec:intro}


The radial velocity method continues to be one of the most productive methods for discovering and characterizing exoplanets.
Many extreme precision radial velocity (EPRV) spectrographs already exist and are taking data with instrumental precisions well below the \ms level \citep{fischer, wright2017, crass2021}.
One barrier to reaching the extreme RV precisions needed to detect terrestrial planets in the Habitable-zone around nearby stars with the RV technique
is telluric contamination \citep{wang2016, neiderror, Allart:2022}.
In ground-based spectroscopic RV observations, the final observed spectrum is a combination of the underlying stellar spectrum and the Earth's atmospheric spectrum.
%
%
If ignored, these telluric lines introduce systematic noise across stellar features and can cause large apparent RV signals.
Oftentimes, parts of the spectrum with deep telluric features are wholly masked out to avoid their effects, but this approach discards a significant amount of spectral information 
that could otherwise be used to increase RV precision. 
This also does not shield against the effects of microtellurics that are hard to reliably find and mask \citep{Allart:2022, cunha2014}.
Similarly, time-variable stellar features can also affect RV estimates (although at a smaller level) by changing line profiles and depths.
There have been many efforts to model telluric features from first principles with line-by-line radiative transfer models \citep[e.g. \texttt{TelFit}, \texttt{TAPAS}, \texttt{Molecfit}, \texttt{TERRASPEC},][]{gullikson2014,clough2005,bertaux2014,smette2015,bender2012}, 
hybrid approaches using synthetic spectra to initialize the fitting of identified lines
\citep[e.g. \emph{blas\'e, etc.,}][]{gully-santiago2022, Allart:2022}, or using data-driven models \citep[e.g. \wobblens, \texttt{YARARA}, etc.,][]{bedell, Cretignier2021, Kjaersgaard2021}
The new abundance of high-quality data from highly-stabilized EPRV instruments can enable us to learn the stellar and telluric spectra directly from the data, relying on the large Doppler shifts from the Earth orbiting the Sun ($\sim$ 30 \kms) which allow telluric and stellar spectral features to be separated with repeated observations. 



\begin{figure*}
    \centering
    \includegraphics[width=18cm]{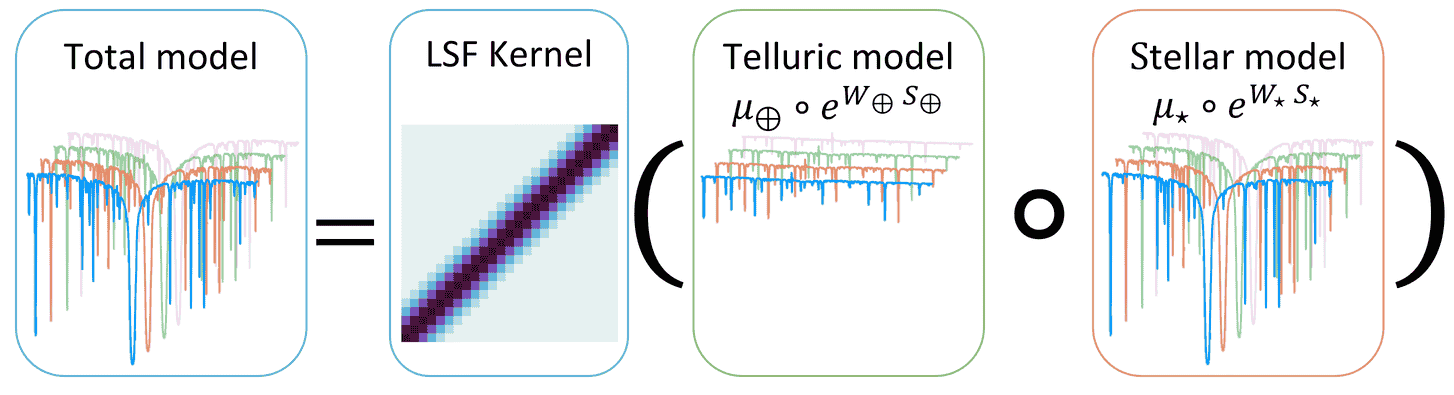}
    \caption{\ssof models are created by element-wise multiplying models for the telluric and stellar spectra and then multiplying by a a sparse matrix which approximates a convolution with the instrumental line-spread function. This flexibility is instrument-agnostic, and results in a robust, automated separation of telluric and stellar components}
    \label{fig:ssof}
\end{figure*}

Here we present \href{https://github.com/christiangil/StellarSpectraObservationFitting.jl}{\ssofjlns} (\ssofns).
\ssof is an open-source package for obtaining precise RV measurements and creating data-driven models (with fast, physically-motivated GP regularization and automated model selection) for the time-variable spectra of both the telluric transmission and the star  (Fig. \ref{fig:ssof}).
While \ssof was designed to provide improved relative radial velocities, it automatically derives a suite of additional data products, including reconstructed spectra in both the observer and barycentric frames, feature vectors that quantify the shapes and amplitudes for the temporal variability of time-variable telluric and stellar features, and weights or scores that can serve as data-drive stellar variability indicators.

\ssof builds on multiple related research and codes for RV extraction, and significantly advances the state of the art in multiple regards.
The three most relevant methods/software packages to compare to are \servalns, \wobblens, and Doppler-constrained Principal Components Analysis (DCPCA).
\serval formalized the approach of combining information from a spectroscopic timeseries for a single star to build a data-driven template spectrum that can produce high-quality RVs (as well as additional ``indicators'' that can be used to diagnose stellar variability).  
However, \serval does not model telluric absorption, so it must mask portions of the spectrum with significant telluric absorption.  
Thus, \serval is limited to analyzing portions of the spectrum where there is is minimal contamination from tellurics \citep{serval}.
As the precission of radial velocity instruments improves,  smaller tellurics become increasingly impactful.  
Avoiding all regions of the spectrum with microtelluric absorption features would result in throwing away large swaths of spectrum and lead to unnecessary loss of precision.  
This motivated the development of \wobble which pioneered the joint fitting of high-SNR stellar templates and time-variable telluric lines.
However, \wobble does not model stellar variability \citep{bedell}.  
Intrinsic stellar variability has become the most important barrier to RV surveys reaching the accuracy needed to characterize Earth-mass planets in the habitable zone of Sun-like stars \citep{crass2021}.
In paralle, Doppler-constrained Principal Components Analysis (DCPCA) provided a framework for analyzing spectral time series with stellar variability \citep{jones,gilbertson2020,Jones2022}.  
However, DCPCA does not account for telluric variability.
We desire to combine the best parts of each of these models to improve the accuracy and robustness of derived RVs and RV uncertainties.
This motivated the development of \ssofns. 


Including both stellar and telluric variability is the primary qualitative advance in \ssof over previous forward modeling approaches such as \servalns, \wobblens, and DCPCA.  
However, implementing such a model poses several significant additional technical challenges.
Indeed, \citet{bedell} identified several of these challenges as avenues for future research which have now been addressed by \ssofns.
First, in order to reduce the computational cost, \wobble assumes Gaussian noise in the logarithm of the observed flux.  \ssof improves on this by assuming Gaussian noise in linear flux.
Second, when combining the telluric and stellar components, \wobblens performs a simple multiplication.  
Since telluric absorption features can be much narrower than the instrument line spread function (LSF), this results in unnecessary inaccuracies.  
\ssof provides the capability to model the interaction of stellar and telluric features more accurately by  performing a convolution with a wavelength-dependent model for the instrumental LSF prior to comparing the model to observed data.
Third, there is a need to model the stellar variability in a way that avoids creating a degeneracy with true Doppler shifts.  
\ssof solves this by forcing the stellar feature vectors to be orthogonal to a Doppler feature vector. 
Thus, the scores for stellar feature vectors are insensitive to the true Doppler shift and can be used as stellar variability indicators for mitigating stellar variability \citep[as in DCPCA][]{gilbertson2020,Jones2022}.
In addition to recommendations from \citet{bedell}, \ssof has also improved the model regularization using fast GP-likelihood terms and improved the model selection process with an information-criterion based model search.
Thanks to the combination of these advances, the \ssof forward model is much more faithful to the physics of the observation process.
Thus, \ssof can be considered a physics-informed machine learning approach that seamlessly combines strong astrophysical knowledge about Doppler shifts, the Earth's rotation and orbit around the Sun, radiative transfer, and instrument optics and with a machine learning approach for characterizing stellar and telluric variability.\footnote{While we regard \ssof as an example of physics-informed machine learning, one could reinterpret the model in the form of a neural network or an autoencoder with non-standard transfer functions that have been customized to faithfully incorporate the aforementioned physics.}
Two major challenges were to implement such a sophisticated model in a computationally efficient manner and to provide the ability to compute gradients (and the Hessian) of the likelihood via auto-differentiation.  
\ssofjl provides a performant, open-source package for meeting all of these needs, performing inference with the \ssof model, and measuring precise RVs. 

There have been several other studies developing advanced models and software for analyzing ensembles of stellar spectra, 
each with their own benefits and tradeoffs.
Several studies have analyzed ensembles of spectra to effectively deconvolve either the stellar or telluric spectrum \citep[e.g.,][]{czekala2015,lienhard2022,gully-santiago2022,Kjaersgaard2021}.
Some of these use synthetic spectral models for line profile shape and/or assumed a shared line profile shape to improve measurement of line parameters \citep[e.g.,]{gully-santiago2022,czekala2015,serval}.
Many other studies have attempted to improve the accuracy of RVs by post-processing the RVs, selected stellar variability indicators, and/or the cross correlation functions (CCF) of the stellar spectrum and a given CCF mask.
For example, \citet{deBeurs2022} compared linear regression, dense, and convolutional neural networks for post-processing of  to reduce stellar variability.  
Similarly, \citet{scalpels} applied singular value decomposition and linear models to the autocorrelation function of the CCF to clean RVs of stellar variability.   
While many of these can improve the RMS RV for observations from an exoplanet survey, different methods can suggest different ``corrections'' \citep{zhao2022}.
Thus, further research is needed to establish the accuracy and reliability of such methods.  
There is also concurrent research on alternative methods which aim to measure RVs in a line-by-line framework \citep{artigau2022,cook2022,Burrows2024,Sigel2024} and show significant promise.
Finally, several other studies have focused on analyzed ensembles of spectra of different stars  \citep[e.g.,][]{Sedaghat2021,SPCANet,deMijolla2021}, rather than prioritizing the highest possible precision RVs for a modest number of stars each targeted many times by precision radial velocity surveys.
While there have been a variety of other studies that apply various forms of statistical inference, machine learning, and neural networks to the analysis of stellar spectroscopic timeseries, \ssof occupies a unique position and provides state-of-the-art capabilities for modeling spectroscopic timeseries including both telluric and stellar variability, in support of precision radial velocity planet surveys. 

This paper begins by describing the mathematical models underlying \ssofns, their implementation, and how all portions of the model can be fit simultaneously to measured spectra (\secref{sec:methods}).
In Section \ref{sec:validation}, we show the ability of the \ssof analysis framework to recover accurate models for self-simulated data. 
In Section \ref{sec:application}, we demonstrate \ssofns's capabilities on data from NEID \citep[R $\sim$ 120,000, 3800-9300 \AA;][]{halverson, schwab}, a stabilized high resolution EPRV spectrograph on the WIYN 3.5m Telescope, for three common EPRV target stars: the moderately active K dwarf HD 26965, planet-hosting K dwarf HD 3651, and the M dwarf Barnard's Star.
In Section \ref{sec:discuss}, we summarize \ssofns's performance and discuss \ssofns's limitations, avenues for improvement, and compare it with some other EPRV pipelines.

\section{Methods}\label{sec:methods}

To build the data-driven model from the as-observed spectra, we take the data to be $Y_D$, which is a $P \times N$ matrix where each entry $Y_{D,n,p}$ is the reduced 1D spectral flux (\secref{ssec:preprocess}) for pixel $p$ of $P$ at observation $n$ of $N$.
In this work each $Y_{D}$ only includes data from a single spectral order, but it is possible to combine the data from multiple orders to simultaneously model larger portions of the observed spectrum. 
The \ssof model assumes that each column (i.e. each observed spectrum), $Y_{D,n}$, can be predicted to be $Y_{M,n}$, the multiplication of a telluric transmission spectrum ($Y_{\oplus,n}$) and a stellar spectrum ($Y_{\star,n}$) combined with a white noise term $\epsilon$. 
\begin{equation}
	Y_{M,n} =  Y_{\oplus,n} \circ Y_{\star,n} + \epsilon
	\label{eq:basic}
\end{equation}
where $\circ$ denotes the Hadamard (i.e. element-wise) product.\footnote{In this form, we are implicitly requiring evaluating the telluric and stellar models at the resolution of the observed data. See \secref{sssec:improvements} for a discussion on improving on this limitation}.
This is formally incorrect for modeling spectra at the resolution of typical EPRV spectrometers as this does not include important instrumental effects such as the line-spread function (LSF) which occur after the spectra are multiplied.
Thus, the base \ssof model (Eq. \ref{eq:basic}) can be expanded to include such instrumental effects by including an additional matrix multiplication as shown below.
\begin{equation}
	Y_{M,n} = T_{I,n} \cdot(Y_{\oplus,n} \circ Y_{\star,n}) + \epsilon
	\label{eq:LSF}
\end{equation}
In this study, we use $T_{I,n}$ to 
approximate the effects of convolving the spectra with the instrumental LSF (see Appendix \secref{app:lsf} for details on NEID's LSF).
These $T_{I}$ can be different for each observation, allowing for time-variable LSF models to be included if desired. 
We have found that the inclusion of this additional sparse matrix operation increases inference times by less than 20\% for NEID-like LSFs, even when $T_{I}$ is different at each time. 

\subsection{Telluric Model}
\ssofns's default telluric transmission spectrum is defined as follows:
\begin{equation}
	Y_{\oplus,n} = T_\oplus(\xi_{D,n}, \xi_\oplus) \cdot Y'_{\oplus,n}
	\label{eq:tel}
\end{equation}
\begin{equation}
    Y'_{\oplus,n}=\mu_\oplus \circ \text{exp.}(W_\oplus \cdot S_{\oplus,n})
    \label{eq:logt}
\end{equation}
where $Y'_{\oplus,n}$ is the telluric model\footnote{\ssof can generally replace log-linear models with linear models (i.e. $Y'_{\oplus,n}=\mu_\oplus + W_\oplus \cdot S_{\oplus,n}$ instead of Eq. \ref{eq:logt})}  at observation $n$ evaluated at a constant set of uniformly-spaced model log wavelengths $\xi_\oplus$.
$T_\oplus$ convolves $Y'_{\oplus,n}$ over the wavelength range of each pixel.
The telluric model has free parameters $\beta_\oplus=\{\mu_\oplus, W_\oplus, S_\oplus\}$.
$\mu_\oplus$ is a spectral template of fluxes, 
$W_\oplus$ is a matrix of feature vectors, $S_{\oplus,n}$ is a vector of scores,
and $\text{exp.}(A)$ denotes element-wise exponentiation (i.e. $(\text{exp.}(A))_{i,j} = \text{exp}(A_{i,j})$).
$W_\oplus$ controls the spectral shape of the temporal variations of the telluric spectrum and is a $J_\oplus \times K_\oplus$ matrix where $J_\oplus$ is the length of the vector of template wavelengths, $\lambda_\oplus$, and $K_\oplus$ is the number of feature vectors being used (if $K_\oplus=0$ then $Y_{\oplus,n}'=\mu_\oplus$ and $\beta_\oplus=\{\mu_\oplus\}$).
$S_{\oplus,n}$ controls the amplitude of the temporal variations of the telluric spectrum and is the $n^{th}$ column of a $K_\oplus \times N$ matrix.
$T_\oplus$ is a constant $P \times J_\oplus$ sparse matrix that is constructed to perform integration of the model contained by each observed pixel assuming a top-hat pixel response\footnote{$T_\oplus$ can alternatively be constructed to perform a linear interpolation} 
\begin{equation}
	Y_{\oplus,n,p} = \int_{\dfrac{\xi_{D,n,p-1}+\xi_{D,n,p}}{2}}^{\dfrac{\xi_{D,n,p}+\xi_{D,n,p+1}}{2}} Y'_{\oplus,n} \,d\xi_\oplus
	\label{eq:tophat}
\end{equation}
which is approximated with trapezoidal integration.
This could be modified to include the actual pixel bounds which, in reality, do not cover the entirety of the detector and/or to account for intrapixel sensitivity variations (if sufficient calibration data are available).

\subsection{Stellar Model}
\ssof can define the stellar contribution (and calculate radial velocity shifts) in two qualitatively different ways which share the same general form:
\begin{equation}
	Y_{\star,n} = T_\star(\xi_{D,n}, \xi_\star, z_n) \cdot Y'_{\star,n}
	\label{eq:star}
\end{equation}
where $Y'_{\star,n}$ is a stellar model evaluated at observation $n$ on a constant set of uniformly-spaced model log wavelengths $\xi_\star$.
$Y'_{\star,n}$ has free parameters $\beta_\star=\{\mu_\star, W_\star, S_\star, z_{\star}\}$ (or $\beta_\star=\{\mu_\star, z_{\star}\}$ if $K_\star=0$), and $T_\star(\xi_{D,n}, \xi_\star, z_n)$ is a linear operator that interpolates the model fluxes $Y'_{\star,n}$ onto the observed log wavelengths, $\xi_{D,n}$, based on the total Doppler shift, $z_n$.

\subsubsection{Variable Interpolation Location}
In the variable interpolation location (VIL) method, we directly model how much we are shifting the stellar wavelengths (similar to \wobblens) by including the free parameter for the system's radial velocity shifts at each time (${z_\star,n}$) with the known observer barycentric-correction shifts \citep[$z_{D,n}$,][]{barycorrpy,barycorr} in the total Doppler shift ($z_n \approx z_{\star,n}+z_{D,n}$). 
The stellar model is defined as
\begin{equation}
    Y'_{\star,n}=\mu_\star \circ \text{exp.}(W_\star \cdot S_{\star,n})
\end{equation}
where $\mu_\star$ is a spectral template of fluxes, $W_\star$ is a matrix of feature vectors, and $S_{\star,n}$ is a vector of scores at observation $n$.
$W_\star$ controls the spectral shape of the temporal variations of the stellar spectrum and is a $J_\star \times K_\star$ matrix where $J_\star$ is the length of the vector of template log wavelengths, $\xi_\star$, and $K_\star$ is the number of feature vectors being used (if $K_\star=0$ then $Y_{\star,n}'=\mu_\star$).
$S_{\star,n}$ controls the amplitude of the temporal variations of the stellar spectrum and is the $n^{th}$ column of a $K_\star \times N$ matrix.
Because it needs to be recomputed for each set of proposed radial velocities, $T_\star(\xi_{D,n}, \xi_\star, z_n)$ is simplified and defined to perform a linear interpolation such that

\begin{equation}
    Y_{\star,n,p} = (1-\alpha) \ Y'_{\star,n,j-1} + \alpha \ Y'_{\star,n,j}
\end{equation}
\begin{equation}
    \alpha=\dfrac{\xi_{D,n,p} + D(z_n)- \xi_{\star,j}}{\Delta\xi_\star}
\end{equation}
\begin{equation}
    \Delta\xi_\star=\xi_{\star,j}-\xi_{\star,j-1}
\end{equation}
where $\xi_{\star,j-1}$ and $\xi_{\star,j}$ are the model log wavelengths that surround $\xi_{D,n,p}$ after accounting for the Doppler shift $D(z_n)=\text{log}\left(1+z_n\right)$, $\Delta\xi_\star$ is the difference between the neighboring model log-wavelengths (which is constant for evenly-spaced $\xi_\star$), $\alpha$ is the fraction of $Y'_{\star,n,j}$ that should contribute to $Y_{\star,n,p}$ 
Using solely $Y_{\star,n,p}$ to infer $z$ (while neglecting telluric transmission and LSF effects) is conceptually similar to template matching \citep{anglada2012, Astudillo-Defru2015, serval, silva2022, almoulla2022}, where we are learning the template directly from the data.
The VIL method is the method that \ssof defaults to using.

\subsubsection{Doppler Component}
In the Doppler component method, we use the fact that spectral changes from small Doppler shifts can be effectively expressed as a Taylor expansion of the mean spectrum \citep{jones,Jones2022}. Therefore we account for the relatively small additional radial velocity shifts ($z_{\star}$) separately from the observer barycentric-correction shifts ($z_D$) with an additional component to the stellar model,
\begin{equation}
    Y'_{\star,n} = \mu_\star \circ \text{exp.}(W_\star \cdot S_{\star,n}) + W_{RV} \cdot S_{RV,n}
    \label{eq:mrv}
\end{equation}
where $W_{RV} = \lambda_\star \circ \dfrac{d\mu_\star}{d\lambda_\star}$ ($\lambda_\star=\text{exp.}(\xi_\star)$ are the model wavelengths) is a $J_\star \times 1$ matrix which approximates the effects of a small RV shift.
$d\mu_\star/d\lambda_\star$ can be effectively estimated with finite differences.
$S_{RV,n}$ controls the amplitude of the approximate Doppler shift at time $n$, is proportional to $z_{\star,n}$, and is the $n^{th}$ element of a $1 \times N$ matrix.
Unlike in the VIL method, this basis vector approximation of Doppler shift only includes information from $\mu_\star$ and thus is, in principle, less accurate when used to measure the RV in parts of the stellar spectrum whose slopes are changing significantly over time.
Because the large, barycentric-correction Doppler shifts are already being accounted for, $T_\star(\xi_{D,n}, \xi_\star, z_n)=T_\star(\xi_{D,n}, \xi_\star, z_D)$ and does not have to be recalculated for different $z_{\star}$. $T_\star$ can either be constructed to perform a linear interpolation (as in the VIL method) or an integration of the model contained by each observed pixel (similar to $T_\oplus$).
\begin{equation}
	Y_{\star,n,p} = \int_{\dfrac{\xi_{D,n,p-1}+\xi_{D,n,p}}{2}+D(z_n)}^{\dfrac{\xi_{D,n,p}+\xi_{D,n,p+1}}{2}+D(z_n)} Y'_{\star,n} \,d\xi_\star
	\label{eq:tophat2}
\end{equation}

\subsection{Model Optimization}\label{ssec:opt}

The \ssof model is optimized by minimizing a modified negative log-likelihood loss (objective) function.

\begin{equation}
\begin{aligned}
	\mathcal{L}(\beta_M) = \sum_{n=1}^N (Y_{D,n} - Y_{M,n})^T \Sigma_n^{-1} (Y_{D,n} - Y_{M,n}) + \textrm{constant}
	\label{eq:loss}
\end{aligned}
\end{equation}
where $\beta_M = \{\beta_\oplus,\beta_\star\}$ is the set of model parameters and $\Sigma_n$ are the covariance matrices of each observation (in this case, diagonal matrices populated by  $\sigma_{D,n}^2$).
This objective function is easily tractable, and we use \nablajl \citep{nabla} one of \julians's many excellent automatic differentiation packages to get gradients of $\mathcal{L}$ with respect to $\beta_M$. 
We chose \nablajl due to its graceful, fast performance in the presence of sparse matrix operations.
We then use a custom implementation of the ADAM optimizer \citep{adam} to efficiently update the model with some additions including preventing $\mu_\oplus$ and $\mu_\star$ from having negative values and ensuring that each vector in $W_\star$ is orthogonal to a Doppler shift (see $W_{RV}$ below \eqref{eq:mrv}).
While there is enough flexibility in the model that finding the globally optimum model is not practical, finding one of the many models that are good enough is quite achievable.
In practice, with a reasonable starting point, most of the model improvement occurs within 100 ADAM update iterations.
We also use L-BFGS \citep{lbfgs} implemented in \optimjl \citep{Mogensen2018} to quickly find a local optimum for the model scores $S_\oplus$, $S_\star$, and $z_{\star}$.
For this step we replace $\sigma_{D}$ with a more conservative version which ensures that any parts of the spectrum that are masked in the stellar frame at any time are always masked (see \secref{ssec:preprocess}).
This prevents the RVs from having a changing bias from seeing different stellar lines at different times.
Whenever we say that we ``optimize $\beta_M$" throughout the paper, we typically mean 100 ADAM update iterations of all model parameters followed by using L-BFGS to finalize $S_\oplus$, $S_\star$, and $z_{\star}$ unless noted otherwise.

\subsection{Model Initialization and Selection}\label{ssec:init}

\begin{figure*}
\centering
\includegraphics[width=18cm]{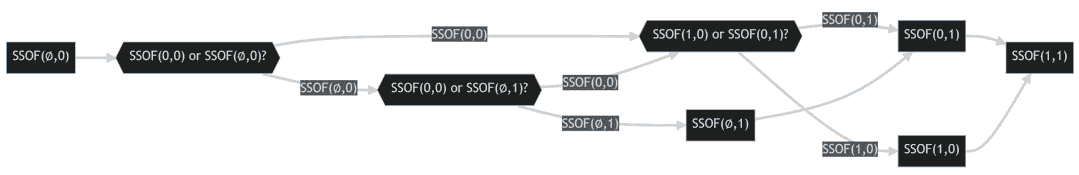}
\caption{Flowchart for the \ssof model searching process up to $(K_\oplus\leq1, K_\star\leq1)$ model components. Each hexagonal cell indicates the comparison of two different proposed \ssof models. Starting from the basic \ssofns($\varnothing$,0) model (with no telluric template or feature vectors and no stellar feature vectors), we repeatedly check whether adding a telluric or stellar model component is ``better" (which in our case we take to be the one with a lower AIC) and add the corresponding component until we reach the desired final model complexity. \ssofns($\varnothing$,1), \ssofns(0,1), and \ssofns(1,0) lead directly to one other \ssof model because they are already at one of the specified feature vectors limits. Once the search is completed up to \ssofns($K_\oplus, K_\star$), the best \ssof model found along the searched path is used for further analyses.}
\label{fig:mermaid}
\end{figure*}

An initial, empty \ssof model is created by assuming evenly sampled (in log-space) template telluric wavelengths $\xi_\oplus$ and template stellar wavelengths $\xi_\star$ that encompass all of the observed wavelengths, and a maximum number of feature vectors to be used in the stellar, $K_\star$, and telluric, $K_\oplus$, components of the model.
We 
searched up to $(K_\oplus\leq5, K_\star\leq5)$ in this work. The amount of useful feature vectors that can be found using the data is limited by the variance in the spectra, the number of observations, and the observation SNR\footnote{In our analysis of three stars (\secref{sec:application}), each with 112 measured spectral orders, we found only 5 orders used in the final RV reduction with $K_\oplus>3$ (all trying to explain H$_2$O line variability from 6821-8362 \AA) and a single order with $K_\star>3$ (for Barnard's Star from 6670-6800 \AA)}. We also investigated the best \ssof models with no stellar variability (searching up to $(K_\oplus\leq5, K_\star=0)$) and no variability (searching up to $(K_\oplus=0, K_\star=0)$). We replaced the normal \ssof model with one of these simpler models (indicated with an asterisk in the Tables \ref{tab:26965}, \ref{tab:3651}, and \ref{tab:barnard}) when the Doppler shifts from the simpler models were more consistent with those measured by all of the other orders.
\ssof models are built up component-by-component, choosing each additional component to maximally improve the model performance.
Fig. \ref{fig:mermaid} outlines the decision logic for selecting model complexity, using quantitative metrics (see below).

The most basic \ssof model uses only $\mu_\star$, having no time variability and no telluric features.
We will indicate this basic model as \ssofns($\varnothing$,0) where \ssofns($K_\oplus$,$K_\star$) is a \ssof model with $K_\oplus$ telluric feature vectors and $K_\star$ stellar feature vectors, and $\varnothing$ further indicates no telluric model (i.e. $\mu_\oplus=\mathbbm{1}$).
We start by estimating an initial guess for $\mu_\star$ by taking the weighed mean of $Y_D$ at $\xi_\star$ (after interpolating\footnote{We perform the interpolations during model initialization by evaluating the posterior mean and variance of a Gaussian Process (GP) with a Mat\'ern $^5/_2$ kernel (quickly using \tgpjlns) conditioned on the observed data at each time. The lengthscale for the kernel defaults to one based on a fit to the quiet solar spectrum provided with \texttt{SOAP 2.0} \citep{Dumusque2014}, although it can be changed by the user.}
each observation from their barycentric corrected $\xi_{D,n}$).
This base model (with only $\mu_\star$) is then optimized (see \secref{ssec:opt}) and several metrics are stored for later model comparison.
These metrics include Akaike information criterion \citep[AIC,][]{aic}, Bayesian Information Criterion \citep[BIC,][]{bic}, RV root mean square (RMS), and intra-night RV RMS.
The metrics are calculated and stored for all models evaluated during the initialization.

Adding some model complexity, we evaluate a constant \ssof model with only $\mu_\oplus$ and $\mu_\star$ (\ssofns(0,0)).
We estimate an initial guess for $\mu_\oplus^{(0)}$ by taking the weighed mean of $Y_D$ at $\xi_\oplus$ (after interpolating each observation from $\xi_{D,n}$).
This initial $\mu_\oplus^{(0)}$ is only used to determine which portions of the measured spectra are dominated by telluric features.
We do this by comparing the pixel-wise loss, $\mathcal{L}$ (from Eqn.~\eqref{eq:loss}) when using only $\mu_\oplus^{(0)}$ or $\mu_\star$ to model the observed spectra.
Pixels with lower $\mathcal{L}$ when using only $\mu_\oplus^{(0)}$ are ``telluric-dominated" and the remaining are ``stellar-dominated". 
The constant \ssof model's $\mu_\star$ is initialized by taking the weighed mean of the stellar-dominated portion of $Y_D$.
This model's $\mu_\oplus$ is initialized by taking the weighed mean of $Y_D \oslash Y_\star$ ($Y'_{\star,n}=\mu_\star$ at this point) after interpolation onto $\xi_\oplus$ where $\oslash$ denotes Hadamard (i.e. element-wise) division. 
Finally, $\mu_\star$ is updated by taking the weighed mean of $Y_D \oslash Y_\oplus$ ($Y'_{\oplus,n}=\mu_\oplus$ at this point) after interpolation onto $\xi_\star$.
The \ssofns(0,0) model is then optimized.

We choose to build on the better of the \ssofns($\varnothing$,0) or \ssofns(0,0) model based on the stored metrics.
We used AIC for this decision making as we have found that models with the lowest AIC tend to adequately balance model complexity and performance, although other metrics could be used for this purpose. 
From this point on we continue to build on the \ssof model by checking if the model is improved more by adding a telluric or stellar component (either a feature vector, or $\mu_\oplus$ if it isn't being used yet), and storing the potential models and their metrics along the way.
For example, assuming we chose to build on the \ssofns(0,0) model in the previous step, the next step would check the performance of the \ssofns(1,0) and \ssofns(0,1) models and then choose to build off whichever of those is better.
This procedure is repeated until we end up with a \ssofns($K_\oplus$,$K_\star$) model that has the amount of feature vectors that it was desired to search up to.
We choose the final \ssof model to be the minimum-AIC model among those searched along the path to \ssofns($K_\oplus$,$K_\star$).

We use noise-weighted Expectation Maximization (EM) Principal Component Analysis (PCA) \citep[][re-implemented in \julia to be 5-10x faster in \texttt{ExpectationMaximizationPCA.jl} \citep{empcajl}]{bailey} to get initial guesses for our feature vectors. 
EMPCA is an iterative algorithm that, like PCA, produces a series of orthogonal basis vectors organized by amount of variance explained in the data but, unlike PCA, is less sensitive to data points whose variance is caused primarily by noise.
This is critical for our application so that our initial guesses for our feature vectors are not distracted by the high variance pixels near the edge of orders.
We have also augmented EMPCA into Doppler-constrained EMPCA (DEMPCA) by adding a preprocessing step of taking out the Doppler basis component described in \citet{jones}.
This allows us to get estimates for the Doppler shift of the spectra and keep the initial feature vectors orthogonal to a Doppler shift.
Each proposed feature vector is initialized by performing (D)EMPCA on the log of $Y_D \oslash Y_B$ interpolated onto $\xi_A$ further divided by $Y'_A$, 
where $Y'_A$ is the current version of the model whose new feature vector is being proposed and $Y_B$ is the current version of the other model (interpolated to the $A$'s log wavelengths) and $A$ and $B$ are either $\oplus$ or $\star$ respectively.

\subsection{Regularization Choice}\label{ssec:reg}

The \ssof model chosen in the procedure from \secref{ssec:init} can be further improved by encouraging sparsity and smoothness of the templates and feature vectors with various regularization terms. We do this by adding a combination of L-norm and GP regularization terms to $\mathcal{L}$.
\begin{equation}
\begin{aligned}
	\mathcal{L}_{R}(\beta_M, \beta_R) & =  \sum_{n=1}^N (Y_{D,n} - Y_{M,n})^T \Sigma_n^{-1} (Y_{D,n} - Y_{M,n}) \\ 
		& + a_1 \ell_{\textrm{LSF}}(\xi_\oplus, \mu_\oplus - 1) + a_2 \ell_{\texttt{SOAP}}(\xi_\star, \mu_\star - 1) \\
		& + a_3||\mu_\oplus||_2^2 + a_4||\mu_\star||_2^2 \\ 
		& + a_5||\mu_\oplus||_1^1 + a_6||\mu_\star||_1^1\\ 
		& + a_7 \sum_i^{K_\oplus} \ell_{\textrm{LSF}}(\xi_\oplus, W_{\oplus,i}) + a_8 \sum_i^{K_\star} \ell_{\texttt{SOAP}}(\xi_\star, W_{\star,i}) \\ 
		& + a_9||W_\oplus||_2^2 + a_{10}||W_\star||_2^2 \\
		& + a_{11}||W_\oplus||_1^1 + a_{12}||W_\star||_1^1 \\
		& + ||S_\oplus||_2^2 + ||S_\star||_2^2
	\label{eq:reg}
\end{aligned}
\end{equation} 
where $\beta_R=\{a_1, a_2, ..., a_{11}, a_{12}\}$ are the model hyperparameters controlling the regularization amplitudes,
$||x||_1^1 \equiv \sum_i |x_i|$ is a L1 regularization which encourages model sparsity for the parameters $p$ being normalized,
$||x||_2^2 \equiv \sum_i x_i^2$ is a L2 regularization which discourages outliers for the parameters $p$ being normalized,
and $\ell(\xi, x)$ is the GP log-likelihood of the vector of parameters $x$ using $\xi$ as the spacing between them which encourages nearby points in $x$ to be correlated and suppresses overall $x$ amplitude.
$\ell_{\textrm{LSF}}$ corresponds to using a GP with a Mat\'ern $^5/_2$ kernel whose hyperparameters were optimized to mimic the wavelength covariance in synthetic LSF lines (and is different for each spectral order) and $\ell_{\texttt{SOAP}}$ corresponds to using a GP with a Mat\'ern $^5/_2$ kernel whose hyperparameters default to ones derived from training on the clean solar spectrum used in \citet{soap} (the same one used for GP interpolation in \secref{ssec:init}), but can be changed to adapt to stars with different rotational broadening than the Sun.
The main difference between these two GPs is their length-scale, which is $\propto$ the solar line width for $\ell_{\texttt{SOAP}}$ and $\propto$ the LSF width ($\sim$ 7 times smaller for NEID) for $\ell_{\textrm{LSF}}$.
Both length-scales can be changed on a per-\ssof model basis by the user, to account for things such as different stellar rotational velocities (and therefore line widths) and changing LSF widths.
Evaluating $\ell$ with naive GP regression algorithms would scale as $\mathcal{O}(N^3)$ where $N$ is the length of the input data, but inference with GPs of certain kernels (including Mat\'ern $^5/_2$) can be done much more quickly exploiting the fact that they can be equivalently expressed as state-space models which solve a certain stochastic differential equation \citep[amongst other ways, ][]{sarkka2019applied,celerite1,celerite2}.
Expressing them this way allows us to use much faster $\mathcal{O}(N)$ Kalman filtering methods for inference.
\ssof has implemented this methodology to efficiently evaluate the GP regularization (see \secref{app:gpreg} for implementation details).

The optimal $\beta_R$ values are found via cross-validation testing.
We start by optimizing the model with all $a_i=0$, and split 67\% of the data into a training set and 33\% into a testing set.
Then, for each $a_i$ in turn starting with $a_1$:
\begin{enumerate}
\item Turn on $a_i$ to a low, high, and moderate values (1e-3, 1e12, and a value in between that is different for each $a_i$ (Tab. \ref{tab:betas}))
\item for each proposed $a_i$, optimize $\beta_M$ using the training data, then optimize $S_\oplus$, $S_\star$, and $z_{\star}$ on the testing data and store the final testing $\mathcal{L}$
\item Starting from the proposed $a_i$ with the lowest testing $\mathcal{L}$, repeat step 2 proposing a higher and lower $a_i$ (by multiplying and dividing $a_i$ by a factor of 10 to quickly get near an optimal value) until a local minimum is found in the testing $\mathcal{L}$, or $a_i$ is outside of the search space
\end{enumerate}

\begin{table*}[]
\begin{tabular}{l|llllllllllll}
 & $a_1$ & $a_2$ & $a_3$ & $a_4$ & $a_5$ & $a_6$ & $a_7$ & $a_8$ & $a_9$ & $a_{10}$ & $a_{11}$ & $a_{12}$ \\ \hline
Default $\beta_R$ & 1e6 & 1e2 & 1e6 & 1e-2 & 1e5 & 1e1 & 1e7 & 1e4 & 1e4 & 1e4 & 1e7 & 1e7
\end{tabular}
\caption{Moderate $\beta_R$ values used in the first stage of regularization optimization via cross-validation. These are typical $\beta_R$ values found in early testing of \ssofns on HD 26965.}
\label{tab:betas}
\end{table*}

\subsection{Error Estimation}\label{ssec:error}
Once the model has been selected and regularized, the last step in the analysis procedure is to determine uncertainty estimates for $\beta_M$. 
A quick way to find variance estimates for small amounts of parameters (like the model scores and RVs) is to look at the curvature of the loss space.
If one assumes that loss is approximately a Gaussian log-likelihood, then the covariance matrix, $\Sigma_{\beta_M}$ can be approximated as 
\begin{equation}
	\Sigma_{\beta_M} \approx (-H(\beta_M))^{-1}
    \label{eq:hessian_to_cov}
\end{equation}
where $H(\beta_M)_{i,j}=\dfrac{\delta^2 \ell(\beta_M)}{\delta \beta_{M,i} \delta \beta_{M,j}}$ is the Hessian matrix and $\ell(\beta_M)$ is the $\approx$ Gaussian log-likelihood (which in our case is $\dfrac{-\mathcal{L}}{2}$).
The variance of each model parameter can be further approximated assuming that the off-diagonal entries of $H(\beta_M)$ are zero (i.e. assuming any $\beta_{M,i}$ is uncorrelated with $\beta_{M,j}$)
\begin{equation}
	\dfrac{1}{\sigma_{\beta_{M,i}}^2} \approx -\dfrac{\delta^2 \ell(\beta_M)}{\delta \beta_{M,i}^2}
\end{equation} 
We effectively approximate $\dfrac{\delta^2 \ell(\beta_M)}{\delta \beta_{M,i}^2}$ with finite differences.
This method is very fast and recommended when performing repeated, iterative analyses (e.g. during data exploration or survey simulation).

Another method available in \ssof for estimating errors is via bootstrap resampling.
In this method, we repeatedly refit the model to the data after adding white noise to each pixel at the reported variance levels.
An estimate for the covariance of $\beta_M$ can then be found by looking at the distribution of the proposed $\beta_M$ after the repeated refittings.
These estimates for the uncertainties tend to be $\sim 1.1-1.5$x higher than the loss space curvature based estimates (likely due to the ignored off-diagonal terms in $H(\beta_M)$).
This method is slower but gives a better overall estimate for the uncertainties (and covariances if desired) and is recommended when finalizing analysis results.
The \ssof model is complete once it has been initialized, regularized, and the uncertainties are estimated. 

\subsection{Combining Order Analyses}\label{ssec:combine}
If we assume that each \ssof order models' Doppler shift measurements at a given time are independent normally distributed estimates of the same Doppler shift (i.e. they have the same mean), then the maximum likelihood estimator of the Doppler shift is their weighted mean.
\begin{equation}
    \widehat{z_{\star,n}} = \dfrac{\sum \dfrac{z_{\star,n,j}}{\sigma^2_{z_{\star},n,j}}}{\sum \dfrac{1}{\sigma^2_{z_\star,n,j}}}, \ 
    \widehat{\sigma_{z_\star,n}} = \sqrt{\dfrac{1}{\sum \dfrac{1}{\sigma^2_{z_\star,n,j}}}}
\end{equation} 
where $z_{\star,n,j}$ is the Doppler shift estimate for observation $n$ from the \ssof model of order $j$, $\sigma_{z_\star,n,j}$ is its corresponding uncertainty, and the sums are over the included orders, $j$.
In principle, this isn't strictly true as spectral orders often overlap and there could be a spurious RVs from model misspecification for a given part of the stellar spectrum that could cause the RVs from \ssof models of these adjacent orders to be correlated.
To maximize the overall RV measurement precision (i.e. minimize $\widehat{\sigma_{z_\star}}$) we wish to include as much information (and therefore as many \ssof order models) as is reasonable.

We start building our RVs by using a subset of all of the orders that have high throughput (and therefore SNR) and are largely devoid of telluric features and stellar variability.
\ssof is robustly able to identify and model these relatively uncomplicated regions of spectra, which we refer to as ``key orders".
We use these ``key orders" to assess how well \ssof modeled the remaining, more complicated regions of the spectra.
We first get an initial, conservative Doppler shift estimate, $z^{KO}_{\star}$, by combining the Doppler shift measurements from the AIC-minimum \ssof models for the ``key orders" (KO).
We then identify the single order \ssof models that produced Doppler shifts with RMS lower than twice the RMS of $z^{KO}_{\star}$ ($\textrm{RMS}(z_{\star,j}) < 2 \ \textrm{RMS}(z^{KO}_{\star})$), uncertainty estimates within a factor of 4 of the median uncertainty in the key orders ($\overline{\sigma_{z_\star,j}} < 4 \ \textrm{med}(\overline{\sigma_{z_\star,i\in KO}})$), 
and are consistent\footnote{where consistent means that, on average, the difference between the single \ssof order model Doppler shifts and the combined ``key order" Doppler shifts was less than two standard errors (i.e. $\overline{\chi^2_j} = \dfrac{\sum_{n=1}^N \left( \dfrac{z^{KO}_{\star,n}-z_{\star,n,j}}{\sigma_{z_\star,n,j}}\right)^2}{n} \leq 4$) where $j$ is the echelle order and $n$ is the observation index} with $z^{KO}_{\star}$ and include them in the final Doppler shift reduction, $z^{\ssofns}_{\star}$ and $\sigma^{\ssofns}_{z_\star}$.
Only one \ssof model is included per order.
It is typically the AIC-minimum \ssof model for a given order but we sometimes replace it with a \ssof model with no stellar variability.

\section{Verification \& Validation}\label{sec:validation}

%
For real-world applications to stars (other than the Sun), the true stellar RV is always unknown.  
This makes many of the typical methods for verification and validation of machine learning models impractical. 
Instead, we perform verification and validation tests on synthetic data sets for which the true parameters are known.  
This allows us to evaluate the performance of SSOF under idealized conditions where the data to be analyzed has been generated from the SSOF model itself.  
We show that \ssof reliably converges to accurate values for the RVs and the reconstructed stellar and telluric spectra.  

First, we generated synthetic observations using the SSOF model fit to actual NEID observations of HD 26965 (see \S\ref{ssec:26965}).  Then, we applied SSOF to the synthetic observations.  
For these tests, we started by simulating 100 observations spread uniformly over one year.  
Each observation is generated using the mean stellar spectrum, mean telluric spectrum, and any feature vectors derived from the SSOF analysis of HD 26965 in \S\ref{ssec:26965}.  
The number of feature vectors to be used for each order is set based on the number of feature vectors which minimized the AIC for the actual data.  
Therefore, many orders include temporal variability, but some others orders do not.
The scores for both telluric and stellar feature vectors were drawn from a normal distribution with mean and variance based on the mean and standard deviation of the scores for the analysis of NEID's observations of HD 26965.
The stellar spectra include $\sim 10$km s$^{-1}$ Doppler shifts due to the Earth's motion around the sun, but no planetary signals.  
We add Gaussian noise to the simulated observations with the SNR as a function of order and pixel based on the SNR for actual NEID observations of HD 26965. 
For orders where the NEID LSF has been measured from LFC calibration data, we use the LSF width model for that order.
For blue orders where the LFC is not available, we adopt the same LSF from Echelle order 118. 

In the next step, we apply SSOF to the synthetic data sets, just as we would apply it to real observations, treating each order independently.  
We assume that the barycentric correction and LSF width are known precisely.
However, we let SSOF estimate the stellar and telluric spectra, as well as a putative RV, for each observation without knowledge of the true input values.   
To summarize the quality of the reconstructed spectra, we compute the RMS residual RV between the input spectrum and RV measured by SSOF.

\begin{figure*}[ht]
\centering
\includegraphics[width=18cm]{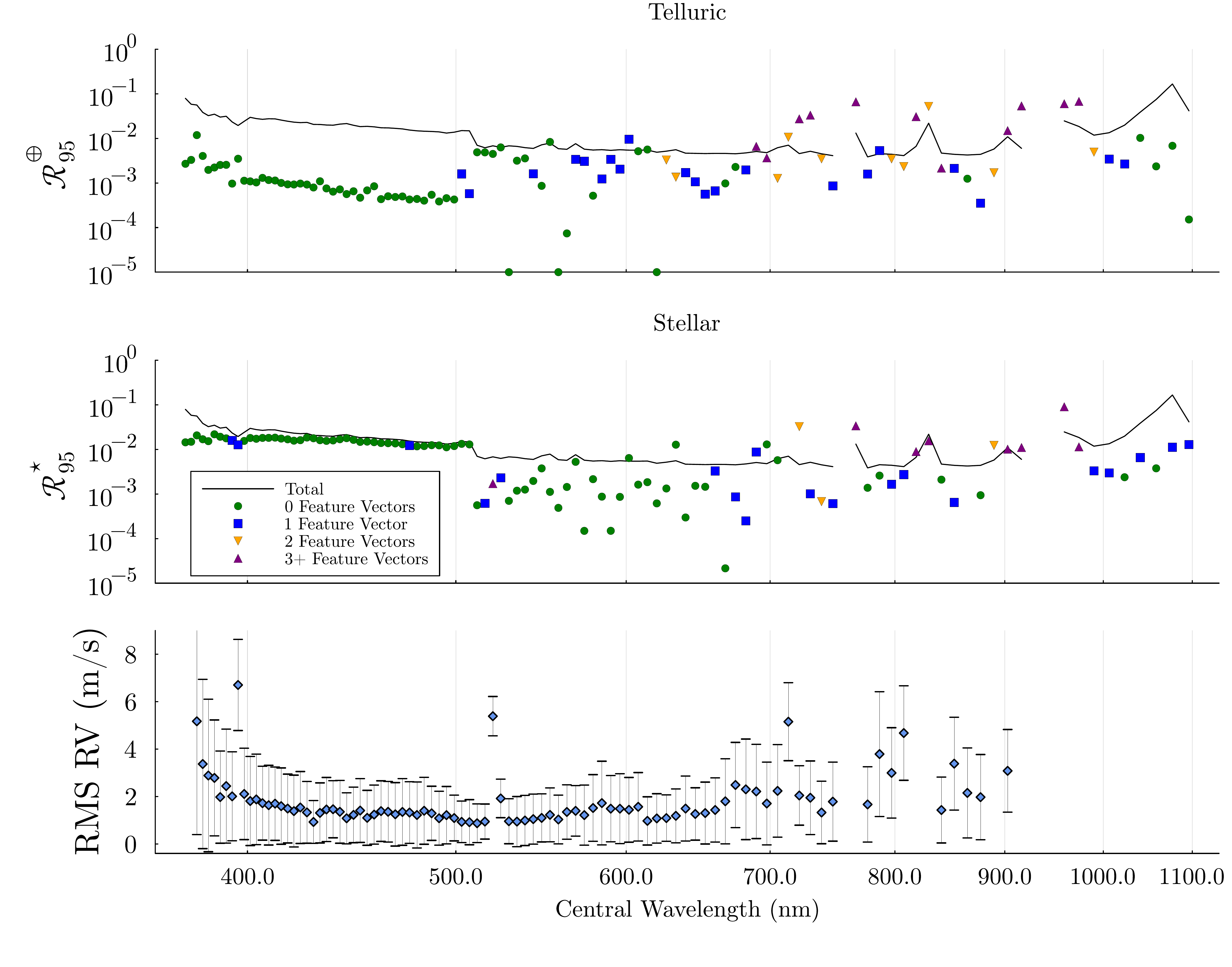}
\caption{Evaluation of SSOF performance on synthetic data set.  
We compare each SSOF-reconstructed spectrum (solid line in top and middle panels), reconstructed telluric transmittance spectrum (points in top panel), and reconstructed stellar spectrum (points in middle panel) to the true values used to generate the input data.  
Each point corresponds to results for one order (that is analyzed independently of other orders).  
The x-axis value indicates the typical central wavelength for that order.  $\mathcal{R}^{\oplus}_{95}$ ($\mathcal{R}^{\star}_{95}$) indicates the median (over 100 observations) of the 95th quantile (over pixel columns 1500-7500) of the residuals between the input telluric (or stellar) input spectrum and reconstructed telluric (or stellar) spectrum, after both have been convolved with a shared line spread function. 
For comparison, the solid line shows $\mathcal{R}^{\mathrm{obs}}_{95}$,  the median of the 95th quantile of the residuals between the input spectrum (before adding photon noise) and reconstructed spectrum, again after both have been convolved with a shared line spread function. 
The point color/style indicates the number of feature vectors used to generate the input data (based on results from analyzing actual observations).  
The lower panel indicates the root mean square deviation (over 100 observations) of the residuals for estimated radial velocities minus the true injected velocities (Earth's motion with no additional planet).  
The RV uncertainty is estimated based on the curvature of the loss function.
The the errorbars show the median (over all observations) of the curvature-based errorbars (see \S\ref{ssec:error}).  
For orders at the blue and red edges plus a few orders in the red portion of the spectrum, the RMS RV exceeds 8 m s$^{-1}$ and is not shown.  
This figure demonstrates that \ssof is accurately estimating each individual component of the model--- the stellar spectrum, the telluric spectrum, and the RV--- for nearly all orders with acceptable SNR.  
The few exceptions are orders with saturated lines (either stellar or telluric), resulting in the SSOF model breaking down and SSOF trying to use more than two feature vectors for at least one component of the model.  
}
\label{fig:validation}
\end{figure*}

The results of our validation tests are summarized in Figure \ref{fig:validation}.  
The primary finding is that \ssof accurately measures the true RV for the vast majority of orders, including orders with significant stellar and/or telluric variability (bottom panel).  
Higher residuals at the extreme blue and red edges are due to low SNR.
For the small number of other orders where SSOF results in a high RMS RV, more than two feature vectors are required for either the stellar model, the telluric model, or both. 
These orders include saturated lines (e.g., Ca II H \& K near 400 nm, O$_2$ bandhead, water bands in the NIR) which are particularly challenging for the SSOF model due to non-linear curve of growth.  
Further, we see that the uncertainty estimates for the RVs are accurate for nearly all orders.  
Most cases where the RV uncertainty is underestimated occur for orders that used three or more feature vectors for either the star or telluric model. 
The one exception is an order with Ca H \& K absorption.   

While the RMS RV provides the primary metric for evaluating the performance of SSOF, it is also interesting to open the ``glass box'', and evaluate how well SSOF is recovering the stellar and telluric spectra separately (top and middle panels of Figure \ref{fig:validation}. 
Comparing the feature vectors and scores can be complicated due to degeneracies.  
For example, multiplying a feature vector by $\alpha$ and dividing corresponding scores by $\alpha$ does not affect the model outputs.
(And for orders with multiple feature vectors, the second analysis can find a rotated version of the feature vectors used to generate the data that results in identical reconstructions to the original .)
Therefore, we evaluate the quality of SSOF's analysis based on the reconstructed stellar spectra, the reconstructed telluric spectra, the total reconstructed spectra, and the measured RVs at each epoch and for each order.  
In principle, we can compare the full spectrum at every observation time.  
In practice, we find it useful to summarize the typical accuracy of each component of the SSOF model using $\mathcal{R}^{\oplus}_{95}$, $\mathcal{R}^{\star}_{95}$, and $\mathcal{R}^{\mathrm{obs}}_{95}$ for each order.  
These correspond to the median (over 100 observations) of the 95th quantile (over pixels) of the residuals for the telluric, stellar and total spectrum, respectively.
 To avoid these statistics being dominated by photon noise at the edges of each order, we compute the 95th quantile over pixels $1500---7500$.
\footnote{This is analogous to how most RV extraction codes exclude data from the edge of each order where the wavelengths overlap with another order that provides higher signal-to-noise data.}
Since our model is computed at a higher-resolution than the observations, a simplistic comparison results in high-frequency noise that overestimates the residuals once the simulated spectra are convolved with the LSF.  
To more accurately assess the agreement between the input and output spectra, we convolve both the input spectra and the reconstructed spectra with the LSF model used for that order before computing $\mathcal{R}^{\oplus}_{95}$, $\mathcal{R}^{\star}_{95}$, and 
$\mathcal{R}^{\oplus}_{95}$, $\mathcal{R}^{\star}_{95}$.  
The original input spectra were normalized so the continuum was near unity.
Therefore, each of three reconstructed spectra have a continuum level near unity.

For many orders in the blue where the generative model uses transmittance of unity across the entire order, the total reconstruction error is dominated by the residuals in the stellar spectrum.
For orders with some telluric absorption (even when SSOF does not use a time-variable component) the reconstruction errors are roughly equally distributed among the stellar and telluric model.
For some orders, $\mathcal{R}^{\oplus}_{95}$ and/or $\mathcal{R}^{\star}_{95}$ exceed
$\mathcal{R}^{\mathrm{obs}}_{95}$ ($\mathcal{R}^{\star}_{95}$, since the two components have anticorrelated errors (e.g., some of the mean telluric absorption is being modeled as a reduction in the stellar continuum).  
Even in these orders, \ssof  can accurately reconstruct the true stellar, true telluric, and total spectra, as well as the true RV.   

In conclusion, we see that the \ssof model and our optimization procedure are effective for the vast majority of orders, and all the orders which are likely to be useful for extremely precise radial velocity surveys.
This test was based on actual amplitude and shape of spectral variability derived from our analysis of HD 26965 in \S\ref{ssec:26965}.  
Thus, the level of difficulty for this test was set by the level of detail in the model for generating synthetic data and indirectly by the number of actual NEID observations of HD 26965.
As the number of observations from environmentally stabilized EPRV spectrographs increases, it will be possible to fit models with additional components for the stellar model in many of the orders.  
At that point, it will be worth updating validation tests with more complex models for stellar variability.

\section{Application to EPRV data}\label{sec:application}

We now demonstrate applying \ssof to real EPRV observations using NEID observations of three stars, HD 26965, HD 3651 and Barnard's Star. While some details such as the orders used, wavelength range of each order, and signal-to-noise are specific to NEID and these targets, the overall process is intended to be applicable to timeseries of spectroscopic observations of other stars from other well-stabilized, high-resolution spectrographs. The main limitation is that \ssof only performs well when there are enough ($\gtrsim$20) observations that cover a range of barycentric velocities. We have also verified that \ssof and the pre-processing described works well using EXPRES observations with data from the EXPRES Stellar Signals Project \citep{zhao2022}.
All differential radial velocities in the following sections are approximated from the differential Doppler shifts as ${v}_{\star} \approx c \ z_\star$.
The 1D spectral data used in this study were reduced using the standard NEID data reduction pipeline (DRP), which computes order-by-order and bulk CCF-based RVs, along with a suite of activity indicators. 

For each stellar target, we additionally extracted RVs with a custom version of the Spectrum Radial Velocity Analyzer
\citep[\texttt{SERVAL};][]{serval} optimized for NEID spectra \citep[see][for further details]{gummi2022}. \texttt{SERVAL} leverages the template-matching technique, where all available spectra are used to build a high S/N as-observed spectral template. In building the template, known regions of telluric and sky emission lines are masked. The template is then used to measure the best-fit Doppler shift through comparing the individual spectra to the shifted optimized template using $\chi^2$ minimization. For the extractions, we followed the same extraction setup as discussed in \cite{gummi2022}, where we extracted orders spanning wavelength regions from 3780-8940\AA. 

\subsection{Data Pre-processing}\label{ssec:preprocess}
The data are transformed in several ways to enable later analyses.
The main data inputs for creating and training a \ssof model are three $P \times N$ matrices where $P$ is the number of pixels in the order and $N$ is the number of observations, the observed 1-D extracted spectral fluxes at each time ($Y_D$), the pixel level white noise variance estimates ($\sigma_D^2$)
, and the observer-frame log-wavelengths ($\xi_D$), as well as the barycentric correction shifts ($z_D$).
We rely on the reliable instrumental calibration and precise reported wavelength calibrations of the EPRV pipelines.
We also flag and remove spectra which have abnormally low SNR or poor wavelength drift solutions.
To simplify \ssofns's spectral model
, we transformed these inputs by:
\begin{enumerate}
  \item Dividing each $Y_{D,n}$ and $\sigma_{D,n}$ by the instrumental blaze function at each time
  \item Totally masking pixels that have already been flagged (e.g. by an instrument's data reduction pipeline)
  \item Totally masking abnormal pixels that change in flux much more rapidly than others in the same order\footnote{more specifically we find pixels where the time-average snap $\left(\overline{\dfrac{\delta^4Y_{D,n}}{\delta\xi_{D,n}^4}}\right)$ is outside 10 standard deviations after winsorizing out the top and bottom 0.5\% pixels} and their neighbors
  \item Totally masking edges of the order with low (i.e. $<5$) mean per-pixel signal-to-noise ratio (SNR)
  \item Masking any unphysically high (i.e. $>2$) fluxes and their neighbors, which happens rarely and may be due to inexact removal of phenomena like cosmic rays. 
  \item Continuum normalizing each $Y_{D,n}$ by fitting and elementwise dividing a (2nd order) polynomial fit on an asymmetrically sigma-clipped subset of $Y_{D,n}$ (that avoids absorption features). Orders without a reliably measured continuum level (i.e. ones with deep stellar lines and little clean continuum region) are not continuum normalized.
\end{enumerate}

\begin{figure*}[ht]
\centering
\includegraphics[width=18cm]{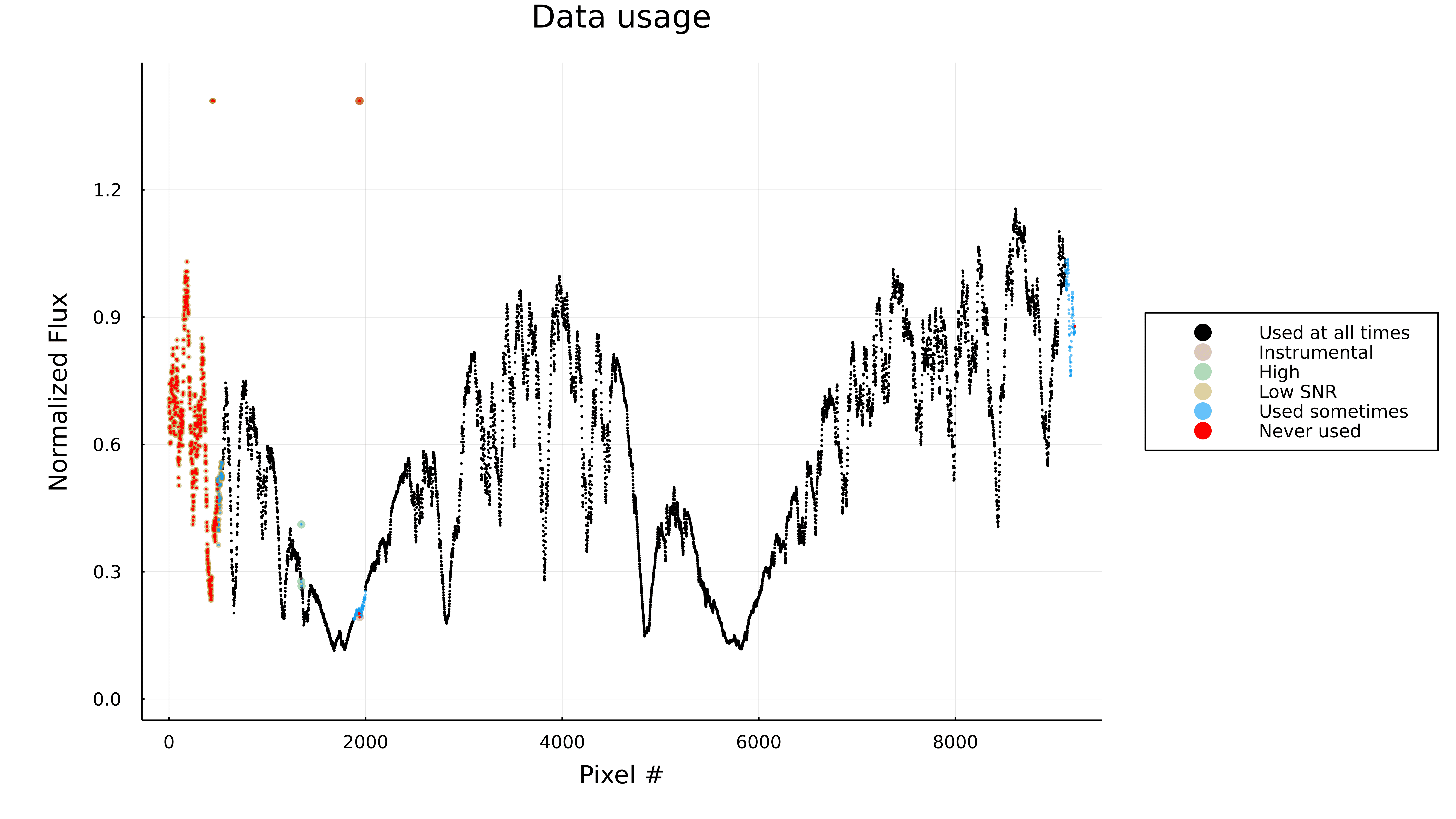}
\caption{An example of a time-averaged spectrum for an order of NEID data (which includes \cahk) showing which parts of the spectrum are being used by \ssof after data pre-processing. Different parts of the spectrum are masked for various reasons. The bluer side of this order (order 155 in Tab. \ref{tab:26965}), had low SNR (yellow border). Some pixels were masked by the instrumental pipeline (orange border) or had anomalously high observations (green border). Pixels which are never used (red) are surrounded by pixels that are sometimes used (blue). The blue pixels are sometimes ignored when fine-tuning $S_\oplus$, $S_\star$, and $v_\star$.}
\label{fig:prep}
\end{figure*}

We use ``totally masking'' to mean setting $\sigma_{D,p}$ to $\infty$ at the given pixel location at all times and ``masking'' to mean setting $\sigma_{D,n,p}$ to $\infty$ at the given pixel location and time.
An example average spectra after pre-processing is shown in Fig. \ref{fig:prep}.

\subsection{HD 26965}\label{ssec:26965}

HD 26965 is a K1V star with a measured rotation period (based on stellar activity indicators) of around 40 days \citep{ma2018}. 
Using either traditional CCF or forward modeling techniques to measure RV, results in a RV RMS $\approx3$ \ms \citep{zhao2022}.
A periodic signal around 42 days (which is consistent with the published rotation period) has been identified in the RVs and activity indicators giving evidence to the fact that the RV signal can be attributed to stellar variability\citep{diaz2018, ma2018,rosenthal2021,zhao2022,Burrows2024,Sigel2024}.
Analysing HD 26965 shows how \ssof performs on a high-quality RV target star likely to exhibit a few \ms signal due to stellar activity.
We analyzed 61 observations of HD 26965 taken by NEID from October 2021 to March 2022 with 112 \ssof models, each fit to a single spectral order.
The RV performance and model complexity for each \ssof model is shown in Tab. \ref{tab:26965}.

\begin{figure*}[ht]
\centering
\includegraphics[width=18cm]{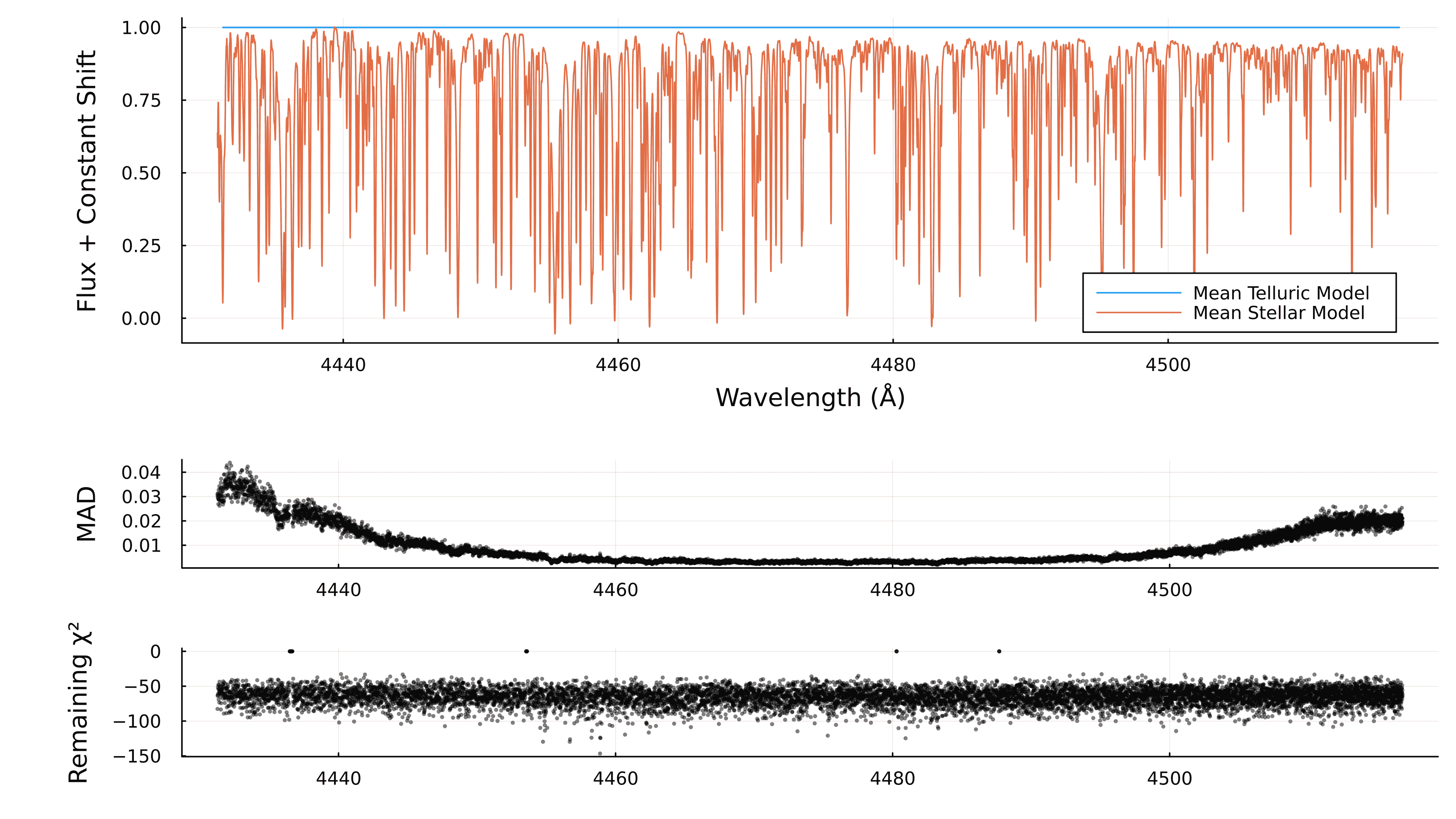}  
\caption{\ssof model of ``key order'' 137 for HD 26965 which can effectively model its portion of the spectrum with only a stellar template. Top: The flux outputs of the model --- $Y'_\oplus$ (blue) which has no telluric transmission (blue) and $Y'_\star$ (orange), a constant stellar model. Middle: The per-pixel mean absolute deviation (MAD) of $Y_D - Y_M$. The majority of the spectra is modeled to $>$99\% accuracy. Bottom: The remaining $\chi^2$ for each observed pixel. All parts of the spectra are modelled equally well in a $\chi^2$-sense. The residuals between $v_{\star,137}$ and $v^{\ssofns}_{\star}$ are shown in Fig. \ref{fig:26965rv}}
\label{fig:26965_ko}
\end{figure*}

A large portion of the RV information that can be extracted from the data is in parts of the measured spectra that have little to no telluric transmission or stellar variability (see orders 149-123 in Tab. \ref{tab:26965}).
With the high SNR of NEID's HD 26965 measurements (SNR $\sim$ 140-440 in the central 1000 pixels of orders used by \ssof to calculate RVs), each order can reach single measurement RV precisions (SMP) of $\sim 1$ \ms using only stellar templates, which are sufficient to model the data to $>$99\% accuracy.
One example of \ssof modeling these less complicated orders is shown in Fig. \ref{fig:26965_ko}.

\begin{figure*}[ht]
\centering
\includegraphics[width=15cm]{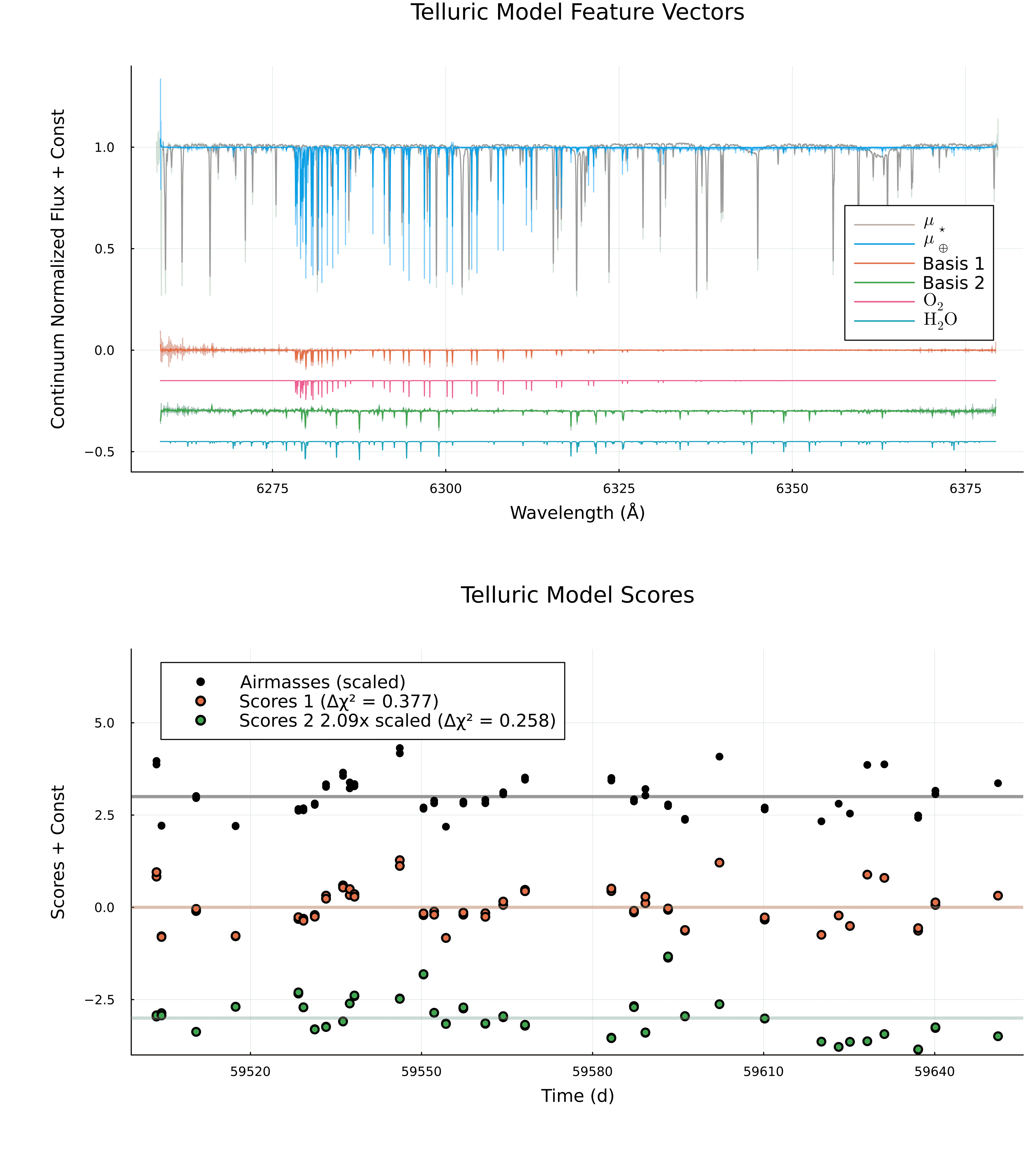}
\caption{\ssof model of order 97 for HD 26965 which separates the two dominant modes of telluric variability. Top: The telluric and stellar templates are shown above in grey and blue while the two telluric feature vectors are shown below in orange and green. For comparison, we also show portions of simulated H$_2$O (turquoise) and O$_2$ (pink) spectra. The simulated lines are those whose line depths are >0.1\% in typical conditions at Kitt Peak. The line locations and emmissivities are from HITRAN2020 \citep{HITRAN2020, HITRAN-850, HITRAN-1179, HITRAN-1387, HITRAN-24}. The translucent portions of the feature vectors show the true values, while the solid portion is a version that has been smoothed by the instrumental LSF used during fitting. The orange feature vector captures the time-variability of O$_2$ lines while the green basis vector captures the time-variability of H$_2$O lines. Bottom: The telluric feature scores as a function of time (in modified Julian date, MJD). The scores for the O$_2$ feature vector (orange) are highly correlated with observation airmass (black) while the scores for the H$_2$O feature vector (green) are more erratic but clearly identified in multiple orders (see Fig. \ref{fig:26965_Halpha}). $\Delta\chi^2$ is one minus the ratio of the final model $\chi^2$ to the model $\chi^2$ if you set all of this feature's scores to zero. Higher values mean the feature vector is more important and capture more variance. The residuals between $v_{\star,97}$ and $v^{\ssofns}_{\star}$ are shown in Fig. \ref{fig:26965rv}}
\label{fig:26965_tel}
\end{figure*}

\begin{figure*}[ht]
\centering
\includegraphics[width=18cm]{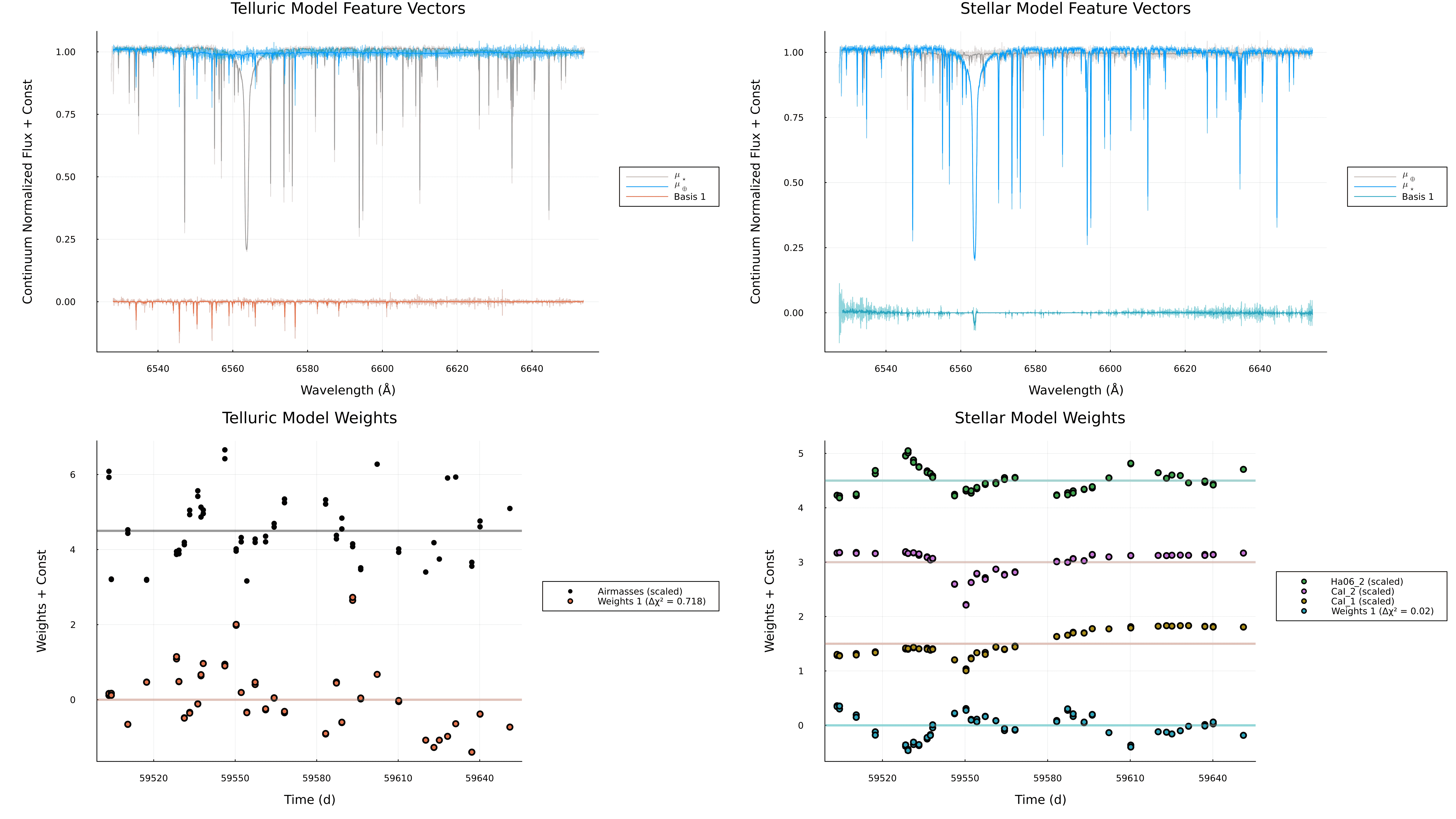}
\caption{\ssof model of order 93 for HD 26965 which identifies H-alpha variation. Top left: The telluric and stellar templates are shown above in grey and blue while the telluric feature vector is shown below (orange). This vector is dominated by changing H$_2$O transmission between observations. Bottom left: The observed airmass (black) and telluric feature scores (orange) as a function of time. The H$_2$O signal is not correlated with airmass but clearly identified in multiple orders (see Fig. \ref{fig:26965_tel}). Top right: The telluric and stellar templates are shown above in grey and blue while the stellar feature vector is shown below (turquoise). This vector is dominated by the changes in H-alpha. Bottom right: The stellar feature scores as a function of time. The scores for the H-alpha feature vector (turquoise) are correlated with the H-alpha variation measured by NEID's DRP (green), although \ssofns's scores are a bit noisier due to the inclusion of the entire order. Some NEID measurements of CaI lines are also shown. The residuals between $v_{\star,93}$ and $v^{\ssofns}_{\star}$ are shown in Fig. \ref{fig:26965rv}}
\label{fig:26965_Halpha}
\end{figure*}

AIC starts to favor models that include a telluric absorption component redwards of 5000 \AA (orders $<$123), which is consistent with the expected telluric absorption. 
Most \ssof models for orders redwards of $<$123 have telluric transmission and most of those have at least one telluric feature vector to explain temporal variation in the transmission.
Almost all of the significant telluric feature vectors found by \ssof can either be traced to known H$_2$O or O$_2$ lines. 
The locations of lines in the feature vectors often match those of the strongest H$_2$O and O$_2$ lines according to HITRAN \citep{HITRAN2020, HITRAN-850, HITRAN-1179, HITRAN-1387, HITRAN-24}.
O$_2$ has relatively slow sources and sinks 
and is distributed relatively uniformly throughout the atmosphere.
Therefore, its column density 
is mostly dependent on airmass (and secondarily on atmospheric pressure).
Indeed, we find that the scores for O$_2$ basis vectors are highly correlated with the observed airmass.
The H$_2$O feature vectors are not directly linked to airmass, 
but still have a shared temporal structure that is measured by many \ssof order models (as can be seen by comparing the H$_2$O scores in Fig. \ref{fig:26965_tel} and Fig. \ref{fig:26965_Halpha}).
\ssof can even separate different telluric species based on uncorrelated variability in the same spectral order, without prior knowledge of line location or observed airmass.
A prime example of this is shown in Fig. \ref{fig:26965_tel}, where \ssof cleanly models both H$_2$O and O$_2$ at the same time and identifies the features with separate feature vectors.
Even in the presence of complicated telluric variability, \ssof can identify its telluric nature and characterize it well enough to get RVs that are consistent with  $v^{\ssofns}_\star$ and with SMP $< \ 1.5$ \ms with no prior knowledge of the location of telluric absorption features.
Achieving this level precision, even in orders contaminated with telluric variability, increases overall precision and could be valuable for recognizing chromatic signatures of stellar activity.

\begin{figure*}[ht]
\centering
\includegraphics[width=8.75cm]{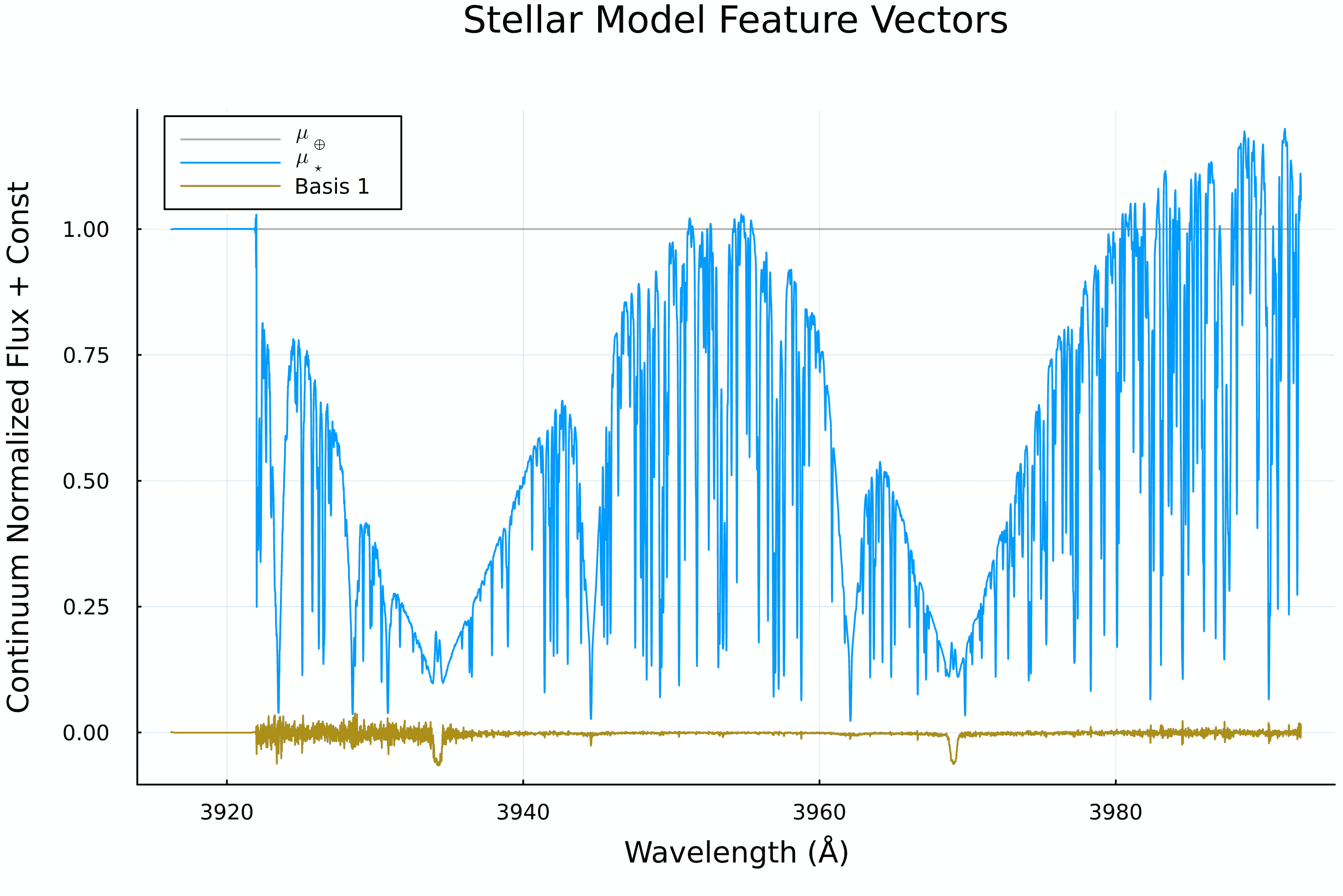}
\includegraphics[width=8.75cm]{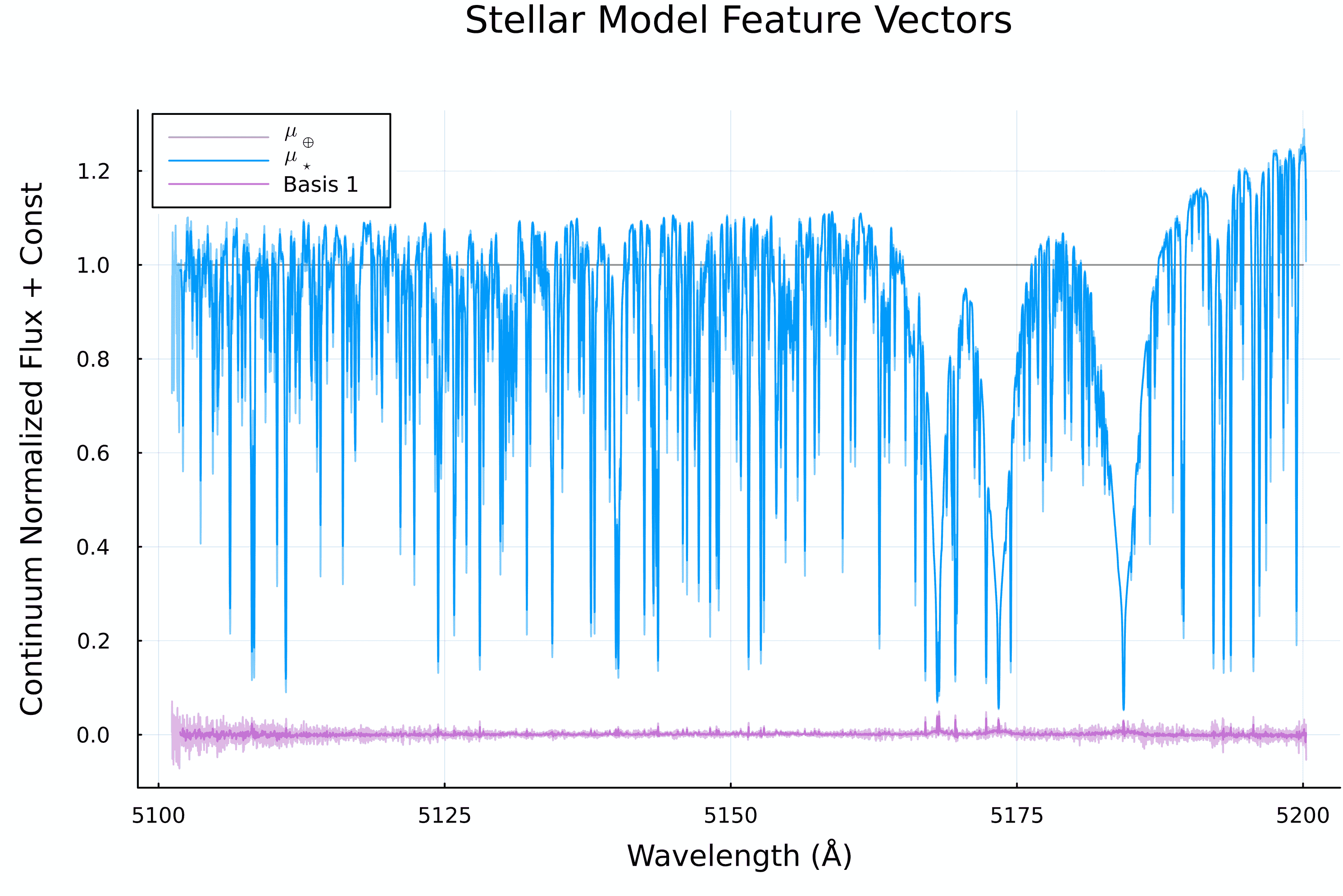}
\includegraphics[width=8.75cm]{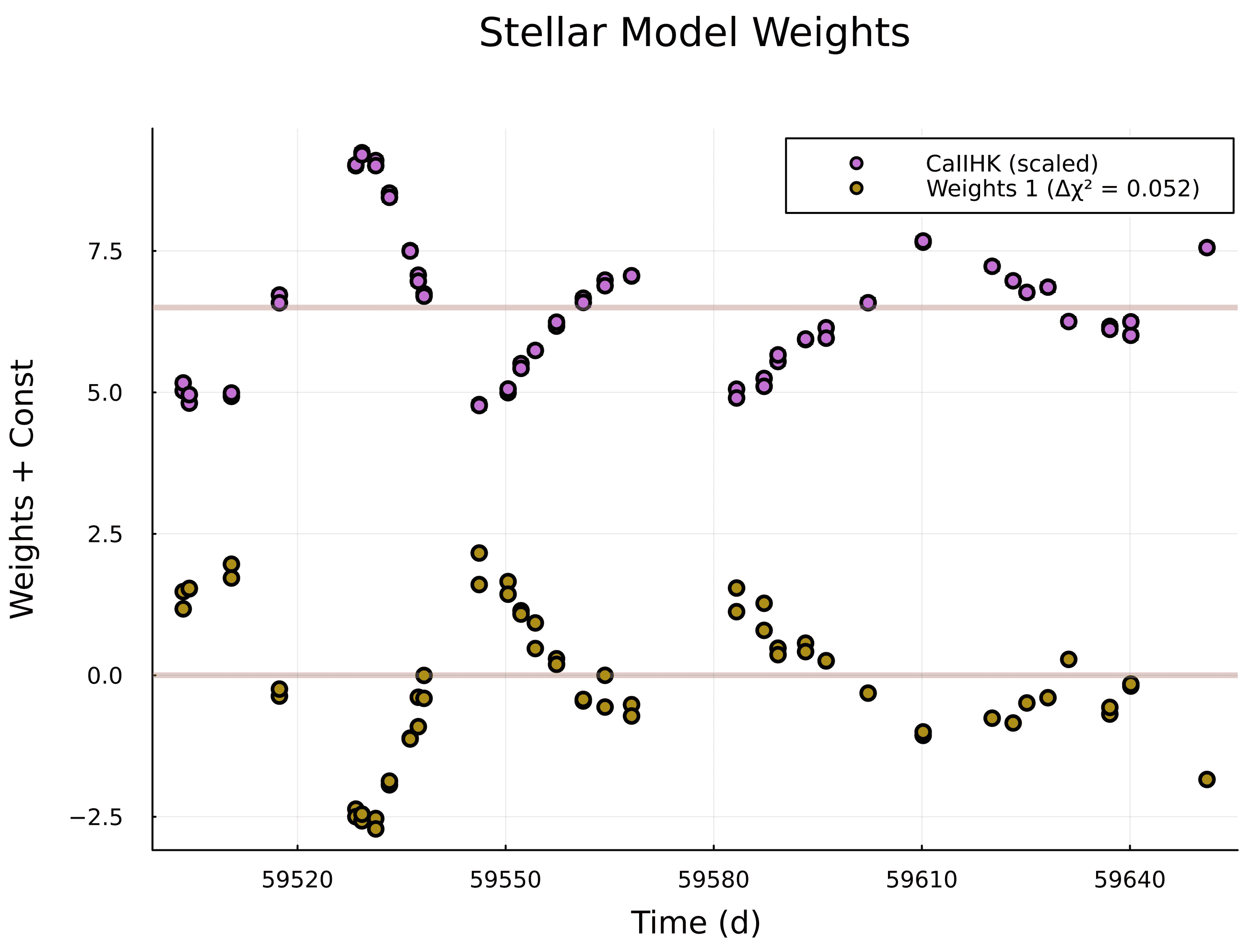}
\includegraphics[width=8.75cm]{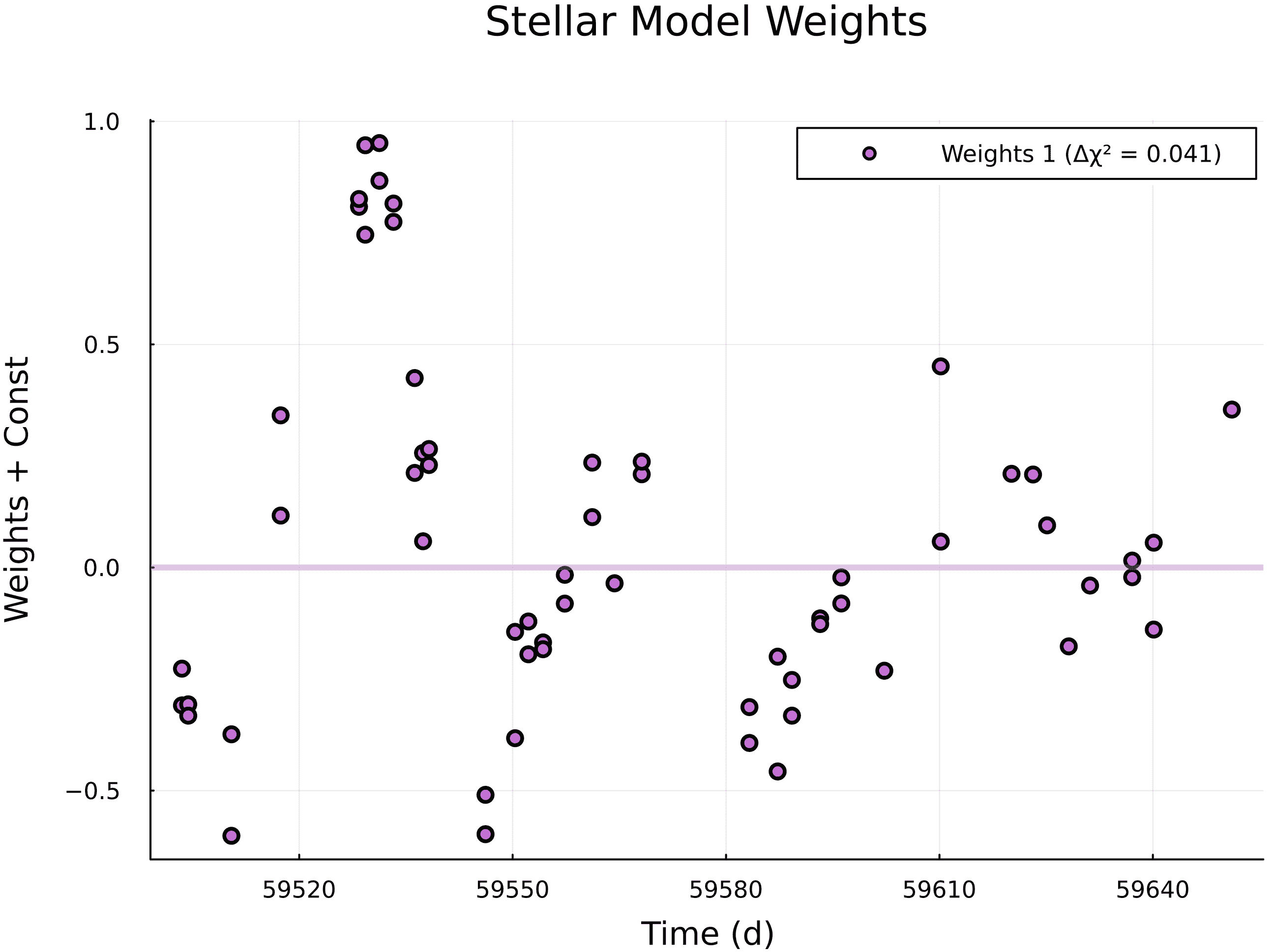}
\caption{\ssof models of order 155 for HD 26965 which identifies \cahk variation and order 119 which identifies Mg triplet and MgH band variation. Top left: The telluric and stellar templates are shown above in grey and blue while the stellar feature vector is shown below (yellow). This vector is dominated by changing emission in the core of \cahk, but also includes variations in other deep lines (including some Fe I lines \citep{vald3}) that are correlated with \cahk changes. Bottom left: The stellar feature scores as a function of time. The scores for the \cahk feature vector (yellow) are correlated with the \cahk variation measured by NEID's DRP (purple), although \ssofns's scores are a bit noisier due to the inclusion of the entire order. These scores show a quasi-periodic, decaying signal with a period of $\sim$ 40 days, consistent with previous measurements of HD 26965's rotation rate. Top right: The telluric and stellar templates are shown above in grey and blue while the stellar feature vector is shown below (purple). This vector is dominated by changing depths in the Mg triplet and MgH bands. Bottom right: The stellar feature scores as a function of time. The scores for the Mg triplet and MgH band feature vector (purple) are also correlated with the \cahk variation measured by both \ssof and NEID's DRP. The residuals between each $v_{\star}$ and $v^{\ssofns}_{\star}$ are shown in Fig. \ref{fig:26965rv}}.
\label{fig:26965_CaHK}
\end{figure*}

\ssof is also able to automatically identify and recover stellar variability in several stellar lines which have classically been used to assess magnetic field variability without any prior knowledge of the shape or location of these variations. 
\ssof recovered signals for \cahk (3968 \AA, 3934 \AA, in order 155), the Mg triplet and MgH band (5184 \AA, 5173 \AA, 5169 \AA, 5167 \AA, in order 119), H-alpha (6563 \AA, in order 93), and the Ca IR triplet (8498 \AA, 8542 \AA, 8662 \AA, orders 72-71).
The \ssof models which detected \cahk and the Mg triplet and MgH band are shown in Fig. \ref{fig:26965_CaHK}.
These feature vectors are dominated by stellar variations in \cahk and the Mg triplet and MgH and their scores are nearly proportional to the \cahk measurements produced by the NEID DRP after centering both the scores and the \cahk measurements.
These scores show a quasi-periodic, decaying signal with a period of $\sim$ 40 days, consistent with previous measurements of HD 26965's rotation rate.
The H-alpha and Ca IR triplet were identified concurrently with telluric variability in the same order.
The \ssof model which detected H-alpha and H$_2$O variations is shown in Fig. \ref{fig:26965_Halpha}.
These models demonstrate \ssof being able to fully realize its potential to simultaneously identify and characterize key forms of variability in EPRV spectra.

\begin{figure*}[ht]
\centering
\includegraphics[width=18cm]{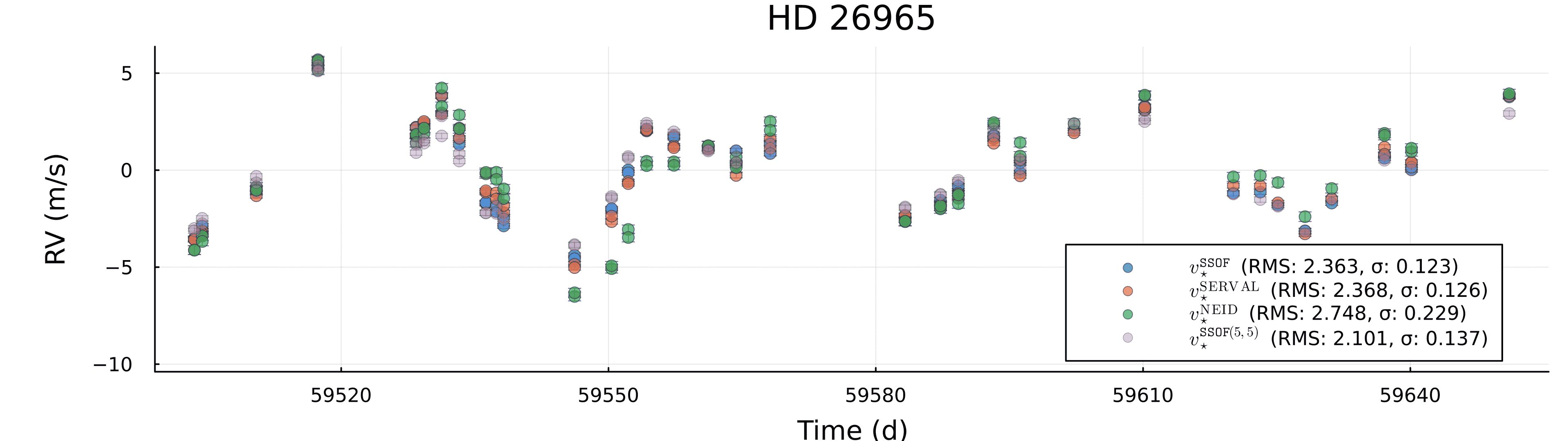}
\includegraphics[width=8.5cm]{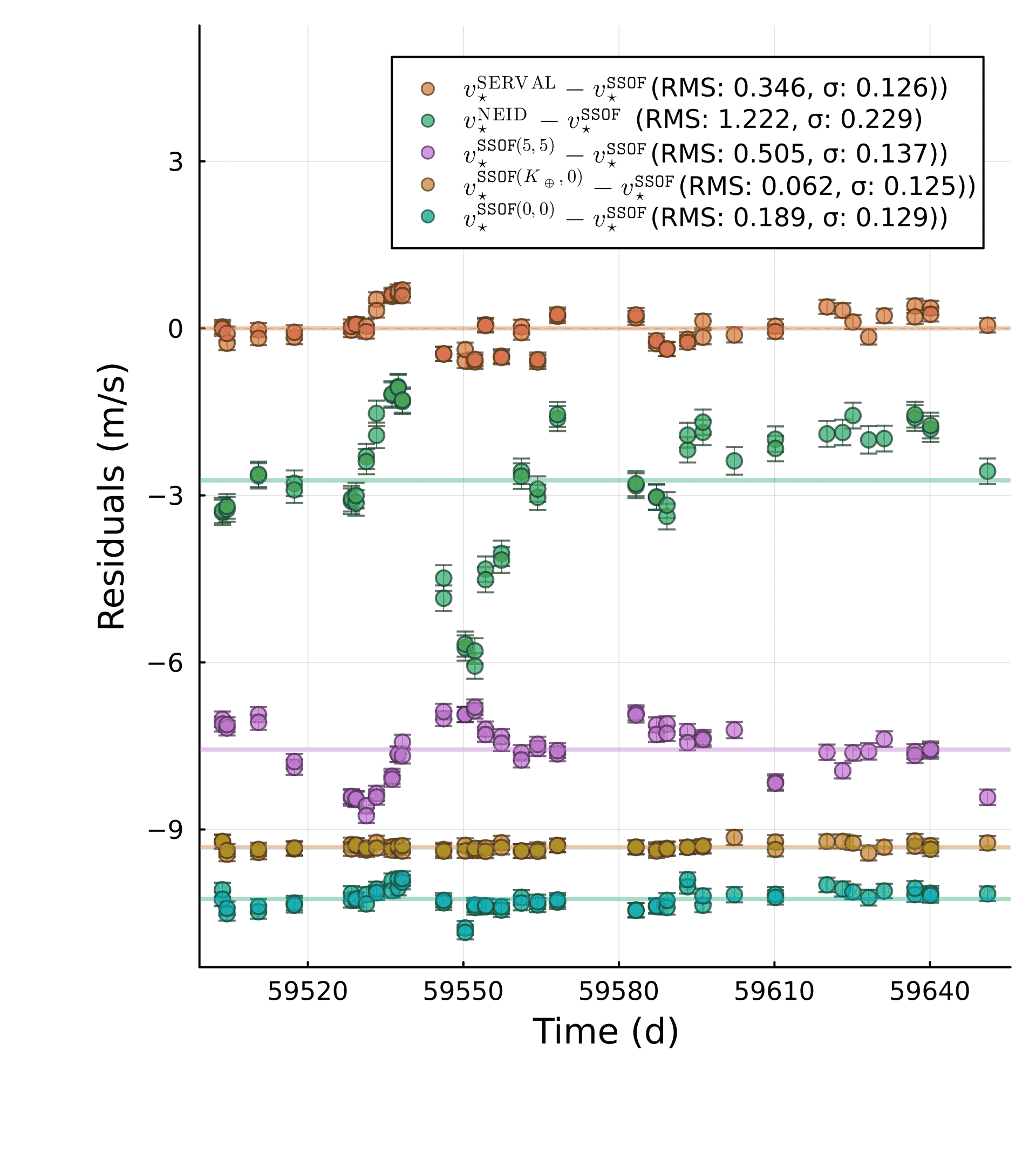}
\includegraphics[width=8.5cm]{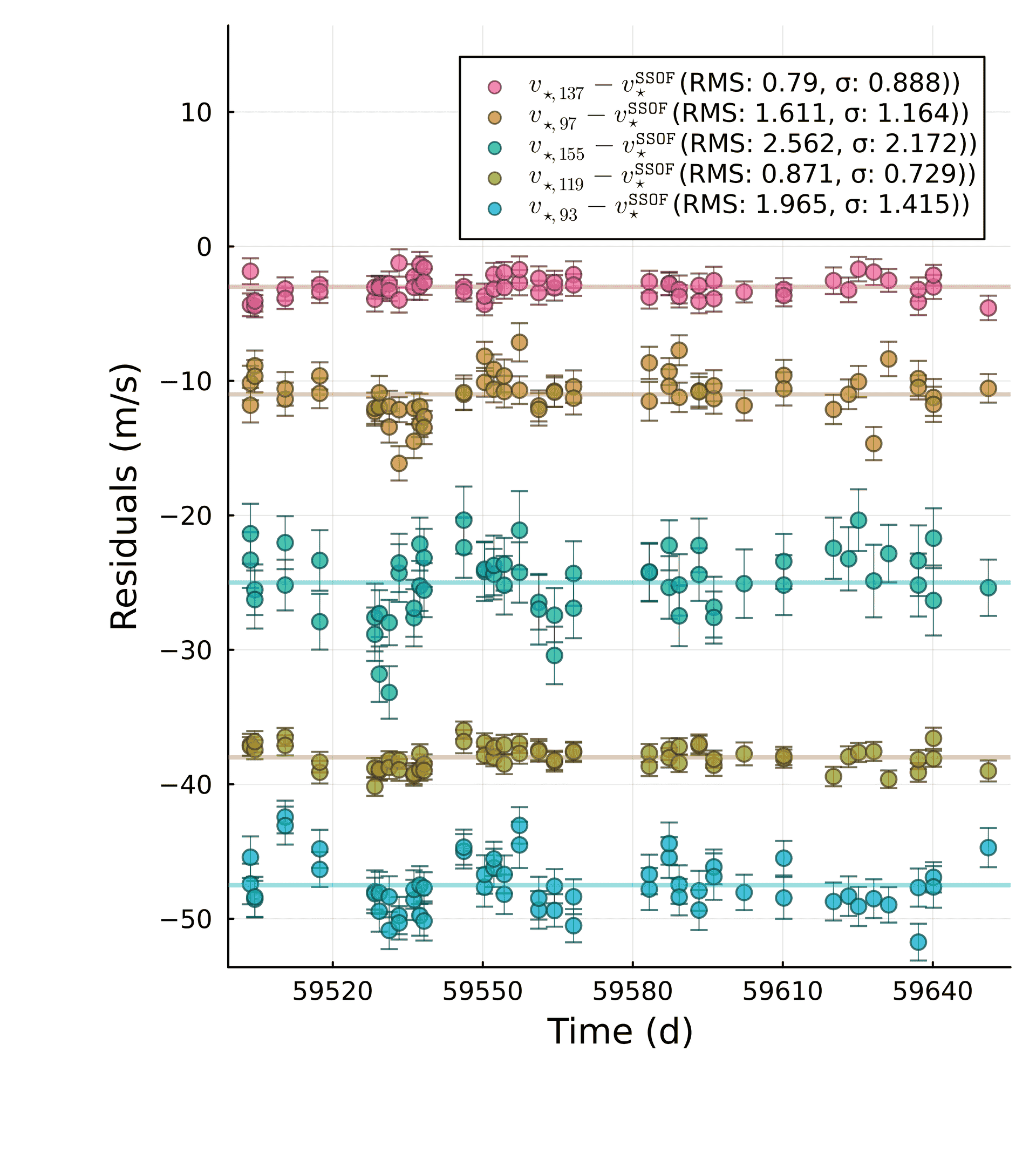}
\caption{Top: The final RV reduction of $v^{\ssofns}_\star$ (blue) for HD 26965 combining the velocity measurements from 87 \ssof order models resulting in a 46\% smaller SMP and 14\% lower RV RMS when compared to the NEID DRP's CCF-based RVs ($v^{\textrm{NEID}}_\star$, green) and a similar SMP and RV RMS when compared to \texttt{SERVAL}'s template-matching RVs ($v^{\textrm{SERVAL}}_\star$, orange). Another \ssof RV reduction , $v^{\ssofns(5,5)}_\star$ (purple), is shown that was constructed using a \ssofns(5,5) model for every order, regardless of AIC. $v^{\ssofns(5,5)}_\star$ had a RMS of 2.1 \ms, even lower than $v^{\ssofns}_\star$. All RV time series are shown with their corresponding 1-$\sigma$ error bars. 
Bottom left: The residual time series for $v^{\textrm{SERVAL}}_\star$ (orange), $v^{\textrm{NEID}}_\star$ (green), and several other versions of $v^{\ssofns}_\star$. The residuals between $v^{\textrm{NEID}}_\star$ and $v^{\ssofns}_\star$ are similar to a time-shifted version of the measured \cahk variations (see Fig. \ref{fig:26965shift}). Interestingly, $v^{\ssofns(5,5)}_\star - v^{\ssofns}_\star$ (purple) is also strikingly similar to the measured \cahk variations (see Fig. \ref{fig:26965shift}). The $v^{\ssofns(K_\oplus,0)}_\star$ reduction (brown) was constructed using the AIC minimum \ssof model without stellar variability for every order. Only a few \ssof models used in $v^{\ssofns}_\star$ used stellar variability so $v^{\ssofns(K_\oplus,0)}_\star$ is practically indistinguishable to $v^{\ssofns}_\star$. The $v^{\ssofns(0,0)}_\star$ reduction (light turquoise) was constructed using the AIC minimum \ssof model without any variability for every order. Somewhat surprisingly given the amount of telluric variability measured by \ssofns, this only adds 20 \cms of variability, but this is because many more orders are rejected (only 68 orders are kept). 
Bottom right: The residual time series for the highlighted \ssof order models. The residual velocities from the ``key order" (pink) are consistent with white noise. The residual velocities from the H$_2$O and O$_2$ separating (dark orange), \cahk (seafoam green), Mg triplet and MgH band (light green), and H-alpha (light blue) orders are largely consistent with white noise, excluding an interesting shared deviation between JD 2459520-2459580.}
\label{fig:26965rv}
\end{figure*}

The final $v^{\ssofns}_{\star}$ time series is shown in Fig. \ref{fig:26965rv}, along with residual RV time series for several different bulk RV reduction schemes and the highlighted \ssof models: (1) $v^{\ssofns(5,5)}_\star$ which was constructed using a \ssofns(5,5) model for every order, regardless of AIC; (2) $v^{\ssofns(K_\oplus,0)}_\star$ which was constructed using the AIC-minimum \ssof models that didn't use stellar variability, and (3) $v^{\ssofns(0,0)}_\star$ which was constructed using the \ssof models with only stellar and (optionally) telluric templates.
\begin{figure*}[ht]
\centering
\includegraphics[width=18cm]{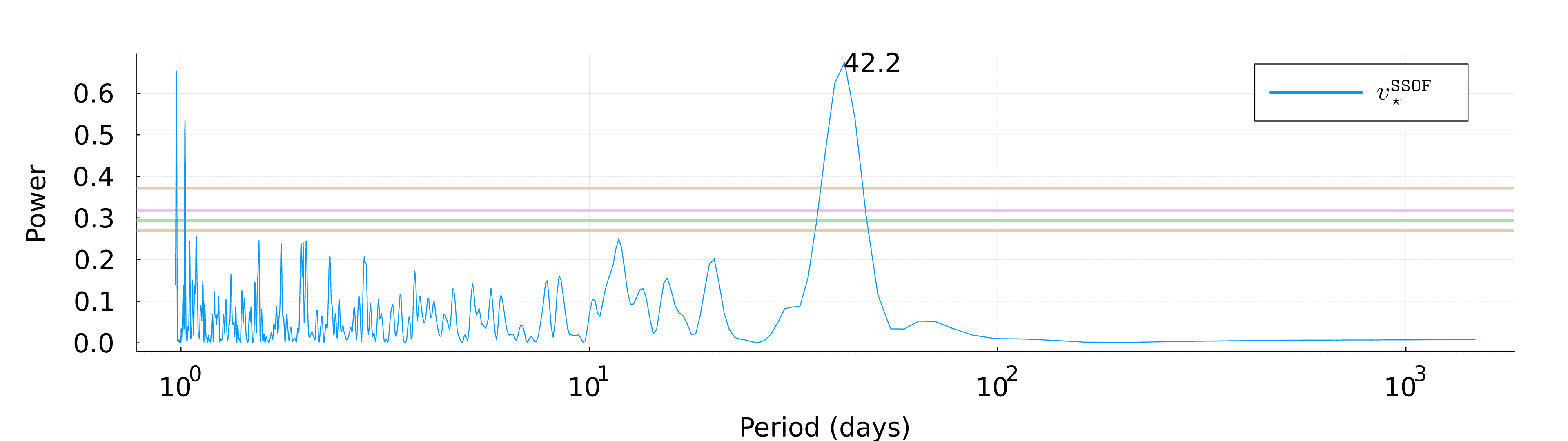}
\caption{Lomb-Scargle periodogram of $v^{\ssofns}_\star$ for HD 26965. The period of the maximum peak is labeled. The horizontal lines are the 95\%, 97.72\% (1-sided 2-$\sigma$), 99\%, and 99.87\% (1-sided 3-$\sigma$) false alarm probability (FAP) powers based on the distribution of maximum powers from many Lomb-Scargle periodograms of bootstrap reorderings of $v^{\ssofns}_\star$. The dominant signal remaining is very significant and at the stellar rotation period.}
\label{fig:26965period}
\end{figure*}
$v^{\ssofns}_{\star}$ have significantly smaller SMP estimates (down 46\% to 12 \cms from the NEID DRP's 23 \cms) and lower RMS (down 14\% to 2.36 \ms from the NEID DRP's 2.75 \ms).
They also have a similar SMP and RV RMS when compared to \texttt{SERVAL}'s template-matching RVs. 
The largest signal in a Lomb-Scargle periodogram of $v^{\ssofns}_\star$ (Fig. \ref{fig:26965period}) is at 42 days, essentially the stellar rotation period.
$v^{\ssofns(5,5)}_\star$ has a RMS of 2.1 \ms, even lower than $v^{\ssofns}_\star$, but have larger measurement uncertainties because fewer orders were selected to be included in the reduction. 
The $\ssofns(5,5)$ models are also much harder to interpret. 
For models with more feature vectors than necessary, information about stellar variability can be distributed across multiple feature vectors. 
Even if this improves the quality of the spectral reconstruction, it results in scores that are not clearly associated with spectral variations that are readily recognized as astrophysicaly significant.
\ssof and \texttt{SERVAL}'s smaller uncertainty estimates compared to NEID's DRP are primarily caused by them using much more of the measured spectra.
The residuals between \ssofns's measured RV values and \texttt{SERVAL}'s are due a combination of the difference in treatment of telluric features (\texttt{SERVAL} cleverly masks regions of the spectra instead of trying to divide them out) and stellar variability (\texttt{SERVAL} uses a constant stellar template).
The difference in residuals between \ssof and \texttt{SERVAL}'s RVs and those from NEID's DRP is not unexpected as it has been shown previously that CCF-based RVs can differ greatly from template-matching style algorithms, which naturally leverage as much of the spectrum as possible (whereas CCF-derived RVs used specific lines)\citep{anglada2012}.
Only a few \ssof models used in $v^{\ssofns}_\star$ used stellar variability, so $v^{\ssofns(K_\oplus,0)}_\star$ is functionally identical to $v^{\ssofns}_\star$.
Somewhat surprisingly, given the amount of telluric variability measured by \ssofns, the $v^{\ssofns(0,0)}_\star$ reduction only differs from $v^{\ssofns}_\star$ by 20 \cms of variability, but this is because many more orders are rejected (only 68 orders are kept).

\begin{figure*}[ht]
\centering
\includegraphics[width=13cm]{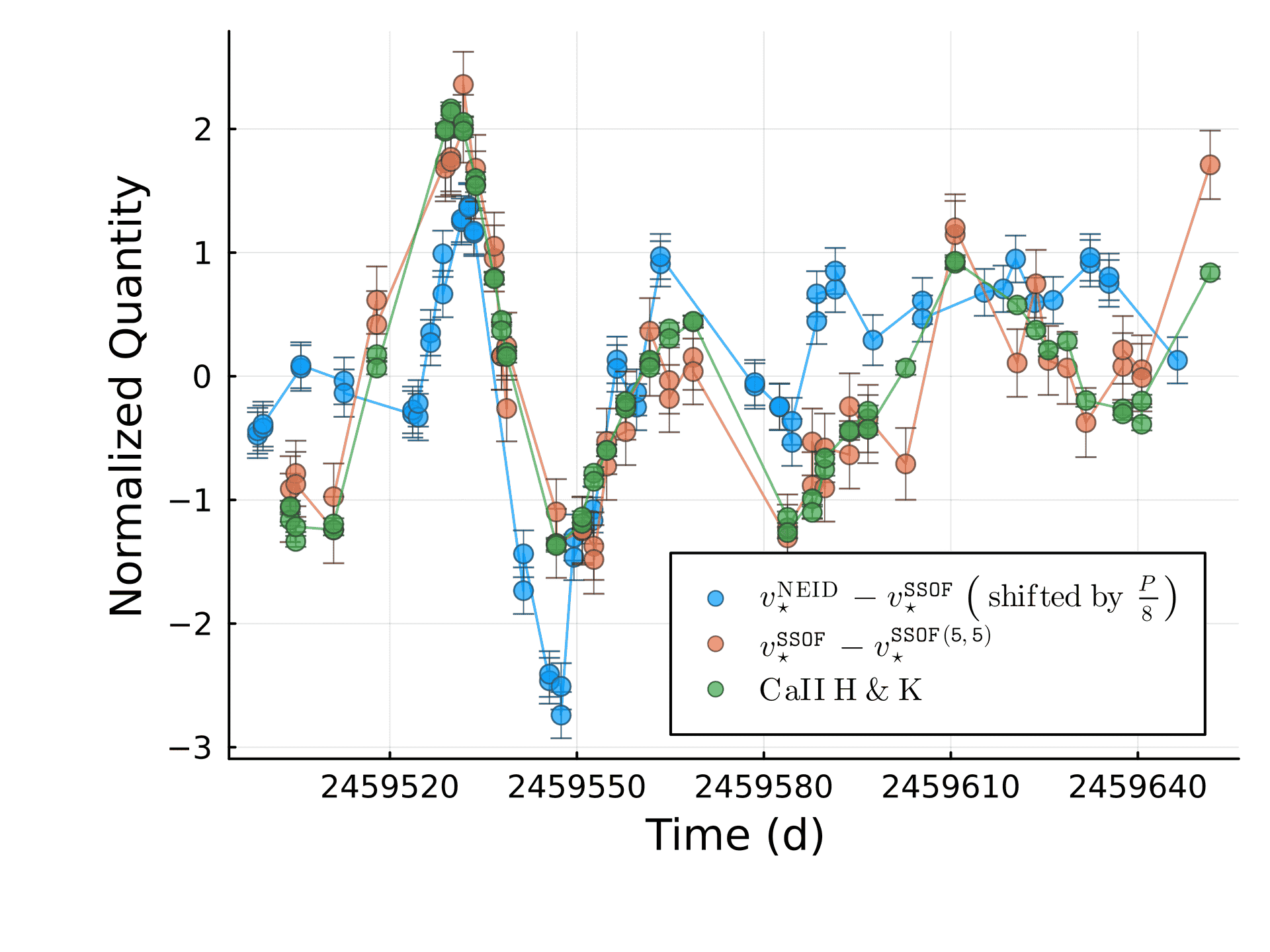}
\caption{Comparing $v^{\textrm{NEID}}_\star-v^{\ssofns}_\star$ (blue) and $v^{\ssofns(5,5)}_\star-v^{\ssofns}_\star$ (orange) with \cahk variations measured by NEID's DRP (green) for HD 26965. All series are shown after subtracting their mean and dividing them element-wise by their respective RMS to bring them to the same scale. Shifting $v^{\textrm{NEID}}_\star-v^{\ssofns}_\star$ back by 5 days (1/8 of the measured orbital period of $\simeq$40 days) makes the signal appear very similar to \cahk. Even more strikingly, $v^{\ssofns(5,5)}_\star-v^{\ssofns}_\star$ traces \cahk almost exactly. The Pearson correlation coefficient between \cahk and $v^{\ssofns(5,5)}_\star-v^{\ssofns}_\star$ is 0.945.}
\label{fig:26965shift}
\end{figure*}

The residual velocities between single \ssof orders and  $v^{\ssofns}_\star$ are largely consistent with white noise, excluding an interesting shared deviation between MJD 59520-59580.
This deviation only appears in models with time-variability and may be the beginnings of the larger effect highlighted in the $v^{\ssofns(5,5)}_\star - v^{\ssofns}_\star$ times-series.
The residuals between $v^{\textrm{NEID}}_\star$ and $v^{\ssofns}_\star$ are similar to a time-shifted version of the measured \cahk variations\footnote{We use NEID's measurements of \cahk as their measurements are slightly cleaner thans \ssofns's due to ignoring the noisy parts of the spectrum outside of the lines of interest}, with \cahk leading the RVs (see Fig. \ref{fig:26965shift}).
It has been noted before that signals in RV measurements can also appear in other activity indicators such as BIS, FWHM, and \cahk with a phase lag (with RVs typically leading the indicators) \citep{Collier-Cameron2019}.
In our case, \cahk appears to lead the RVs residuals by 1/8 of the stellar rotation period. 
Interestingly, $v^{\ssofns(5,5)}_\star - v^{\ssofns}_\star$ is also strikingly similar to the measured \cahk variations (see Fig. \ref{fig:26965shift}).
This effect is observed in nearly all of the \ssof ``key order" RV time series, even those that do not include \cahk or other canonical stellar activity indicators.
We interpret this as evidence that the mechanism causing variations in \cahk are also measurably affecting lines (and their respective RVs) in these other orders. 
The correlation of $v^{\ssofns(5,5)}_\star - v^{\ssofns}_\star$  with \cahk could be a sign that even though the \ssofns(5,5) models are not favored in an information-criterion sense and are more challenging to interpret, the added components may still be providing information about higher-order spectral effects and remain a promising avenue for further reducing the effects of stellar variability in RV time series.
However, we did not observe a similar relationship in the analyses of either HD 3651 or Barnard's Star.

\afterpage{
\begin{longtable*}[p]{|l|l|l|l|l|p{1.15cm}|c|c|c|p{3cm}|}
    \caption{\ssof model results for each tested NEID order for 61 observations of HD 26965 including: echelle grating diffraction order; the observed wavelength range used by \ssof (\AA); the RMS of \ssof order RVs over time, average of RV uncertainties over time, and RMS of the RV residuals comparing each \ssof order RV to the combined \ssof RVs (\ms) at the same time; The number of feature vectors used (* indicates that the model is not the minimum-AIC model); whether or not $v_{\star,j}$ was used in the calculation of $v^{\ssofns}_{\star}$, was a ``key order" (KO), and/or was used by NEID's DRP. A number indicates the reason why the order wasn't used to calculate $v^{\ssofns}_{\star}$ (1 indicates that $\overline{\sigma_{v_\star,j}}$  was too high, 2 indicates that $\textrm{RMS}(v_{\star,j})$  was too high, and 3 indicates that $v_{\star,j}$ was inconsistent with $v^{KO}_{\star}$); any notable comments about the order (H$_2$O indicates the presence of a water-like \ssof basis vector and O$_2$ indicates the presence of a oxygen-like \ssof basis vector). \label{tab:26965}}\\

    \hline
    \multicolumn{10}{|c|}{Beginning of Results Table for HD 26965 (Tab. \ref{tab:26965})}\\
    \hline
    Order & $\textrm{min}(\lambda_{D})$ & $\textrm{max}(\lambda_{D})$ & $\textrm{RMS}(v_{\star,j})$ & $\overline{\sigma_{v_\star,j}}$ & Residual RMS & $K_\oplus$ & $K_\star$ & Used & Comments \\
    \hline
    \endfirsthead

    \hline
    \multicolumn{10}{|c|}{Continuation of Results Table for HD 26965 (Tab. \ref{tab:26965})}\\
    \hline
    Order & $\textrm{min}(\lambda_{D})$ & $\textrm{max}(\lambda_{D})$ & $\textrm{RMS}(v_{\star,j})$ & $\overline{\sigma_{v_\star,j}}$ & Residual RMS & $K_\oplus$ & $K_\star$ & Used & Comments \\
    \hline
    \endhead

    \hline
    \endfoot

    \hline
    \multicolumn{10}{| c |}{End of Results Table for HD 26965 (Tab. \ref{tab:26965})}\\
    \hline
    \endlastfoot
    
        167 & 3668.17 & 3684.7 & 41.68 & 25.5 & 42.29 & 0 & 0 & 1, 2 & ~ \\ \hline
        166 & 3693.78 & 3704.52 & 25.44 & 21.48 & 25.16 & $\varnothing$ & 0 & 1, 2 & ~ \\ \hline
        165 & 3703.84 & 3724.64 & 16.29 & 17.53 & 15.83 & $\varnothing$ & 0 & 1, 2 & ~ \\ \hline
        164 & 3723.52 & 3758.33 & 11.11 & 10.8 & 10.35 & $\varnothing$ & 0 & 1, 2 & ~ \\ \hline
        163 & 3744.5 & 3796.34 & 7.59 & 6.38 & 7.6 & $\varnothing$ & 0 & 1, 2 & ~ \\ \hline
        162 & 3764.62 & 3819.77 & 4.67 & 3.84 & 4.26 & $\varnothing$ & 0 & \ssofns, NEID & RVs start to be useful \\ \hline
        161 & 3784.12 & 3843.49 & 4.61 & 3.67 & 4.01 & $\varnothing$ & 0 & \ssofns, NEID & ~ \\ \hline
        160 & 3805.56 & 3867.51 & 5.14 & 4.35 & 5.14 & $\varnothing$ & 0 & \ssofns, NEID & ~ \\ \hline
        159 & 3833.24 & 3891.85 & 3.46 & 2.75 & 2.83 & $\varnothing$ & 0 & \ssofns, NEID & ~ \\ \hline
        158 & 3851.59 & 3916.49 & 3.19 & 2.18 & 2.51 & $\varnothing$ & 0 & \ssofns, NEID & ~ \\ \hline
        157 & 3874.06 & 3941.45 & 3.16 & 1.92 & 2.19 & $\varnothing$ & 0 & \ssofns, NEID & CaII K \\ \hline
        156 & 3896.54 & 3966.69 & 3.37 & 2.76 & 3.22 & $\varnothing$ & 1 & \ssofns, NEID & CaII K (variance detected) \\ \hline
        155 & 3921.81 & 3992.28 & 2.74 & 2.17 & 2.61 & $\varnothing$ & 1 & \ssofns, NEID & \cahk (variance detected) \\ \hline
        154 & 3945.75 & 4018.2 & 2.86 & 1.63 & 1.76 & $\varnothing$ & 0 & \ssofns, NEID & CaII H \\ \hline
        153 & 3972.69 & 4044.47 & 2.76 & 1.29 & 1.41 & $\varnothing$ & 0 & \ssofns, NEID & ~ \\ \hline
        152 & 3993.91 & 4071.07 & 2.56 & 1.33 & 1.45 & $\varnothing$ & 0 & \ssofns, NEID & ~ \\ \hline
        151 & 4020.37 & 4098.04 & 2.91 & 1.28 & 1.39 & $\varnothing$ & 0 & \ssofns, NEID & ~ \\ \hline
        150 & 4047.16 & 4125.35 & 2.81 & 1.17 & 1.42 & $\varnothing$ & 0 & \ssofns, NEID & ~ \\ \hline
        149 & 4074.35 & 4153.07 & 2.71 & 1.11 & 1.13 & $\varnothing$ & 0 & \ssof (KO), NEID & ~ \\ \hline
        148 & 4101.87 & 4181.11 & 2.74 & 1.16 & 1.4 & $\varnothing$ & 0 & \ssof (KO), NEID & ~ \\ \hline
        147 & 4129.77 & 4209.54 & 2.79 & 1.08 & 1.44 & $\varnothing$ & 0 & \ssof (KO), NEID & ~ \\ \hline
        146 & 4158.06 & 4238.37 & 2.98 & 1.1 & 1.37 & $\varnothing$ & 0 & \ssof (KO), NEID & ~ \\ \hline
        145 & 4186.74 & 4267.61 & 2.94 & 1.13 & 1.5 & $\varnothing$ & 0 & \ssof (KO), NEID & ~ \\ \hline
        144 & 4215.83 & 4297.25 & 2.68 & 1.06 & 1.24 & $\varnothing$ & 0 & \ssof (KO), NEID & ~ \\ \hline
        143 & 4245.3 & 4327.08 & 2.49 & 0.91 & 1.08 & $\varnothing$ & 0 & \ssof (KO), NEID & ~ \\ \hline
        142 & 4275.21 & 4357.77 & 2.84 & 1.05 & 1.28 & $\varnothing$ & 0 & \ssof (KO), NEID & ~ \\ \hline
        141 & 4305.52 & 4388.66 & 2.58 & 1 & 1.03 & $\varnothing$ & 0 & \ssof (KO), NEID & ~ \\ \hline
        140 & 4336.29 & 4420.02 & 2.52 & 0.96 & 1.22 & $\varnothing$ & 0 & \ssof (KO), NEID & ~ \\ \hline
        139 & 4367.49 & 4451.81 & 2.73 & 0.97 & 1.18 & $\varnothing$ & 0 & \ssof (KO), NEID & ~ \\ \hline
        138 & 4399.14 & 4484.07 & 2.62 & 0.79 & 0.95 & $\varnothing$ & 0 & \ssof (KO), NEID & ~ \\ \hline
        137 & 4431.26 & 4516.81 & 2.44 & 0.89 & 0.83 & $\varnothing$ & 0 & \ssof (KO), NEID & ~ \\ \hline
        136 & 4463.84 & 4549.9 & 2.62 & 0.98 & 1.1 & $\varnothing$ & 0 & \ssof (KO), NEID & ~ \\ \hline
        135 & 4496.92 & 4583.72 & 2.6 & 0.9 & 0.86 & $\varnothing$ & 0 & \ssof (KO), NEID & ~ \\ \hline
        134 & 4530.48 & 4617.92 & 2.44 & 0.94 & 1.05 & $\varnothing$ & 0 & \ssof (KO), NEID & ~ \\ \hline
        133 & 4564.55 & 4652.31 & 2.37 & 0.98 & 0.99 & $\varnothing$ & 0 & \ssof (KO), NEID & ~ \\ \hline
        132 & 4599.13 & 4687.89 & 2.32 & 0.97 & 0.88 & $\varnothing$ & 0 & \ssof (KO), NEID & ~ \\ \hline
        131 & 4634.25 & 4723.67 & 2.73 & 0.99 & 1.15 & $\varnothing$ & 0 & \ssof (KO), NEID & ~ \\ \hline
        130 & 4669.9 & 4760.01 & 2.58 & 0.94 & 1.17 & $\varnothing$ & 0 & \ssof (KO), NEID & ~ \\ \hline
        129 & 4706.11 & 4796.91 & 2.55 & 0.88 & 0.93 & $\varnothing$* & 0* & \ssof (KO), NEID & ~ \\ \hline
        128 & 4742.88 & 4834.38 & 2.5 & 1.06 & 0.99 & $\varnothing$ & 0 & \ssof (KO), NEID & ~ \\ \hline
        127 & 4780.22 & 4872.03 & 2.62 & 0.95 & 1 & $\varnothing$ & 0 & \ssof (KO), NEID & ~ \\ \hline
        126 & 4818.17 & 4910.24 & 2.4 & 0.87 & 0.93 & $\varnothing$ & 0 & \ssof (KO), NEID & ~ \\ \hline
        125 & 4856.72 & 4950.4 & 2.69 & 0.82 & 1 & $\varnothing$ & 0 & \ssof (KO), NEID & ~ \\ \hline
        124 & 4895.89 & 4990.35 & 2.67 & 0.89 & 1.23 & $\varnothing$ & 0 & \ssof (KO), NEID & ~ \\ \hline
        123 & 4935.7 & 5030.89 & 2.28 & 0.92 & 0.87 & $\varnothing$ & 0 & \ssof (KO), NEID & ~ \\ \hline
        122 & 4976.17 & 5072.13 & 2.48 & 0.74 & 0.76 & 1 & 0 & \ssof (KO), NEID & First telluric features, H$_2$O \\ \hline
        121 & 5017.3 & 5114.08 & 2.69 & 0.81 & 1.16 & 1 & 0 & \ssof (KO), NEID & H$_2$O \\ \hline
        120 & 5059.11 & 5156.63 & 2.64 & 0.76 & 0.79 & 0 & 0 & \ssof (KO), NEID & ~ \\ \hline
        119 & 5101.63 & 5199.99 & 2.25 & 0.73 & 0.93 & $\varnothing$ & 1 & \ssof (KO), NEID & Mg triplet and MgH band (variance detected) \\ \hline
        118 & 5144.99 & 5244.06 & 16.37 & 0.98 & 15.89 & 5* & 0* & NEID, 2, 3 & Mg triplet and MgH band \\ \hline
        117 & 5188.86 & 5288.88 & 4.52 & 0.89 & 3.68 & $\varnothing$* & 0* & NEID, 3 & ~ \\ \hline
        116 & 5233.58 & 5334.47 & 2.26 & 0.83 & 1.13 & $\varnothing$ & 0 & \ssof (KO), NEID & ~ \\ \hline
        115 & 5279.1 & 5380.85 & 2.4 & 0.88 & 0.87 & $\varnothing$ & 0 & \ssof (KO), NEID & ~ \\ \hline
        114 & 5325.44 & 5428.05 & 2.6 & 1 & 0.95 & $\varnothing$ & 0 & \ssof (KO), NEID & ~ \\ \hline
        113 & 5373.73 & 5476.09 & 2.91 & 1.01 & 1.6 & 1 & 0 & \ssof (KO), NEID & H$_2$O \\ \hline
        112 & 5420.52 & 5524.98 & 2.97 & 1.03 & 1.33 & 0 & 0 & \ssof (KO), NEID & ~ \\ \hline
        111 & 5469.36 & 5574.75 & 2.38 & 1.06 & 1.25 & $\varnothing$ & 0 & \ssof (KO), NEID & ~ \\ \hline
        110 & 5519.09 & 5625.43 & 2.3 & 0.96 & 0.98 & $\varnothing$ & 0 & \ssof (KO), NEID & ~ \\ \hline
        109 & 5569.73 & 5677.04 & 2.69 & 1.04 & 1.13 & 0 & 0 & \ssof (KO), NEID & ~ \\ \hline
        108 & 5621.31 & 5729.6 & 2.65 & 1 & 1.21 & 1 & 0 & \ssof (KO), NEID & H$_2$O \\ \hline
        107 & 5673.85 & 5783.15 & 2.8 & 1.27 & 1.31 & 1 & 0 & \ssof (KO), NEID & H$_2$O \\ \hline
        106 & 5727.38 & 5837.7 & 3.01 & 1.53 & 1.71 & 0 & 0 & \ssof (KO), NEID & ~ \\ \hline
        105 & 5781.93 & 5893.3 & 3.01 & 1.64 & 1.81 & 1 & 0 & \ssof (KO), NEID & H$_2$O \\ \hline
        104 & 5837.53 & 5949.96 & 2.71 & 1.34 & 1.56 & 1 & 0 & \ssof (KO) & H$_2$O \\ \hline
        103 & 5894.21 & 6007.73 & 2.94 & 1.38 & 1.51 & 1 & 0 & \ssof (KO) & H$_2$O \\ \hline
        102 & 5952.01 & 6066.63 & 2.91 & 1.29 & 1.57 & 1 & 0 & \ssof (KO), NEID & H$_2$O \\ \hline
        101 & 6010.94 & 6126.69 & 2.7 & 1.33 & 1.46 & $\varnothing$ & 0 & \ssof (KO), NEID & ~ \\ \hline
        100 & 6071.06 & 6187.95 & 2.47 & 1.01 & 1.13 & $\varnothing$ & 0 & \ssof (KO), NEID & ~ \\ \hline
        99 & 6132.39 & 6250.46 & 2.54 & 1.01 & 1.09 & $\varnothing$ & 0 & \ssof (KO), NEID & ~ \\ \hline
        98 & 6194.97 & 6314.24 & 2.74 & 1.03 & 1.53 & 2 & 0 & \ssofns, NEID & H$_2$O, O$_2$ \\ \hline
        97 & 6258.84 & 6379.33 & 2.85 & 1.16 & 1.62 & 2 & 0 & \ssofns, NEID & H$_2$O, O$_2$ \\ \hline
        96 & 6324.04 & 6445.78 & 2.37 & 1.33 & 1.29 & 1 & 0 & \ssofns, NEID & H$_2$O \\ \hline
        95 & 6390.62 & 6513.29 & 2.71 & 1.45 & 1.26 & 1 & 0 & \ssofns, NEID & H$_2$O \\ \hline
        94 & 6458.61 & 6582.58 & 2.96 & 1.46 & 1.78 & 1 & 0 & \ssofns & H-alpha, H$_2$O \\ \hline
        93 & 6528.06 & 6653.36 & 2.97 & 1.41 & 2 & 1 & 1 & \ssofns, NEID & H-alpha (variance detected), H$_2$O \\ \hline
        92 & 6599.03 & 6725.67 & 3.17 & 1.77 & 2.29 & 0 & 0 & \ssofns, NEID & ~ \\ \hline
        91 & 6671.55 & 6799.58 & 3.13 & 1.83 & 2.35 & $\varnothing$* & 0* & \ssofns, NEID & ~ \\ \hline
        90 & 6745.69 & 6875.13 & 3.73 & 2.19 & 2.93 & 1 & 1 & \ssofns, NEID & O$_2$ \\ \hline
        89 & 6821.49 & 6952.37 & 4.33 & 2.49 & 3.59 & 5* & 0* & \ssofns, NEID & H$_2$O, O$_2$ \\ \hline
        88 & 6899.01 & 7031.37 & 3.31 & 1.71 & 2.17 & 3 & 0 & \ssofns & H$_2$O, O$_2$ \\ \hline
        87 & 6978.32 & 7112.56 & 3.98 & 2.2 & 2.65 & 2 & 0 & \ssofns & H$_2$O, O$_2$ \\ \hline
        86 & 7059.47 & 7195.26 & 3.68 & 1.96 & 2.68 & 4* & 0* & \ssofns & H$_2$O, O$_2$? \\ \hline
        85 & 7142.53 & 7279.91 & 3.26 & 1.4 & 2.1 & 5 & 2 & \ssofns & H$_2$O \\ \hline
        84 & 7227.56 & 7366.57 & 3.58 & 2.13 & 2 & 5* & 0* & \ssofns & H$_2$O \\ \hline
        83 & 7314.65 & 7455.32 & 3.53 & 1.62 & 1.91 & 2 & 2 & \ssofns, NEID & H$_2$O \\ \hline
        82 & 7403.86 & 7546.24 & 3.51 & 2.1 & 1.93 & 2* & 0* & \ssofns, NEID & H$_2$O \\ \hline
        81 & 7495.27 & 7639.4 & 15.15 & 1.71 & 15.22 & 0* & 0* & NEID, 2, 3 & ~ \\ \hline
        80 & 7588.97 & 7734.89 & 5.6 & 1.63 & 5.33 & 5* & 0* & 2, 3 & ~ \\ \hline
        79 & 7685.04 & 7832.8 & 3.85 & 2.17 & 2.5 & 1 & 0 & \ssofns, NEID & O$_2$? \\ \hline
        78 & 7783.58 & 7933.22 & 3.94 & 3.14 & 3.92 & 1 & 0 & \ssofns, NEID & H$_2$O \\ \hline
        77 & 7884.67 & 8036.24 & 3.74 & 1.97 & 3.51 & 2 & 1 & \ssofns & H$_2$O \\ \hline
        76 & 7988.42 & 8141.98 & 4.28 & 2.46 & 3.61 & 2 & 1 & \ssofns & H$_2$O \\ \hline
        75 & 8094.95 & 8250.53 & 5.26 & 1.89 & 4.44 & 5* & 0* & 3 & H$_2$O \\ \hline
        74 & 8204.35 & 8362.02 & 4.77 & 2.65 & 3.88 & 5* & 0* & \ssofns & H$_2$O \\ \hline
        73 & 8316.74 & 8476.57 & 3.73 & 1.43 & 2.74 & 3 & 0 & \ssofns & H$_2$O \\ \hline
        72 & 8432.26 & 8594.29 & 4.58 & 2.23 & 4.65 & 1 & 1 & 3 & Ca IR triplet 1 \& 2 (variance detected), H$_2$O \\ \hline
        71 & 8551.04 & 8715.34 & 4.3 & 2.07 & 4.79 & 0 & 3 & 3 & Ca IR triplet 3 (variance detected, \ssof decomposes it into several features) \\ \hline
        70 & 8673.2 & 8839.84 & 5.01 & 2.15 & 3.71 & 1 & 0 & \ssofns & H$_2$O \\ \hline
        69 & 8798.91 & 8967.95 & 5.66 & 3.06 & 5.31 & 4* & 0* & 2 & RVs stop being useful, H$_2$O \\ \hline
        68 & 8928.32 & 9099.82 & 5.78 & 2.48 & 5.22 & 5 & 4 & 2, 3 & H$_2$O \\ \hline
        67 & 9061.59 & 9235.64 & 5.41 & 3.08 & 4.85 & 5* & 0* & 2 & H$_2$O \\ \hline
        66 & 9198.89 & 9375.57 & 336.8 & 11.64 & 337.28 & 5* & 0* & 1, 2, 3 & H$_2$O \\ \hline
        65 & 9340.42 & 9519.8 & 362.77 & 25.57 & 362.55 & 5* & 0* & 1, 2, 3 & H$_2$O \\ \hline
        64 & 9486.37 & 9668.54 & 38.43 & 5.78 & 38.11 & 5 & 5 & 1, 2, 3 & H$_2$O \\ \hline
        63 & 9636.97 & 9822.01 & 94 & 9.36 & 93.74 & 5* & 0* & 1, 2, 3 & H$_2$O \\ \hline
        62 & 9792.43 & 9980.42 & 17.9 & 11.14 & 18.34 & 3* & 0* & 1, 2 & H$_2$O \\ \hline
        61 & 9952.97 & 10143.94 & 17.52 & 14.97 & 17.12 & 1 & 1 & 1, 2 & H$_2$O \\ \hline
        60 & 10118.94 & 10313.31 & 26.75 & 19.58 & 25.82 & 1 & 0 & 1, 2 & H$_2$O \\ \hline
        59 & 10290.31 & 10487.71 & 26.72 & 22.28 & 27.04 & 0 & 1 & 1, 2 & ~ \\ \hline
        58 & 10467.81 & 10668.97 & 70.23 & 60 & 70.4 & $\varnothing$ & 0 & 1, 2 & ~ \\ \hline
        57 & 10651.78 & 10855.95 & 119.21 & 95.34 & 119.61 & $\varnothing$* & 0* & 1, 2 & ~ \\ \hline
        56 & 10842.54 & 11003.36 & 96.96 & 78.71 & 97.14 & 1* & 0* & 1, 2 & ~ \\ \hline
\end{longtable*}
\clearpage 
}
\subsection{HD 3651}\label{ssec:3651}
HD 3651 is a K dwarf star with a known eccentric sub-Saturn \citep{fischer2003}. The massive, eccentric planet severely limits the possibility of the existence of any dynamically stable shorter period planets.
This makes HD 3651 is an excellent test for the long term precision of extreme precision radial velocity (EPRV) programs \citep{brewer2020}. 
Following the removal of the planet signal, HD 3651 has a RV RMS of $40-60$ \cms \citep{brewer2020,seifahrt2022} as measured by two other high-resolution, fiber-fed, optical spectrographs, MAROON-X \citep[R $\sim$ 85,000, 5000-9200 \AA;][]{seifahrt2016} and EXPRES \citep[R $\sim$ 137,000, 3900-7800 \AA;][]{jurgenson}.

Analysing HD 3651 shows how \ssof performs on a quiet star whose RV signal is dominated by the effects of a planet.
We analyzed 42 observations of HD 3651 taken by NEID from January 2021 to January 2022 with 112 \ssof models, each fit to a single spectral order. 
The RV performance and model complexity for each \ssof model is shown in Tab. \ref{tab:3651}.
Again, a large (but smaller compared to HD 26965) portion of the RV information that can be extracted from the data is in parts of the measured spectra that have little to no telluric transmission or stellar variability (see orders 143-123 in Tab. \ref{tab:3651}).
The SNR of NEID's HD 3651 measurements (SNR $\sim$ 90-280 in the central 1000 pixels of orders used by \ssof to calculate RVs) allows for each of these orders to achieve RV single measurement precisions (SMP) $\sim 1.5$ \ms using only stellar templates.

\begin{figure*}[ht]
\centering
\includegraphics[width=15cm]{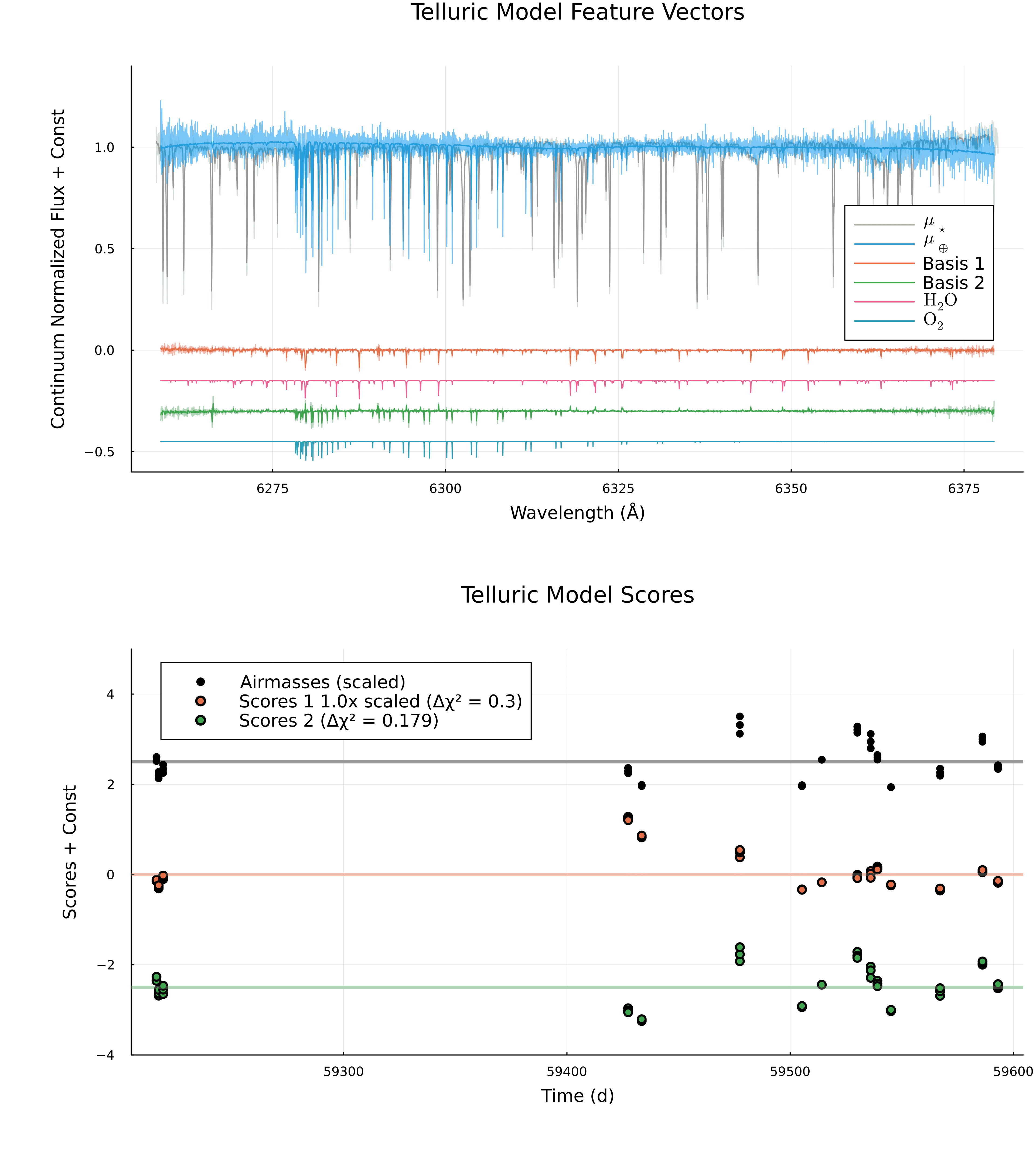}
\caption{\ssof model of order 97 for HD 3561 which separates the two dominant modes of telluric variability. Top: The telluric and stellar templates are shown above in grey and blue while the two telluric feature vectors are shown below in orange and green. For comparison, we also show portions of simulated H$_2$O (pink) and O$_2$ (turquoise) spectra. The simulated lines are those whose line depths are >0.1\% in typical conditions at Kitt Peak. The line locations and emmissivities are from HITRAN2020 \citep{HITRAN2020, HITRAN-850, HITRAN-1179, HITRAN-1387, HITRAN-24}. The green feature vector captures the time-variability of O$_2$ lines while the orange basis vector captures the time-variability of H$_2$O lines. Bottom: The telluric feature scores as a function of time. The scores for the O$_2$ feature vector (green) are highly correlated with observation airmass (black) while the scores for the H$_2$O feature vector (orange) are more erratic but clearly identified in multiple orders (see Fig. \ref{fig:26965_Halpha}). The residuals between $v_{\star,97}$ and $v^{\ssofns}_{\star}$ are shown in Fig. \ref{fig:3651rv}}
\label{fig:3651_tel}
\end{figure*}

Like with HD 26965, most \ssof models of orders $<$122 have telluric transmission and most of those have at least one telluric feature vector to explain temporal variation in the transmission, which can usually be tied to either H$_2$O or O$_2$ lines.
Another example of \ssof separating different telluric species is shown in Fig. \ref{fig:3651_tel}.

\begin{figure*}[ht]
\centering
\includegraphics[width=8.5cm]{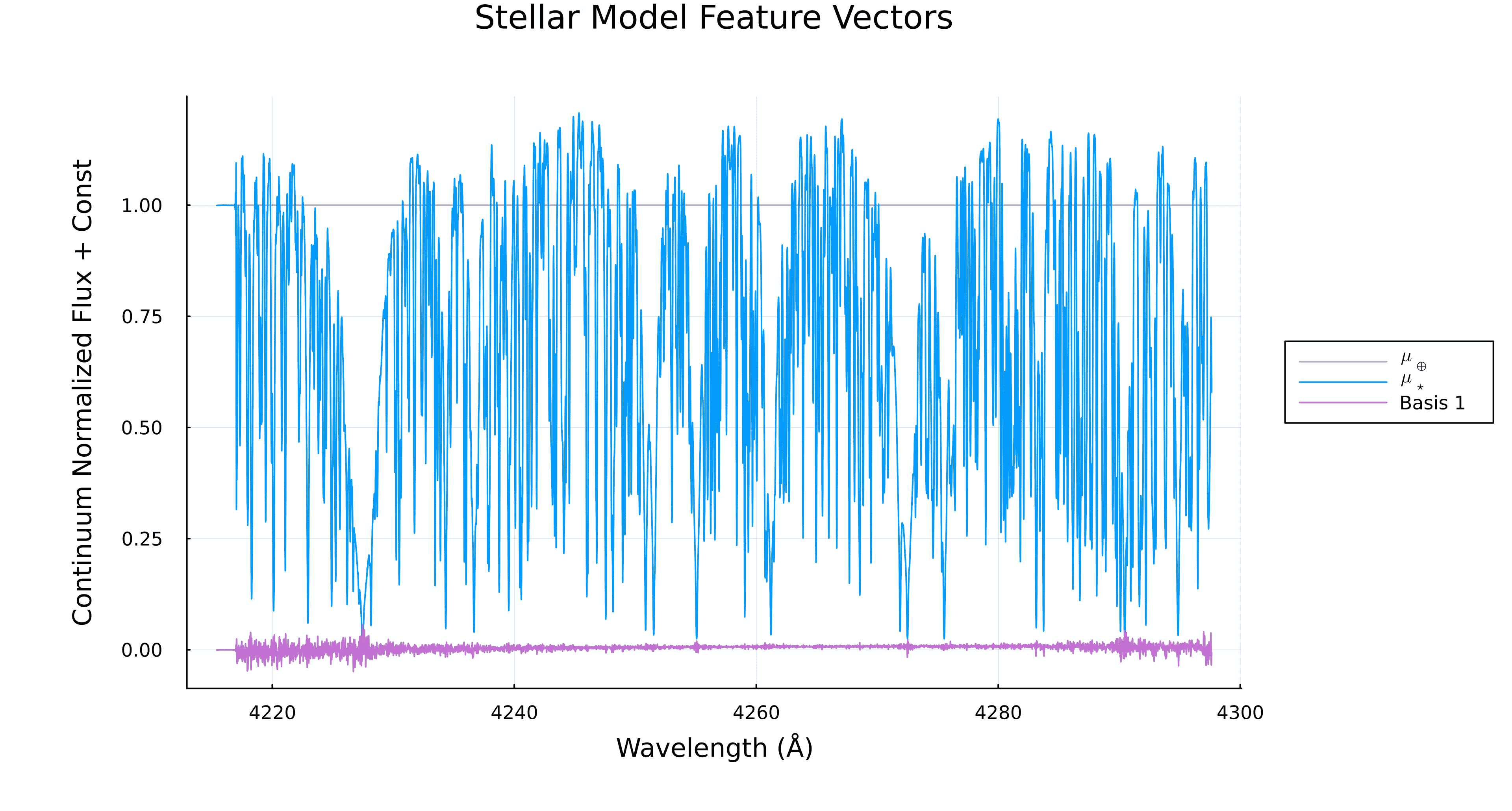}
\includegraphics[width=8.5cm]{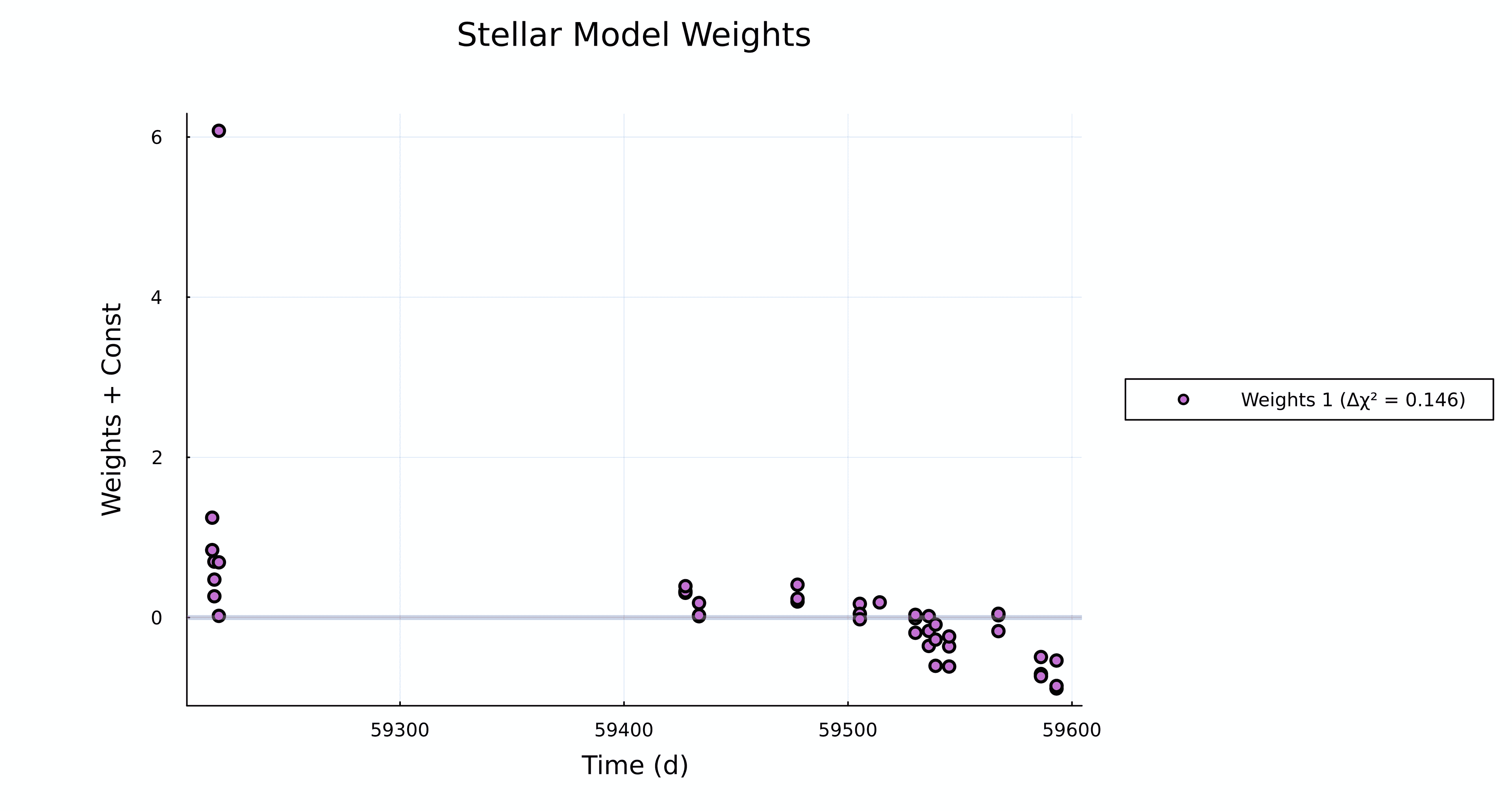}
\caption{\ssof model of order 144 for HD 3561 which identifies Fe I and Ca I variation. Left: The telluric and stellar templates are shown above in grey and blue while the stellar feature vector is shown below (purple). This vector is dominated by variations in deep lines. Right: The stellar feature scores as a function of time. The residuals between $v_{\star,144}$ and $v^{\ssofns}_{\star}$ are shown in Fig. \ref{fig:3651rv}}.
\label{fig:3651_Fe}
\end{figure*}

HD 3651 is a notably quiescent star.
\ssof only detected variability that was plausibly stellar in origin in some deep Fe I and Ca I lines (orders 149-144) and in the Ca IR triplet (orders 72-71). 
The \ssof model which detected variability in Fe I and Ca I is shown in Fig. \ref{fig:3651_Fe}.

\begin{figure*}[ht]
\centering
\includegraphics[width=18cm]{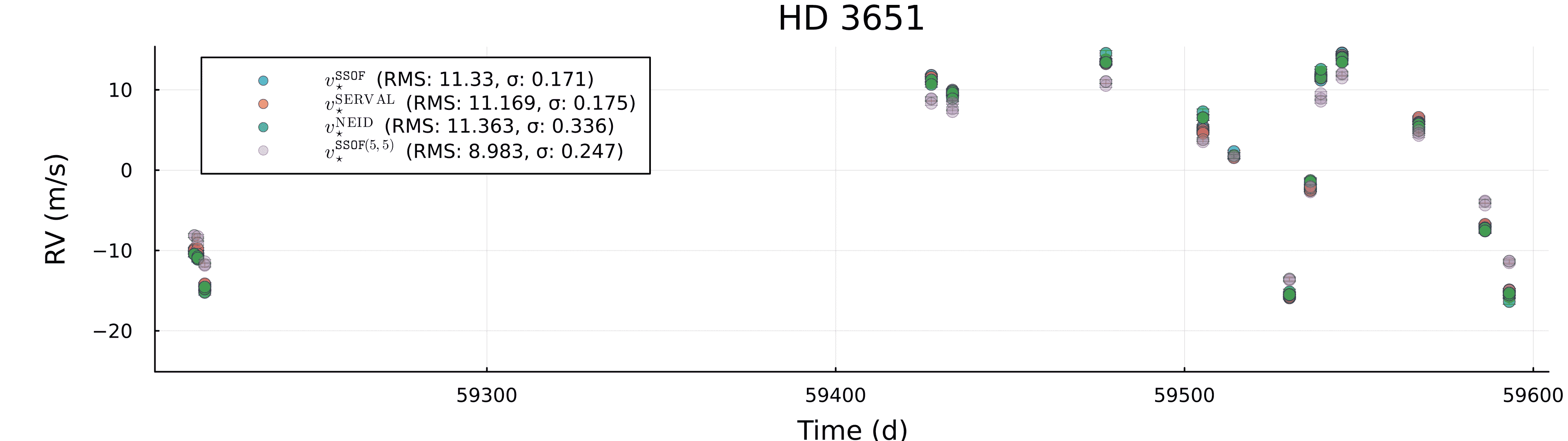}
\includegraphics[width=8.5cm]{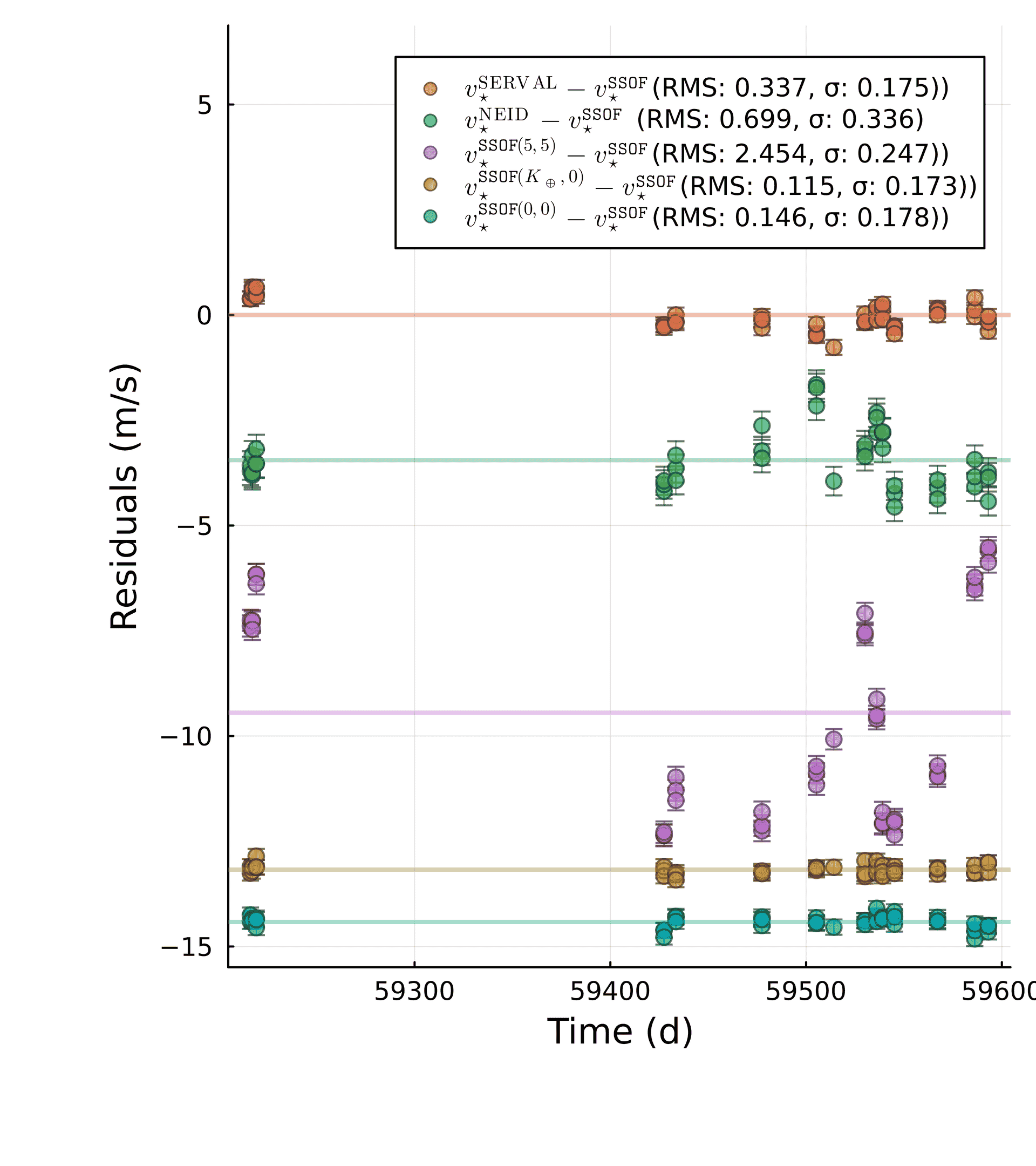}
\includegraphics[width=8.5cm]{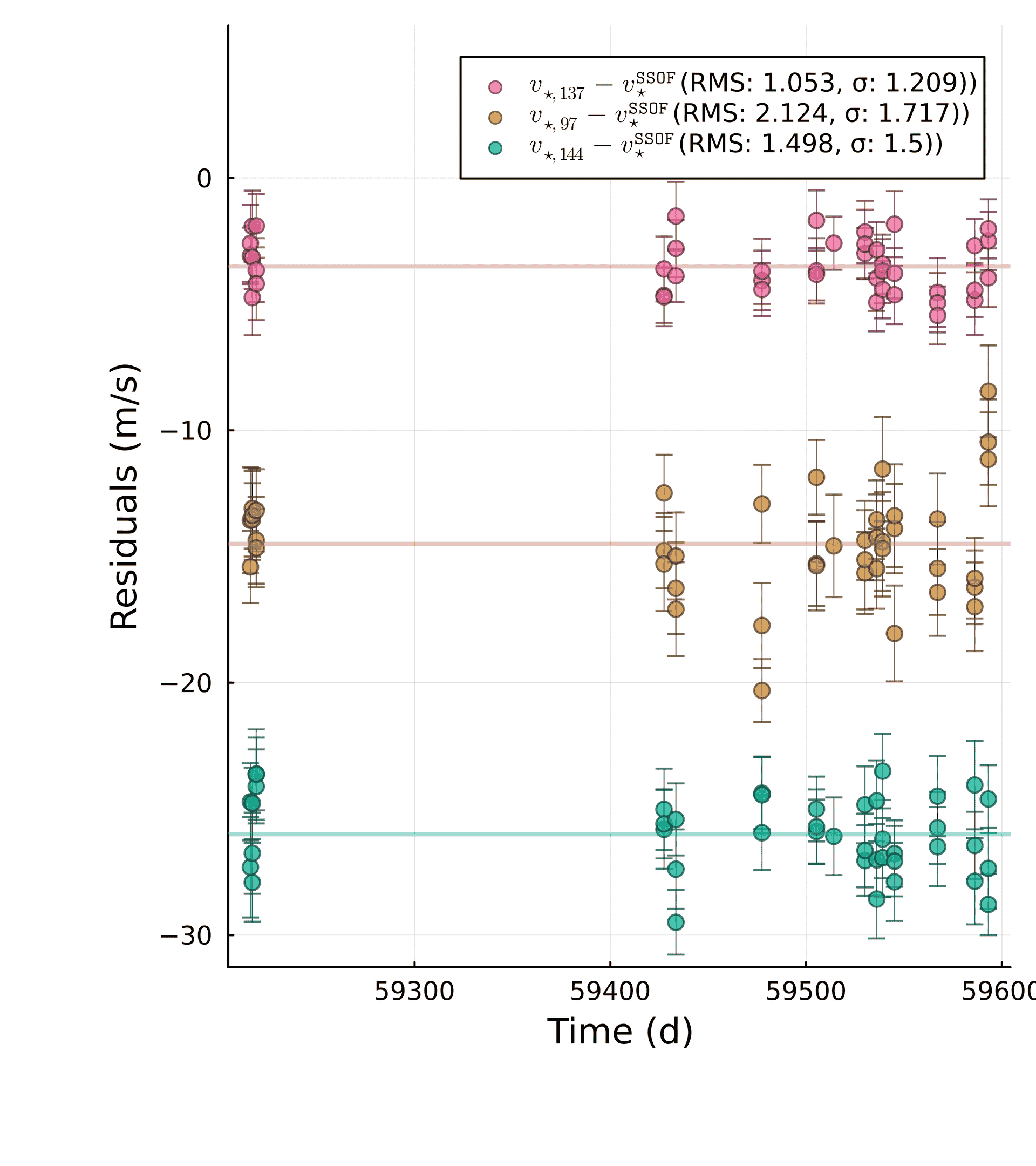}
\caption{Top: The final RV reduction of $v^{\ssofns}_\star$ (blue) for HD 3561 combining the velocity measurements from 77 \ssof order models resulting in a 49\% tighter SMP when compared to the NEID DRP's CCF-based RVs ($v^{\textrm{NEID}}_\star$, green) and a similar SMP and RV RMS when compared to \texttt{SERVAL}'s template-matching RVs ($v^{\textrm{SERVAL}}_\star$, orange). $v^{\ssofns(5,5)}_\star$ (purple) has a lower RMS, but only by taking out a portion of the planetary signal. All RV time series are shown with their corresponding 1-$\sigma$ error bars.
Bottom left: The residual time series for $v^{\textrm{SERVAL}}_\star$ (orange), $v^{\textrm{NEID}}_\star$ (green), and several other versions of $v^{\ssofns}_\star$. Only a few \ssof models used in $v^{\ssofns}_\star$ used stellar variability so $v^{\ssofns(K_\oplus,0)}_\star$ is functionally identical to $v^{\ssofns}_\star$. Ignoring telluric variability only adds 15 \cms of variability. Bottom right: The residual time series for the highlighted \ssof order models. The residual velocities from the ``key order" (pink) H$_2$O and O$_2$ separating order (dark orange), and Fe I variability order (seafoam green) are all essentially consistent with white noise.}
\label{fig:3651rv}
\end{figure*}

The final $v^{\ssofns}_{\star}$ time series is shown in Fig. \ref{fig:3651rv}, along with residual RV time series for several different bulk RV reduction schemes and the highlighted \ssof models including: (1) $v^{\ssofns(5,5)}_\star$ which was constructed using a \ssofns(5,5) model for every order, regardless of AIC; (2) $v^{\ssofns(K_\oplus,0)}_\star$ which was constructed using the AIC-minimum \ssof models that didn't use stellar variability, and (3) $v^{\ssofns(0,0)}_\star$ which was constructed using the \ssof models with only stellar and (optionally) telluric templates.
$v^{\ssofns}_{\star}$ have significantly smaller error estimates than NEID's DRP (down 49\% to 17 \cms from the NEID DRP's 34 \cms) while leaving the planet signal entirely intact. 
They also have a similar SMP and RV RMS when compared to \texttt{SERVAL}'s RVs. 
$v^{\ssofns(5,5)}_\star$ has a lower RMS but only by taking out a portion of the planetary signal and again has larger error bars because fewer orders were selected to be included in the reduction. 
This provides a cautionary counter-example for the use of \ssof models that are not favored in an information-criterion sense.
$v^{\ssofns(K_\oplus,0)}_\star$ is functionally identical to $v^{\ssofns}_\star$ and $v^{\ssofns(0,0)}_\star$ only differs from $v^{\ssofns}_\star$ by 15 \cms of variability.
The residual velocities between single \ssof orders and $v^{\ssofns}_\star$ are largely consistent with white noise.
%
\afterpage{
\begin{longtable*}[p]{|l|l|l|l|l|p{1.15cm}|c|c|c|p{3cm}|}
    \caption{\ssof model results for each tested NEID order for 42 observations of HD 3651 including: echelle grating diffraction order; the observed wavelength range used by \ssof (\AA); the RMS of \ssof order RVs over time, average of RV uncertainties over time, and RMS of the RV residuals comparing each \ssof order RV to the combined \ssof RVs (\ms) at the same time; The number of feature vectors used (* indicates that the model is not the minimum-AIC model); whether or not $v_{\star,j}$ was used in the calculation of $v^{\ssofns}_{\star}$, was a ``key order" (KO), and/or was used by NEID's DRP. A number indicates the reason why the order wasn't used to calculate $v^{\ssofns}_{\star}$ (1 indicates that $\overline{\sigma_{v_\star,j}}$  was too high, 2 indicates that $\textrm{RMS}(v_{\star,j})$  was too high, and 3 indicates that $v_{\star,j}$ was inconsistent with $v^{KO}_{\star}$); any notable comments about the order (H$_2$O indicates the presence of a water-like \ssof basis vector and O$_2$ indicates the presence of a oxygen-like \ssof basis vector). \label{tab:3651}}\\

    \hline
    \multicolumn{10}{|c|}{Beginning of Results Table for HD 3651 (Tab. \ref{tab:3651})}\\
    \hline
    Order & $\textrm{min}(\lambda_{D})$ & $\textrm{max}(\lambda_{D})$ & $\textrm{RMS}(v_{\star,j})$ & $\overline{\sigma_{v_\star,j}}$ & Residual RMS & $K_\oplus$ & $K_\star$ & Used & Comments \\
    \hline
    \endfirsthead

    \hline
    \multicolumn{10}{|c|}{Continuation of Results Table for HD 3651 (Tab. \ref{tab:3651})}\\
    \hline
    Order & $\textrm{min}(\lambda_{D})$ & $\textrm{max}(\lambda_{D})$ & $\textrm{RMS}(v_{\star,j})$ & $\overline{\sigma_{v_\star,j}}$ & Residual RMS & $K_\oplus$ & $K_\star$ & Used & Comments \\
    \hline
    \endhead

    \hline
    \endfoot

    \hline
    \multicolumn{10}{| c |}{End of Results Table for HD 3651 (Tab. \ref{tab:3651})}\\
    \hline
    \endlastfoot
    
        167 & 3667.05 & 3698.85 & 68.95 & 95.32 & 70.21 & $\varnothing$ & 0 & 1, 2 & ~ \\ \hline
        166 & 3691.91 & 3727.55 & 52.28 & 43.91 & 51.67 & $\varnothing$ & 0 & 1, 2 & ~ \\ \hline
        165 & 3712.42 & 3719.12 & 44.98 & 23.59 & 45.62 & $\varnothing$* & 0* & 1, 2 & ~ \\ \hline
        164 & 3730.01 & 3746.22 & 27.84 & 25.55 & 30.91 & $\varnothing$ & 0 & 1, 2 & ~ \\ \hline
        163 & 3751.65 & 3779.51 & 17.72 & 13.08 & 15.3 & $\varnothing$ & 0 & 1 & ~ \\ \hline
        162 & 3769.26 & 3805.53 & 12.77 & 7.14 & 6.9 & $\varnothing$ & 0 & 1 & ~ \\ \hline
        161 & 3791.57 & 3825.46 & 11.1 & 6.75 & 7.98 & $\varnothing$ & 0 & 1 & ~ \\ \hline
        160 & 3812.86 & 3853.91 & 15.82 & 8.3 & 9.19 & $\varnothing$ & 0 & 1 & ~ \\ \hline
        159 & 3839.6 & 3878.96 & 12.98 & 5.01 & 5.96 & $\varnothing$ & 0 & \ssofns, NEID & RVs start to be useful \\ \hline
        158 & 3861.99 & 3916.49 & 11.71 & 3.61 & 3.82 & $\varnothing$ & 0 & \ssofns, NEID & ~ \\ \hline
        157 & 3883.6 & 3929.1 & 11 & 3.19 & 3.95 & $\varnothing$ & 0 & \ssofns, NEID & CaII K \\ \hline
        156 & 3906.73 & 3961.35 & 12.72 & 4.48 & 5.05 & $\varnothing$ & 0 & \ssofns, NEID & CaII K \\ \hline
        155 & 3936.37 & 3992.28 & 11.75 & 3.39 & 3.43 & $\varnothing$ & 0 & \ssofns, NEID & CaII H and K \\ \hline
        154 & 3952.26 & 4018.2 & 12.04 & 2.38 & 2.67 & $\varnothing$ & 0 & \ssofns, NEID & CaII H \\ \hline
        153 & 3977.72 & 4044.47 & 11.5 & 1.98 & 2.19 & $\varnothing$ & 0 & \ssofns, NEID & ~ \\ \hline
        152 & 3999.6 & 4071.07 & 11.87 & 2.12 & 2.25 & $\varnothing$ & 0 & \ssofns, NEID & ~ \\ \hline
        151 & 4025.84 & 4098.04 & 11.53 & 1.9 & 2.27 & $\varnothing$ & 0 & \ssofns, NEID & ~ \\ \hline
        150 & 4052.12 & 4125.35 & 12.17 & 1.83 & 2.24 & $\varnothing$ & 0 & \ssofns, NEID & ~ \\ \hline
        149 & 4079.85 & 4153.07 & 12.05 & 1.73 & 1.98 & $\varnothing$ & 1 & \ssof (KO), NEID & Fe I (variance detected) \\ \hline
        148 & 4105.61 & 4181.11 & 11.59 & 1.79 & 2.03 & $\varnothing$ & 1 & \ssof (KO), NEID & Fe I (variance detected) \\ \hline
        147 & 4130.99 & 4209.54 & 11.49 & 1.67 & 1.52 & $\varnothing$ & 1 & \ssof (KO), NEID & Fe I (variance detected) \\ \hline
        146 & 4159.15 & 4238.37 & 11.39 & 1.62 & 1.89 & $\varnothing$ & 1 & \ssof (KO), NEID & Fe I (variance detected) \\ \hline
        145 & 4188.93 & 4267.61 & 11.64 & 1.68 & 2.06 & $\varnothing$ & 0 & \ssof (KO), NEID & ~ \\ \hline
        144 & 4216.66 & 4297.25 & 11.37 & 1.5 & 1.5 & $\varnothing$ & 1 & \ssof (KO), NEID & Fe I and Ca I (variance detected) \\ \hline
        143 & 4245.3 & 4327.29 & 11.51 & 1.35 & 1.49 & $\varnothing$ & 0 & \ssof (KO), NEID & ~ \\ \hline
        142 & 4275.77 & 4357.77 & 11.19 & 1.57 & 1.51 & $\varnothing$ & 0 & \ssof (KO), NEID & ~ \\ \hline
        141 & 4306.09 & 4388.66 & 11.27 & 1.51 & 1.47 & $\varnothing$ & 0 & \ssof (KO), NEID & ~ \\ \hline
        140 & 4336.29 & 4420.02 & 11.6 & 1.42 & 1.36 & $\varnothing$ & 0 & \ssof (KO), NEID & ~ \\ \hline
        139 & 4367.49 & 4451.81 & 11.3 & 1.4 & 1.33 & $\varnothing$ & 0 & \ssof (KO), NEID & ~ \\ \hline
        138 & 4399.14 & 4484.07 & 11.89 & 1.11 & 1.23 & $\varnothing$ & 0 & \ssof (KO), NEID & ~ \\ \hline
        137 & 4431.26 & 4516.81 & 11.06 & 1.21 & 1.05 & $\varnothing$ & 0 & \ssof (KO), NEID & ~ \\ \hline
        136 & 4463.84 & 4550.01 & 11.42 & 1.36 & 1.13 & $\varnothing$ & 0 & \ssof (KO), NEID & ~ \\ \hline
        135 & 4496.92 & 4583.72 & 11.26 & 1.17 & 1.29 & $\varnothing$ & 0 & \ssof (KO), NEID & ~ \\ \hline
        134 & 4530.48 & 4617.92 & 11.5 & 1.23 & 1.49 & $\varnothing$ & 0 & \ssof (KO), NEID & ~ \\ \hline
        133 & 4564.55 & 4652.65 & 11.67 & 1.22 & 1.3 & $\varnothing$ & 0 & \ssof (KO), NEID & ~ \\ \hline
        132 & 4599.13 & 4687.89 & 11.07 & 1.26 & 1.3 & $\varnothing$ & 0 & \ssof (KO), NEID & ~ \\ \hline
        131 & 4634.25 & 4723.67 & 11.31 & 1.25 & 1.39 & $\varnothing$ & 0 & \ssof (KO), NEID & ~ \\ \hline
        130 & 4669.9 & 4760.01 & 11.44 & 1.33 & 1.18 & $\varnothing$ & 0 & \ssof (KO), NEID & ~ \\ \hline
        129 & 4706.11 & 4796.91 & 11.37 & 1.22 & 1.21 & $\varnothing$ & 0 & \ssof (KO), NEID & ~ \\ \hline
        128 & 4742.88 & 4834.38 & 11.37 & 1.31 & 1.19 & $\varnothing$ & 0 & \ssof (KO), NEID & ~ \\ \hline
        127 & 4780.22 & 4872.45 & 11.37 & 1.35 & 1.29 & $\varnothing$ & 0 & \ssof (KO), NEID & ~ \\ \hline
        126 & 4818.17 & 4911.12 & 11.86 & 1.18 & 1.25 & $\varnothing$ & 0 & \ssof (KO), NEID & ~ \\ \hline
        125 & 4856.72 & 4950.4 & 11.51 & 1.16 & 1.21 & $\varnothing$ & 0 & \ssof (KO), NEID & ~ \\ \hline
        124 & 4895.89 & 4990.35 & 10.67 & 1.24 & 1.48 & $\varnothing$ & 0 & \ssof (KO), NEID & ~ \\ \hline
        123 & 4935.7 & 5030.89 & 11.3 & 1.04 & 1.06 & $\varnothing$ & 0 & \ssof (KO), NEID & ~ \\ \hline
        122 & 4976.17 & 5072.13 & 10.73 & 0.98 & 1.55 & 2* & 0* & \ssof (KO), NEID & First telluric features, H$_2$O  \\ \hline
        121 & 5017.3 & 5114.08 & 10.41 & 1.19 & 2.56 & 2* & 0* & \ssof (KO), NEID & H$_2$O \\ \hline
        120 & 5059.11 & 5156.36 & 10.72 & 1.1 & 1.83 & $\varnothing$* & 0* & \ssof (KO), NEID & ~ \\ \hline
        119 & 5101.63 & 5199.99 & 11.49 & 1.28 & 1.02 & $\varnothing$* & 0* & \ssof (KO), NEID & ~ \\ \hline
        118 & 5144.87 & 5243.79 & 22.02 & 1.13 & 16.31 & 1* & 0* & NEID, 3 & ~ \\ \hline
        117 & 5188.85 & 5288.88 & 11.65 & 1.05 & 2.56 & $\varnothing$ & 0 & NEID, 3 & ~ \\ \hline
        116 & 5233.59 & 5334.19 & 11.72 & 1.13 & 1.09 & $\varnothing$ & 0 & \ssof (KO), NEID & ~ \\ \hline
        115 & 5279.1 & 5380.85 & 11.61 & 1.32 & 1.37 & $\varnothing$ & 0 & \ssof (KO), NEID & ~ \\ \hline
        114 & 5325.42 & 5428.05 & 11.69 & 1.28 & 1.35 & $\varnothing$ & 0 & \ssof (KO), NEID & ~ \\ \hline
        113 & 5372.55 & 5476.09 & 12.31 & 1.3 & 1.96 & 0 & 0 & \ssof (KO), NEID & ~ \\ \hline
        112 & 5420.52 & 5524.98 & 11.44 & 1.29 & 1.51 & 0 & 0 & \ssof (KO), NEID & ~ \\ \hline
        111 & 5469.36 & 5574.46 & 11.11 & 1.35 & 1.54 & $\varnothing$ & 0 & \ssof (KO), NEID & ~ \\ \hline
        110 & 5519.09 & 5625.43 & 11.33 & 1.3 & 1.43 & $\varnothing$ & 0 & \ssof (KO), NEID & ~ \\ \hline
        109 & 5569.73 & 5676.74 & 11.37 & 1.37 & 1.54 & $\varnothing$ & 0 & \ssof (KO), NEID & ~ \\ \hline
        108 & 5621.31 & 5729.3 & 10.76 & 1.43 & 2.09 & 1 & 0 & \ssof (KO), NEID & H$_2$O \\ \hline
        107 & 5673.85 & 5782.85 & 11.61 & 1.53 & 1.57 & 1 & 0 & \ssof (KO), NEID & H$_2$O \\ \hline
        106 & 5727.38 & 5837.4 & 11.29 & 1.63 & 1.83 & 0 & 0 & \ssof (KO), NEID & ~ \\ \hline
        105 & 5781.93 & 5892.99 & 11.56 & 1.97 & 1.94 & 1* & 0* & \ssof (KO), NEID & H$_2$O \\ \hline
        104 & 5837.53 & 5949.66 & 11.27 & 1.94 & 2.08 & 2* & 0* & \ssof (KO), NEID & H$_2$O \\ \hline
        103 & 5894.21 & 6007.42 & 12.08 & 1.85 & 2.06 & 1 & 0 & \ssof (KO) & H$_2$O \\ \hline
        102 & 5952.01 & 6066.31 & 12.42 & 1.63 & 2.07 & 1 & 0 & \ssof (KO), NEID & H$_2$O \\ \hline
        101 & 6010.94 & 6126.37 & 11.23 & 1.74 & 1.45 & $\varnothing$ & 0 & \ssof (KO), NEID & ~ \\ \hline
        100 & 6071.06 & 6187.63 & 11.22 & 1.34 & 1.43 & $\varnothing$ & 0 & \ssof (KO), NEID & ~ \\ \hline
        99 & 6132.39 & 6250.13 & 11.03 & 1.34 & 1.56 & $\varnothing$ & 0 & \ssof (KO), NEID & ~ \\ \hline
        98 & 6194.97 & 6313.91 & 11.53 & 1.71 & 1.53 & 2 & 0 & \ssofns, NEID & H$_2$O, O$_2$  \\ \hline
        97 & 6258.84 & 6379 & 10.8 & 1.72 & 2.12 & 2* & 0* & \ssofns, NEID & H$_2$O, O$_2$  \\ \hline
        96 & 6324.04 & 6445.44 & 12.63 & 2.46 & 2.49 & 1 & 0 & \ssofns, NEID & H$_2$O \\ \hline
        95 & 6390.62 & 6513.63 & 12.11 & 1.79 & 2 & 1 & 0 & \ssofns, NEID & H$_2$O \\ \hline
        94 & 6458.61 & 6582.92 & 12.66 & 1.69 & 2.41 & 1 & 0 & \ssofns & H-alpha, H$_2$O  \\ \hline
        93 & 6528.06 & 6653.7 & 11.23 & 1.89 & 2.03 & 1 & 0 & \ssofns, NEID & H-alpha, H$_2$O  \\ \hline
        92 & 6599.03 & 6726.02 & 10.94 & 2.22 & 3 & $\varnothing$ & 0 & \ssofns, NEID & ~ \\ \hline
        91 & 6671.55 & 6799.93 & 11.73 & 2.26 & 2.43 & $\varnothing$ & 0 & \ssofns, NEID & ~ \\ \hline
        90 & 6745.69 & 6875.48 & 11.73 & 2.94 & 2.76 & 2* & 0* & \ssofns, NEID & O$_2$ \\ \hline
        89 & 6821.49 & 6952.73 & 12.48 & 2.89 & 3.11 & 3* & 0* & \ssofns, NEID & H$_2$O, O$_2$  \\ \hline
        88 & 6899.01 & 7031.74 & 12.46 & 2.56 & 8.23 & 4* & 0* & 3 & H$_2$O, O$_2$  \\ \hline
        87 & 6978.31 & 7112.56 & 13.21 & 2.34 & 3.55 & 1 & 0 & \ssofns & H$_2$O \\ \hline
        86 & 7059.47 & 7194.89 & 9.77 & 2.25 & 12.73 & 5* & 0* & 3 & H$_2$O, O$_2$?  \\ \hline
        85 & 7142.53 & 7279.53 & 39.32 & 3.6 & 38.26 & 4* & 0* & 2, 3 & H$_2$O \\ \hline
        84 & 7227.56 & 7366.19 & 40.15 & 3.34 & 38.32 & 3 & 0 & 2, 3 & H$_2$O \\ \hline
        83 & 7314.65 & 7454.94 & 11.35 & 1.74 & 3.89 & 2* & 0* & NEID, 3 & H$_2$O \\ \hline
        82 & 7403.86 & 7545.85 & 11.75 & 2.25 & 2.45 & 2* & 0* & \ssofns, NEID & H$_2$O \\ \hline
        81 & 7495.27 & 7639.4 & 28.24 & 2.27 & 19.14 & 0* & 0* & NEID, 2, 3 & ~ \\ \hline
        80 & 7588.97 & 7734.89 & 10.05 & 2.47 & 5.14 & 5* & 0* & 3 & O$_2$ \\ \hline
        79 & 7685.04 & 7832.8 & 11.44 & 2.25 & 2.87 & 1 & 0 & \ssofns, NEID & O$_2$ \\ \hline
        78 & 7783.58 & 7933.21 & 12.77 & 3.36 & 5.57 & 2* & 0* & \ssofns, NEID & H$_2$O \\ \hline
        77 & 7884.67 & 8036.24 & 11.82 & 2.41 & 2.71 & 2* & 0* & \ssofns & H$_2$O \\ \hline
        76 & 7988.43 & 8141.98 & 12.66 & 2.3 & 3.98 & 3* & 0* & \ssofns & H$_2$O \\ \hline
        75 & 8094.95 & 8250.53 & 11.07 & 2.47 & 6.29 & 5* & 0* & 3 & H$_2$O \\ \hline
        74 & 8204.35 & 8362.02 & 40.6 & 3.27 & 35.39 & 5* & 0* & 2, 3 & H$_2$O \\ \hline
        73 & 8316.74 & 8476.57 & 11.51 & 2.07 & 3.94 & 3* & 0* & 3 & H$_2$O \\ \hline
        72 & 8432.26 & 8594.29 & 16.38 & 2.82 & 11.24 & 1 & 1 & 3 & Ca IR triplet 1 \& 2 (variance detected), H$_2$O  \\ \hline
        71 & 8551.04 & 8715.34 & 13.61 & 2.85 & 4.51 & 0 & 1 & \ssofns & Ca IR triplet 3 (variance detected, \ssof decomposes it into several features)  \\ \hline
        70 & 8673.2 & 8839.84 & 10.65 & 2.55 & 2.67 & 1* & 0* & \ssofns & H$_2$O \\ \hline
        69 & 8798.91 & 8967.95 & 13.75 & 3.37 & 5.27 & 3* & 0* & \ssofns & RVs stop being useful, H$_2$O  \\ \hline
        68 & 8928.32 & 9099.82 & 61.56 & 6.44 & 54.53 & 2* & 0* & 1, 2, 3 & H$_2$O \\ \hline
        67 & 9061.59 & 9235.64 & 25.29 & 2.16 & 26.87 & 3* & 0* & 2, 3 & H$_2$O \\ \hline
        66 & 9198.89 & 9375.57 & 915.7 & 6.82 & 912.09 & 0* & 0* & 1, 2, 3 & H$_2$O \\ \hline
        65 & 9340.43 & 9519.8 & 1646.16 & 204.08 & 1640.38 & 1* & 0* & 1, 2, 3 & H$_2$O \\ \hline
        64 & 9486.37 & 9668.54 & 1254.58 & 67.05 & 1253.52 & 1* & 0* & 1, 2, 3 & H$_2$O \\ \hline
        63 & 9636.97 & 9822.01 & 20.68 & 6.52 & 23.76 & 3* & 0* & 1, 3 & H$_2$O \\ \hline
        62 & 9792.43 & 9980.42 & 30.84 & 16.95 & 27.58 & 1* & 0* & 1, 2, 3 & H$_2$O \\ \hline
        61 & 9952.97 & 10143.94 & 28.72 & 25.92 & 27.22 & 2* & 0* & 1, 2 & H$_2$O \\ \hline
    60 & 10118.94 & 10313.31 & 26.16 & 21.89 & 28 & 0* & 0* & 1, 2 & H$_2$O \\ \hline
        59 & 10290.31 & 10487.71 & 37.32 & 28.96 & 34.12 & $\varnothing$ & 0 & 1, 2 & ~ \\ \hline
        58 & 10467.81 & 10657.76 & 74.59 & 97.74 & 77.23 & $\varnothing$ & 0 & 1, 2 & ~ \\ \hline
        57 & 10722.11 & 10759.09 & 537.8 & 345.89 & 538.85 & $\varnothing$ & 0 & 1, 2 & ~ \\ \hline
        56 & 10911.82 & 10916.96 & 3595.43 & 1903.25 & 3594.28 & $\varnothing$ & 0 & 1, 2 & ~ \\ \hline
\end{longtable*}
\clearpage 
}
We analyzed HD 3651 to evaluate \ssofns's ability to provide precise RVs 
without interfering with planet signals in a well-understood system.
We fit a linearized version of the Keplerian equations \citep{wright2009} (with a velocity offset) and their analytical gradients (in \julians) to the RVs using the L-BFGS optimization algorithm \citep{lbfgs} implemented in \optimjl \citep{Mogensen2018}.
We minimized a negative log-likelihood assuming the residuals between $v_\star$ and the Keplerian model at each time were Gaussian independent with uncertainties $\sigma_{v_\star,n}$ (i.e. we did not account for any stellar variability) including priors on the Keplerian parameters (see Appendix B of \citet{gilbertson2020}).
We finalized the fit by fitting to a fully non-linear infinitely-differentiable version of the Keplerian equations reparameterized so none of its variables yield discontinuities \citep{pal2009}.
We performed this Keplerian fitting on $v^{\ssofns}_\star$, $v^{\ssofns(5,5)}_\star$, the NEID DRP RVs, and the EXPRES RVs (taken from August 2019 to February 2020) published in \citet{brewer2020}.
The resulting maximum a posteriori (MAP) estimates for the planet parameters, $\theta$, are shown in Tab. \ref{tab:3651_orb} along with a replication of the results from \citet{brewer2020}.
The MAP Keplerian model of $v^{\ssofns}_\star$ is shown in Fig. \ref{fig:3651_kep}.

\begin{deluxetable*}{l|ccccc}
\tablecaption{MAP Keplerian models for HD 3651 b\label{tab:3651_orb} using the combined RVs from the AIC-minimum \ssof models, \ssofns(5,5) models, NEID DRP RVs, and 61 EXPRES RVs from \citep{brewer2020}. These include the orbital period, time of periastron passage, eccentricity, argument of periapsis, and velocity semi-amplitude. We have also reproduced the column from Tab. 4 in \citet{brewer2020}. Their uncertainties were obtained by looking at the distribution of results from 1000 bootstrap Monte Carlo (MC) trials where they fit a Keplerian orbit to the data, subtracted the maximum likelihood estimate (MLE) model, scrambled the residuals, and added the scrambled residuals back to the MLE Keplerian model velocities.}
\tablewidth{0pt}
\tablehead{
\colhead{} & \colhead{\ssofns}  & \colhead{\ssofns(5,5)} & \colhead{NEID DRP} & \colhead{EXPRES (re-analysis)} & \colhead{EXPRES \citep{brewer2020}}}
\startdata
$P$ (d)   & $62.171  \pm 0.010$ & $62.233  \pm 0.017$  & $62.172 \pm 0.019$  & $61.881 \pm 0.033$ & $61.88 \pm 0.55$ \\
$T_p$ (d) & $58727.00 \pm 0.02 $ & $58726.20 \pm 0.04 $ & $58726.83 \pm 0.04$ & $58726.55 \pm 0.03$ & $58726.2 \pm 1.2$ \\
$e$       & $0.603   \pm 0.003$ & $0.628   \pm 0.006$  & $0.620 \pm 0.006$   & $0.607 \pm 0.004$ & $0.606 \pm 0.09$ \\
$\omega$  & $242.49   \pm 0.39$  & $240.51   \pm 0.74$   & $241.2 \pm 0.76$    & $243.8 \pm 0.40$ & $243.8 \pm 23.4$ \\
$K$ (\ms) & $16.20   \pm 0.07 $ & $13.18   \pm 0.13 $  & $16.44 \pm 0.17 $   & $16.33 \pm 0.08 $ & $16.93 \pm 0.22 $ \\
${\rm RMS}$ (\ms) & 0.60 & 0.85 & 0.71 & 0.56 & 0.58 \\
\enddata
\end{deluxetable*}

\begin{figure*}[ht]
\centering
\includegraphics[width=8.5cm]{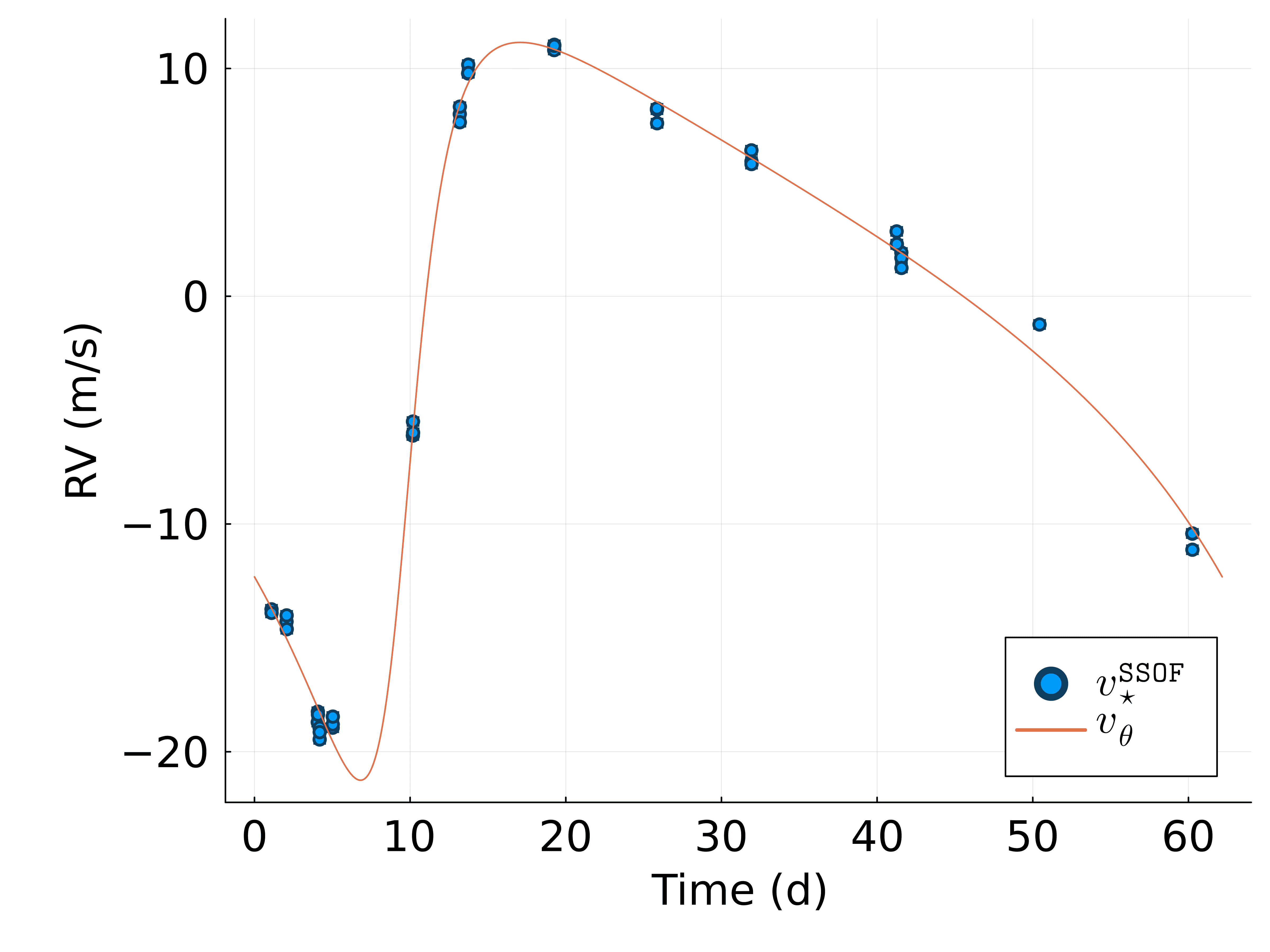}
\includegraphics[width=8.5cm]{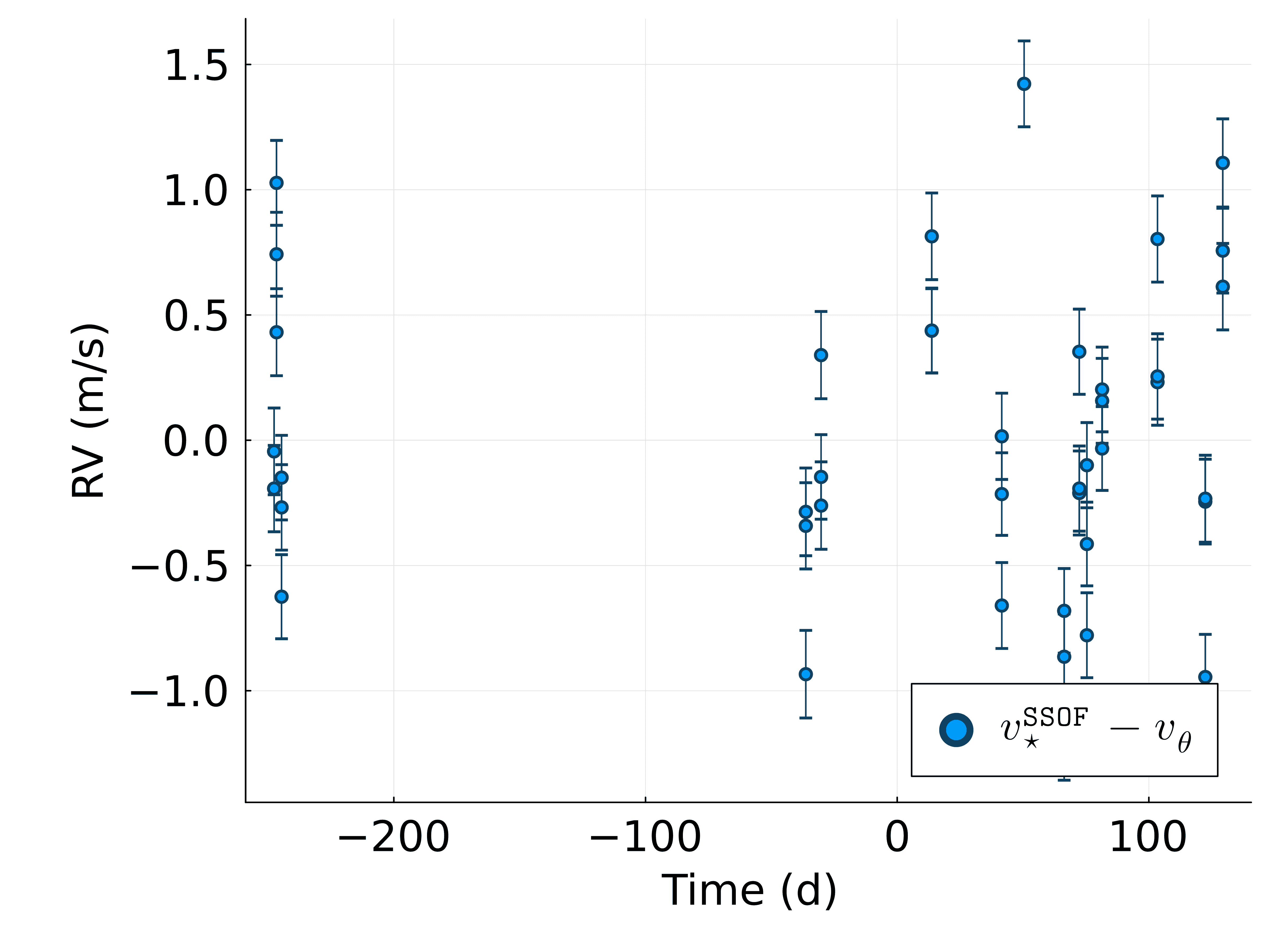}
\caption{Keplerian model of $v^{\ssofns}_\star$. Left: $v^{\ssofns}_\star$ after phase folding at the planet period (blue) and the MAP Keplerian RVs (orange). Center: $v^{\ssofns}_\star - $ MAP Keplerian RVs residuals (blue). Much more remains in the residuals than can be adequately explained by the measurement noise alone.}
\label{fig:3651_kep}
\end{figure*}

Our variances on the orbital parameters, $\sigma_\theta^2$, are taken from the diagonal of the covariance matrix estimated from the inverse of the analytical Hessian of the log-likelihood (like in \eqref{eq:hessian_to_cov}).
The \citet{brewer2020} uncertainties were obtained in a fundamentally different manner by looking at the distribution of results from 1000 bootstrap Monte Carlo (MC) trials where they fit a Keplerian orbit to the data, subtracted the maximum likelihood estimate (MLE) model, scrambled the residuals, and added the scrambled residuals back to the MLE Keplerian model velocities. 
The uncertainties estimated from our method are much smaller than those reported in \citet{brewer2020}, and are certainly smaller than they should be because our model specification is inadequate.
Our model assumes that the all of the residuals are due solely to measurement uncertainty which is demonstrably untrue as the RMS scatter of our residuals is $\sim60$ \cms while our measurement uncertainties are only 17 \cms. 
We reanalysed the \citet{brewer2020} RVs to enable a methodologically consistent comparison.

The orbital parameters for HD 3651 b obtained from analyzing $v^{\ssofns}_\star$, $\theta^{\ssofns}$, are, except for time of periastron passage, consistent within 3-$\sigma$ to those found from analyzing the NEID DRP RVs.
The uncertainties on $\theta^{\ssofns}$ are lower than those on on $\theta^{\textrm{NEID}}$ proportional to the reduction between their respective $\sigma_{v_\star}$.
$\theta^{\ssofns(5,5)}$ are also similar, with the notable exception of a much lower velocity semi-amplitude and a larger residual RMS.
This again suggests that the \ssofns(5,5) models can overfit the spectra and remove a portion of the planetary RV signal. 
The low RMS of 60 \cms after removing the planet signal is consistent with the previous results analyzing MAROON-X and EXPRES data \citep{seifahrt2022,brewer2020}.
The current implementation of \ssof with the extant coverage of HD 3651 with NEID suggests that NEID's long-term precision on stars, in real life conditions, is better than 60 \cms.

\begin{figure*}[ht]
\centering
\includegraphics[width=8.5cm]{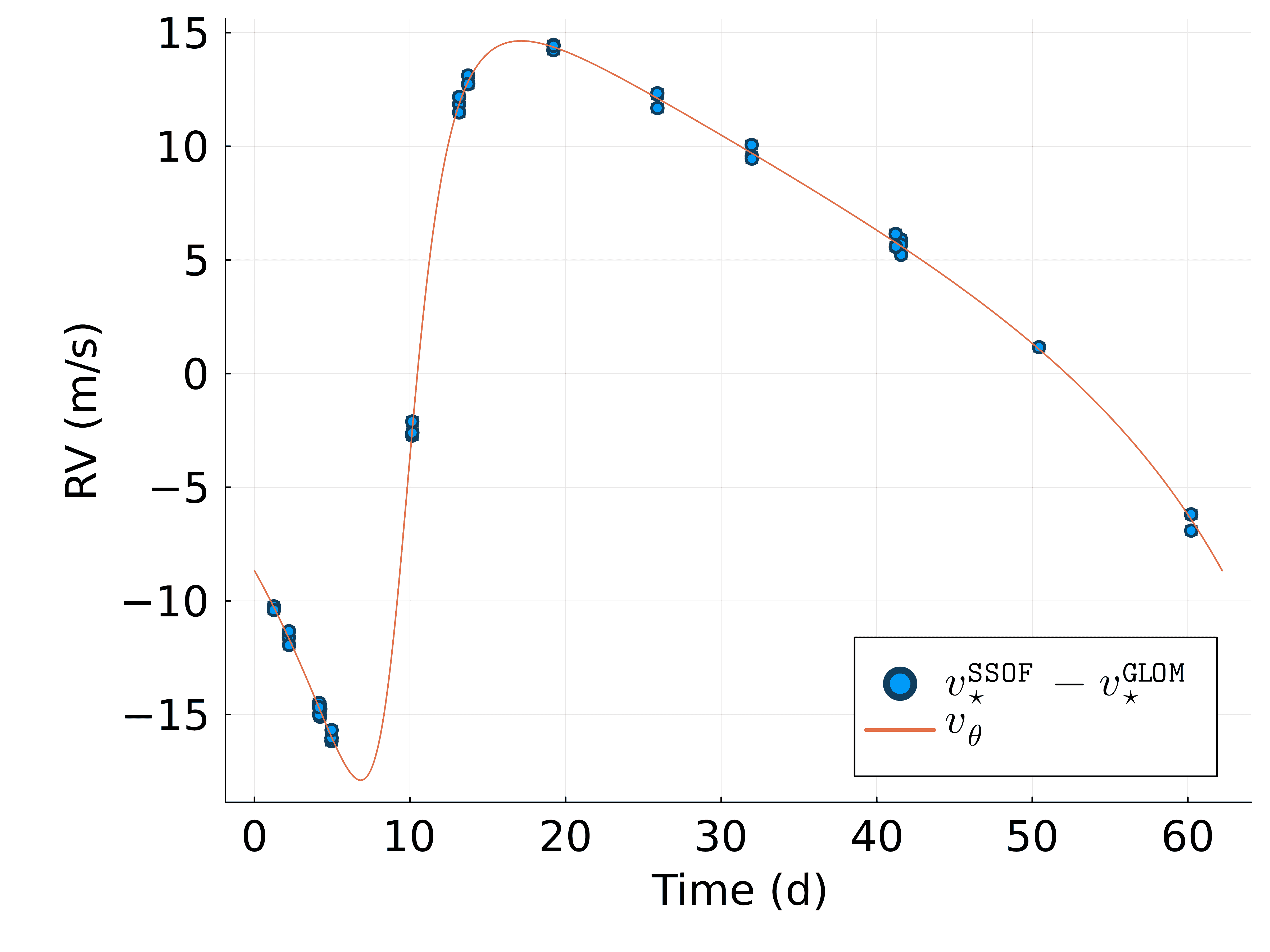}
\includegraphics[width=8.5cm]{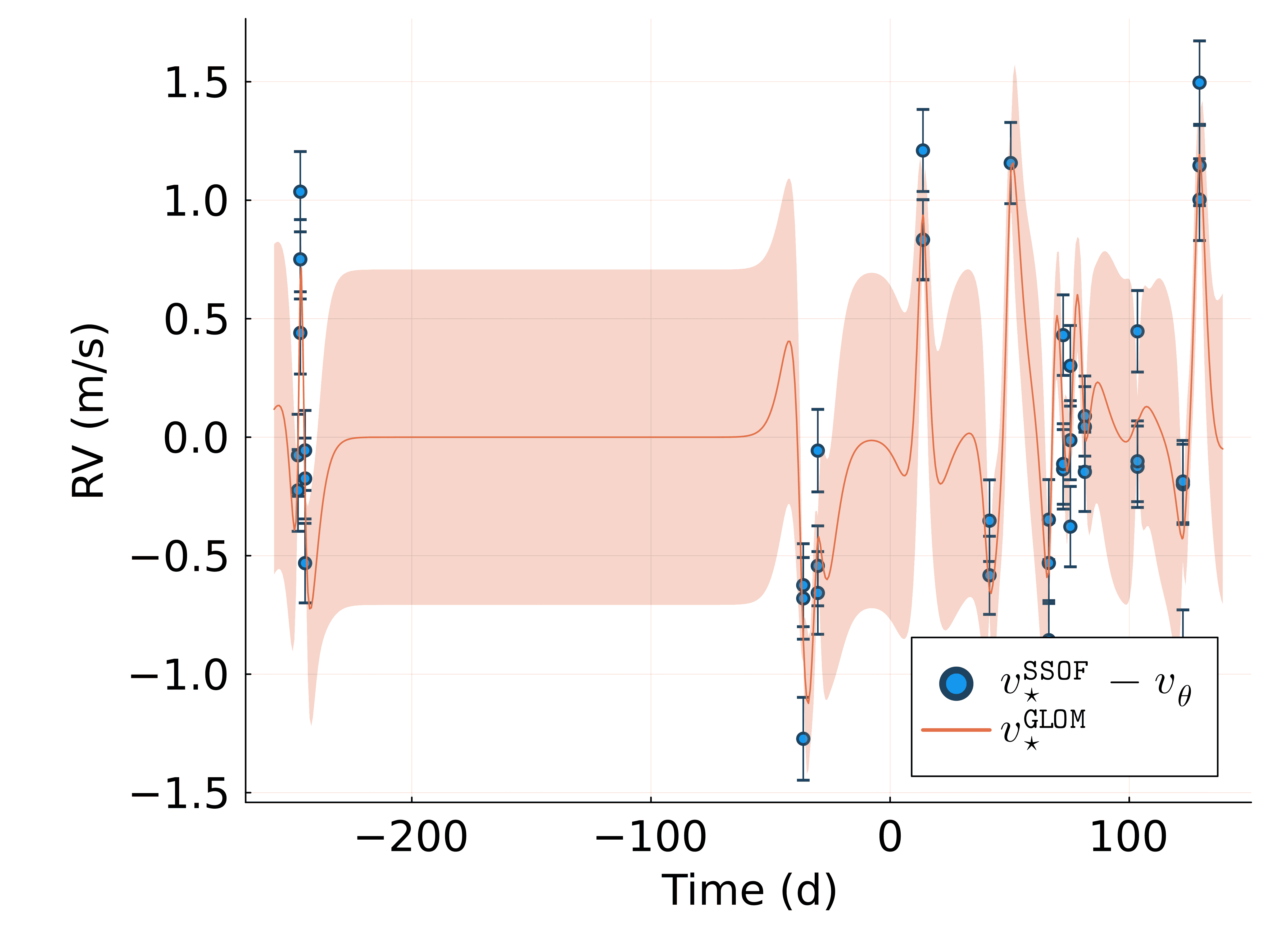}
\caption{\texttt{GLOM} and Keplerian model of $v^{\ssofns}_\star$. Left: $v^{\ssofns}_\star$ after phase folding at the planet period (blue) and the MAP Keplerian RVs (orange). Center: $v^{\ssofns}_\star - $ MAP Keplerian RVs residuals (blue) and the posterior mean \texttt{GLOM} RVs (orange).}
\label{fig:3651_glom}
\end{figure*}
\begin{figure*}[ht]
\centering
\includegraphics[width=5.5cm]{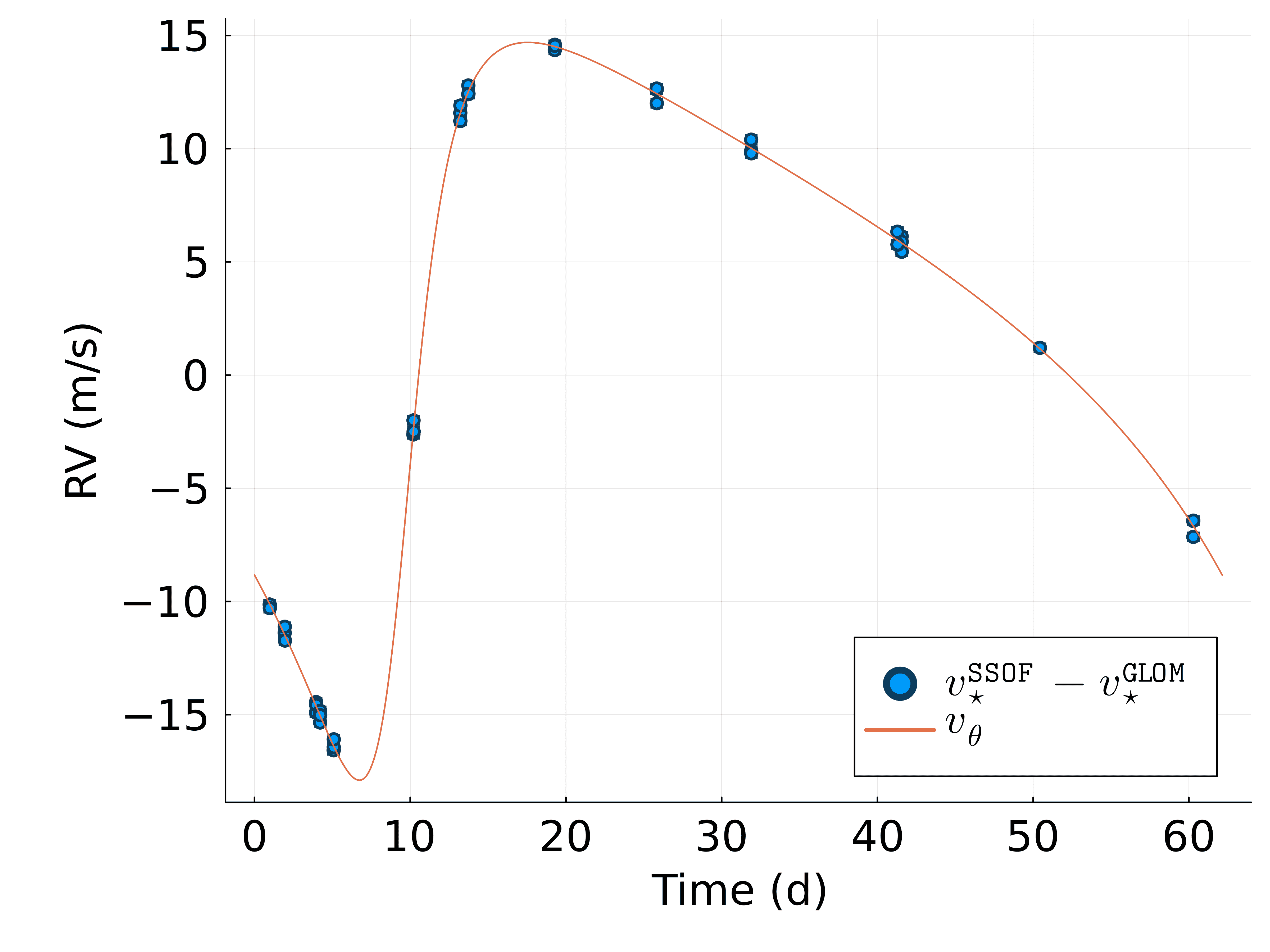}
\includegraphics[width=5.5cm]{HD_3651_Ha_fit3_rv_nokep}
\includegraphics[width=5.5cm]{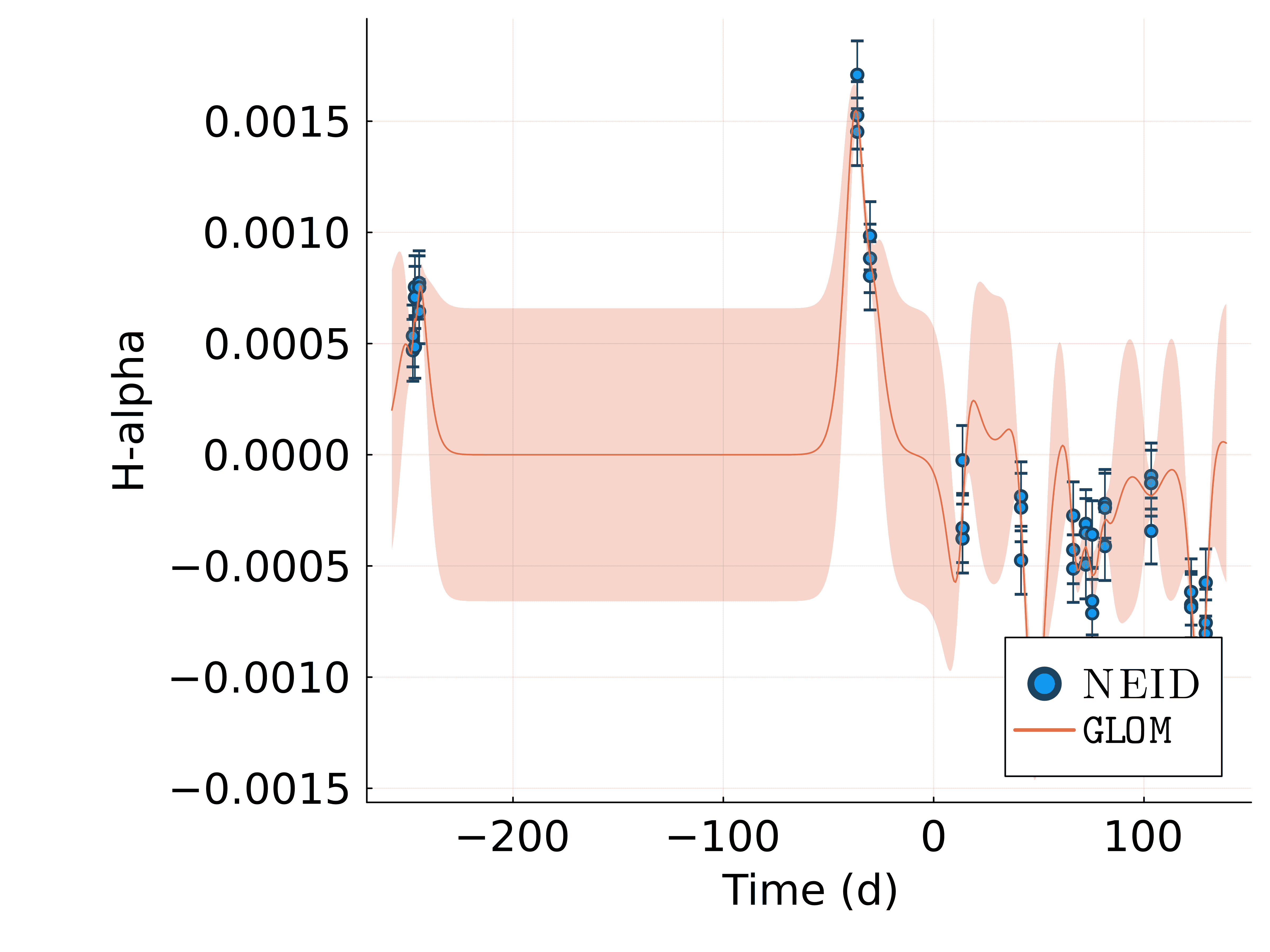}
\caption{\texttt{GLOM} and Keplerian model of $v^{\ssofns}_\star$ and H-alpha. Left: $v^{\ssofns}_\star$ after phase folding at the planet period (blue) and the MAP Keplerian RVs (orange). Center: $v^{\ssofns}_\star - $ MAP Keplerian RVs residuals (blue) and the posterior mean \texttt{GLOM} RVs (orange). Right: NEID's H-alpha measurements and posterior mean \texttt{GLOM} H-alpha model(orange).}
\label{fig:3651_glom_ha}
\end{figure*}

We also investigated including a Gaussian process (GP) model to account for stellar variability.
We used \texttt{GPLinearODEMaker.jl} (\texttt{GLOM}) \citep{gilbertson2020} to fit a linear combination of a latent GP (with a Mat\'ern $^5/_2$ kernel) and its derivatives to the RVs and an indicator simultaneously.
We modeled the RVs as the Keplerian RV signal plus the latent GP and its derivative and H-alpha (when included) as the latent GP and its second derivative (see Eq. 12-13 in \citet{gilbertson2020})).
We performed this analysis using $v^{\ssofns}_\star$, $v^{\ssofns(5,5)}_\star$, and the NEID DRP RVs.
The resulting MAP estimates for the planet parameters, $\theta$, and the GP lengthscale, $l_{GP}$, are shown in Tab. \ref{tab:3651_orb_gp}.
The \texttt{GLOM} and Keplerian models of $v^{\ssofns}_\star$ (with and without H-alpha) are shown in Figs. \ref{fig:3651_glom}-\ref{fig:3651_glom_ha}.

The orbital parameters for HD 3651 b obtained from the \texttt{GLOM} analyses are consistent with their non-\texttt{GLOM} counterparts, largely due to the stellar variability being dwarfed by the planetary signal.
Modeling just the RVs with \texttt{GLOM} only had the effect of slightly increasing orbital periods.
Modeling the RVs and H-alpha with \texttt{GLOM} had the effect of decreasing orbital periods and eccentricities and increasing the MAP GP lengthscale, $l_{GP}$.
The uncertainties on $\theta$ when performing the \texttt{GLOM} analyses are 5-10x larger when compared to assuming that no stellar variability remained in the RV time series.
We believe that these are the more credible uncertainties as the residual RMS of the Keplerian fits is greater than twice the uncertainties reported by either \ssof or NEID's DRP.
This indicates that spectral variability is still impacting the RVs and should not be totally ignored.

\begin{deluxetable*}{l|cccccc}
\tablecaption{MAP Keplerian models for HD 3651 b\label{tab:3651_orb_gp} using \texttt{GLOM} to model the combined RVs from the AIC-minimum \ssof models, \ssofns(5,5) models, and NEID DRP RVs with and without H-alpha. These include the orbital period, time of periastron passage, eccentricity, argument of periapsis, and velocity semi-amplitude. The orbital parameters for HD 3651 b obtained from the \texttt{GLOM} analyses are consistent with their non-\texttt{GLOM} counterparts, largely due to the stellar variability being dwarfed by the planetary signal. The uncertainties on $\theta$ when performing the joint analyses are 5-10x larger when compared to assuming that no stellar variability remained in the RV time series.}
\tablewidth{0pt}
\tablehead{
\colhead{} & \colhead{\ssofns}  & \colhead{\ssofns(5,5)} & \colhead{NEID DRP} & \colhead{\ssofns + H-alpha}  & \colhead{\ssofns(5,5) + H-alpha} & \colhead{NEID DRP + H-alpha}}
\startdata
$P$ (d)   & $62.208 \pm 0.068$ & $62.24 \pm 0.126$ & $62.193 \pm 0.067$ & $62.141 \pm 0.088$ & $62.130 \pm 0.211$ & $61.962 \pm 0.103$ \\
$T_p$ (d) & $58726.48 \pm 0.16$ & $58725.89 \pm 0.18$ & $58726.52 \pm 0.16$ & $58727.26 \pm 0.19$ & $58727.30 \pm 0.25$ & $58729.62 \pm 0.18$ \\
$e$       & $0.605 \pm 0.018$ & $0.615 \pm 0.026$ & $0.619 \pm 0.022$ & $0.592 \pm 0.020$ & $0.605 \pm 0.028$ & $0.596 \pm 0.017$ \\
$\omega$  & $241.44 \pm 2.42$ & $239.83 \pm 2.96$ & $240.48 \pm 2.73$ & $240.56 \pm 2.62$ & $240.37 \pm 3.11$ & $242.64 \pm 2.22$ \\
$K$ (\ms) & $16.26 \pm 0.47$ & $12.95 \pm 0.46$ & $16.45 \pm 0.61$ & $16.29 \pm 0.39$ & $13.19 \pm 0.45$ & $16.24 \pm 0.34$ \\
$\rm RMS$ (\ms) & 0.62 & 0.92 & 0.72 & 0.67 & 0.97 & 0.93 \\
$l_{GP}$ (d) & $3.20 \pm 1.75$ & $17.51 \pm 6.26$ & $3.47 \pm 2.25$ & $5.38 \pm 1.38$ & $27.38 \pm 7.45$ & $12.13 \pm 3.26$ \\
\enddata
\end{deluxetable*}

\subsection{Barnard's Star}\label{ssec:411}

Barnard's Star \citep{barnard1916} is a M4 dwarf with a measured RV RMS of 1-2 \ms \citep{silva2022}.
It is a commonly measured RV standard star that has been used to characterize the performance of many RV instruments \citep{cale2019, metcalf2019} 
and pipelines \citep{artigau2022, silva2022, bedell,serval, anglada2012} because of its high apparent brightness and convenient observability from both hemispheres \citep{lubin2021}.
Rotationally modulated changes have been measured in the time-series spectra \citep{terrien2022} and the measured RV variability has periodic signals that are aliased with the rotational period \citep{lubin2021}.
Analyzing Barnard's Star shows how \ssof performs on a M dwarf whose RV information content is primarily in the redder wavelengths which NEID's DRP (v1.1) struggles to fully utilize.
We analyzed 45 observations of Barnard's Star taken by NEID from February 2021 to June 2022 with 112 \ssof models, each fit to a single spectral order.
The RV performance and model complexity for each \ssof model is shown in Tab. \ref{tab:barnard}.

Although Barnard's star is bright for a mid-M dwarf due to its proximity to Earth, its output is low in the optical wavelengths measured by NEID ($V\sim$9.5).
Using \ssofns, time-variable telluric activity is not identified until wavelengths redward of 5900 \AA (order 103), as opposed to 5000 \AA (order 122) for both HD 26965 and HD 3651.
The lower amount of optical flux causes \ssof to prefer simplified models with many more models lacking telluric transmission or stellar variability (see most orders 135-104 in Tab. \ref{tab:barnard}).
The SNR of NEID's Barnard Star measurements (SNR $\sim$ 30-90 in the central 1000 pixels of orders used by \ssof to calculate RVs) allows for each of these orders to achieve RV single measurement precisions (SMP) $< 2.5$ \ms.

\begin{figure*}[ht]
\centering
\includegraphics[width=15cm]{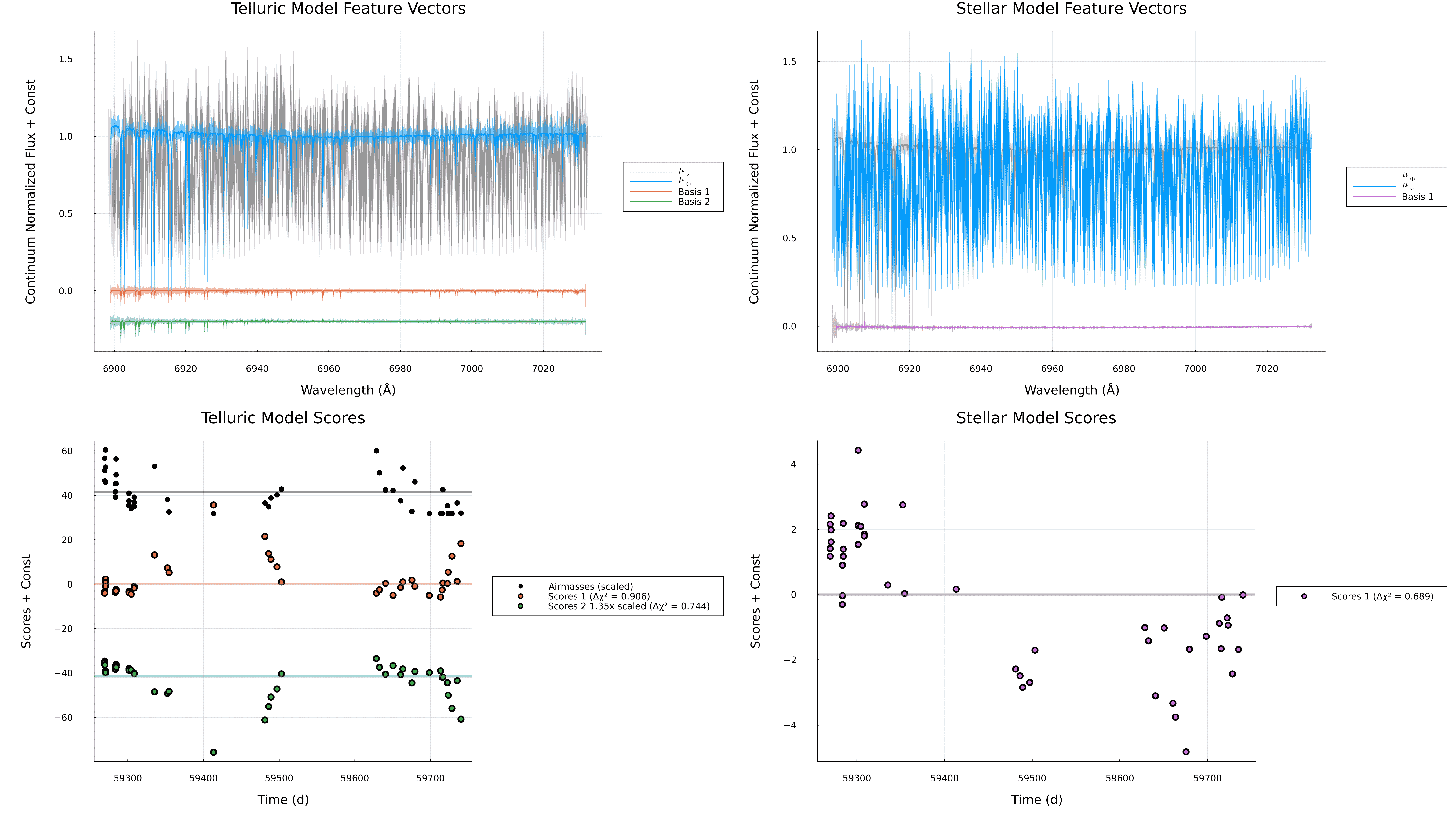}
\caption{\ssof model of order 88 for Barnard's Star which separates the two dominant modes of telluric variability. Top left: The telluric and stellar templates are shown above in grey and blue while the two telluric feature vectors are shown below in orange and green. Both the green and orange feature vectors capture the time-variability of O$_2$ and H$_2$O lines. Bottom left: The telluric feature scores as a function of time. The both scores are somewhat correlated with observation airmass (black). The residuals between $v_{\star,88}$ and $v^{\ssofns}_{\star}$ are shown in Fig. \ref{fig:barnardrv}. Top right: The stellar and telluric templates are shown above in grey and blue while the one stellar feature vector is shown below in purple. The stellar feature captures many, small-scale variations. Bottom right: The stellar feature scores as a function of time. The residuals between $v_{\star,88}$ and $v^{\ssofns}_{\star}$ are shown in Fig. \ref{fig:barnardrv}}
\label{fig:barnard_tel}
\end{figure*}

Only some of Barnard's Star's \ssof models of orders $<$103 have telluric transmission and at least one telluric feature vector to explain temporal variation in the transmission.
The only clear example of \ssof beginning to separate different telluric species in the Barnard's Star data is shown in Fig. \ref{fig:barnard_tel}.

\begin{figure*}[ht]
\centering
\includegraphics[width=18cm]{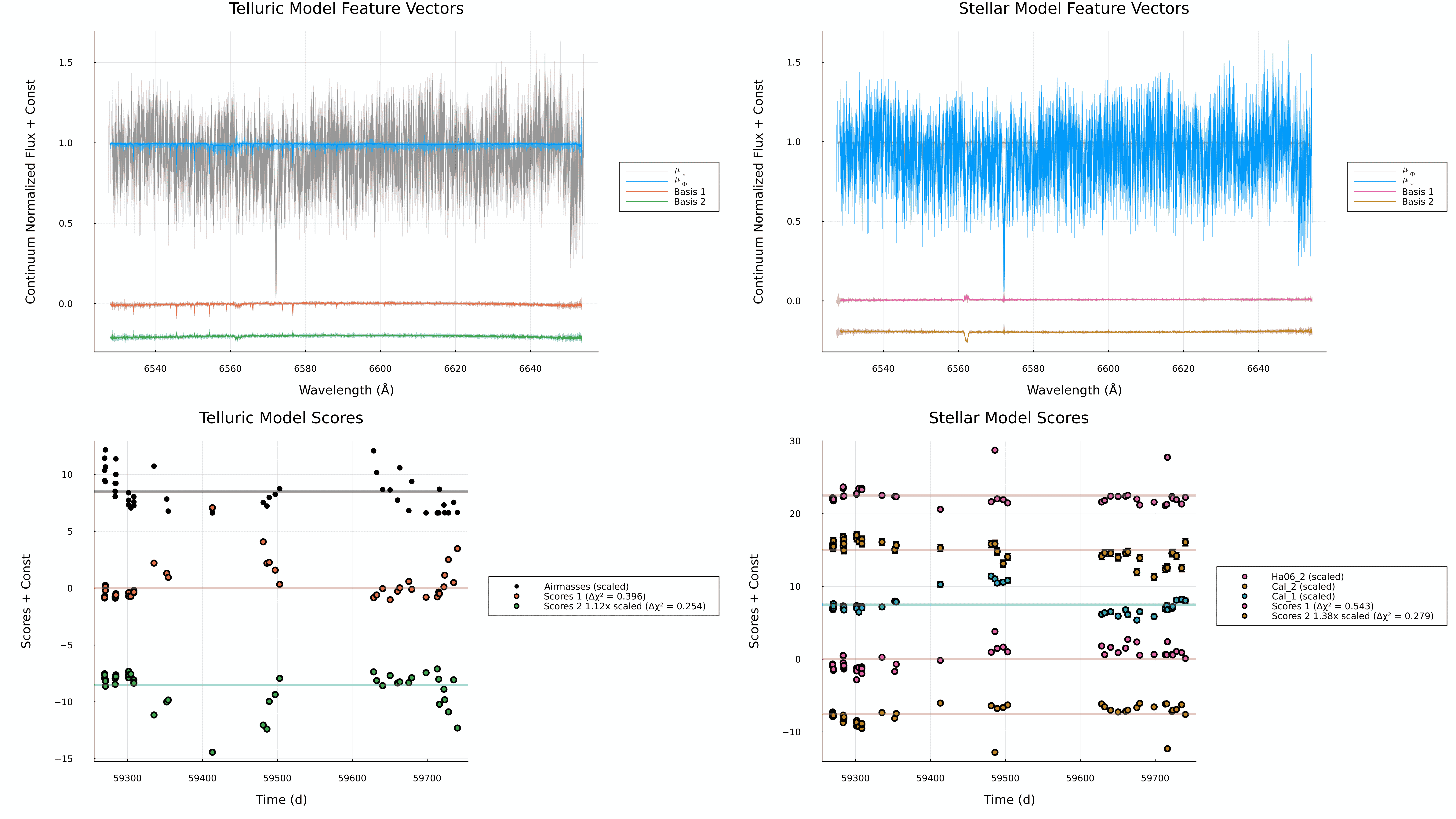}
\caption{\ssof model of order 93 for Barnard's Star which identifies H-alpha variation. Top left: The telluric and stellar templates are shown above in grey and blue while the telluric feature vectors are shown below (orange and green). These vectors are dominated by changing H$_2$O transmission between observations. Bottom left: The observed airmass (black) and telluric feature scores (orange and green) as a function of time. Top right: The stellar and telluric templates are shown above in grey and blue while the stellar feature vectors are shown below (pink and brown). These vectors are dominated by the changes in H-alpha. Bottom right: The stellar feature scores as a function of time. The scores for the H-alpha feature vectors (pink and brown) are correlated with the H-alpha variation measured by NEID's DRP (purple). Some NEID measurements of CaI lines are also shown. The residuals between $v_{\star,93}$ and $v^{\ssofns}_{\star}$ are shown in Fig. \ref{fig:barnardrv}}.
\label{fig:barnard_ha}
\end{figure*}

\ssof only detected variability that was plausibly stellar in origin in H-alpha (order 93) and in the Ca IR triplet (orders 72-71). 
The \ssof model which detected variability in H-alpha is shown in Fig. \ref{fig:barnard_ha}.

\begin{figure*}[ht]
\centering
\includegraphics[width=18cm]{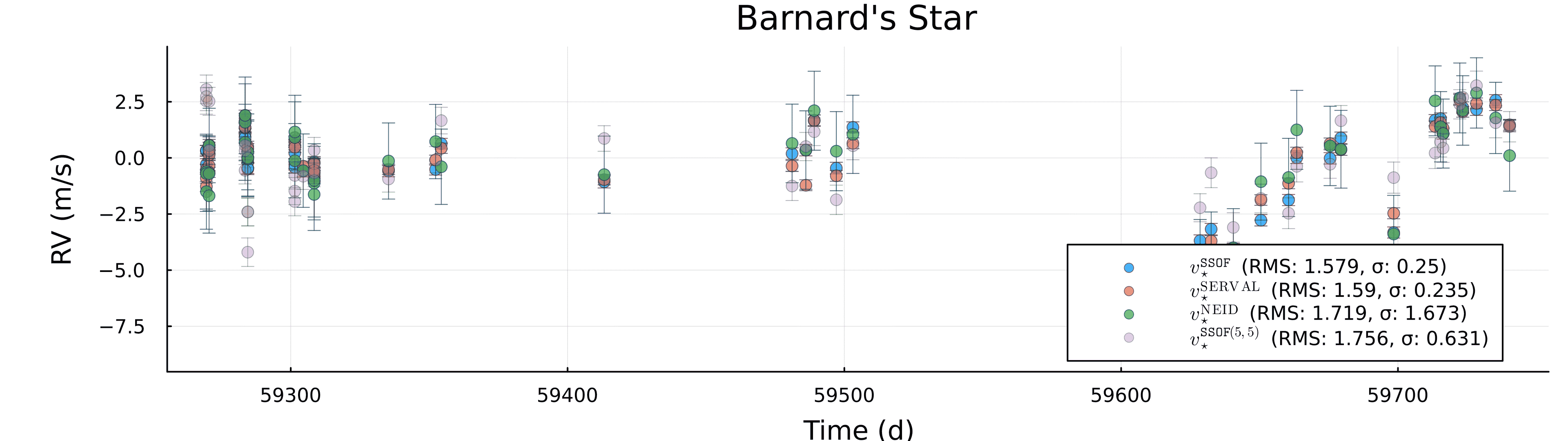}
\includegraphics[width=8.5cm]{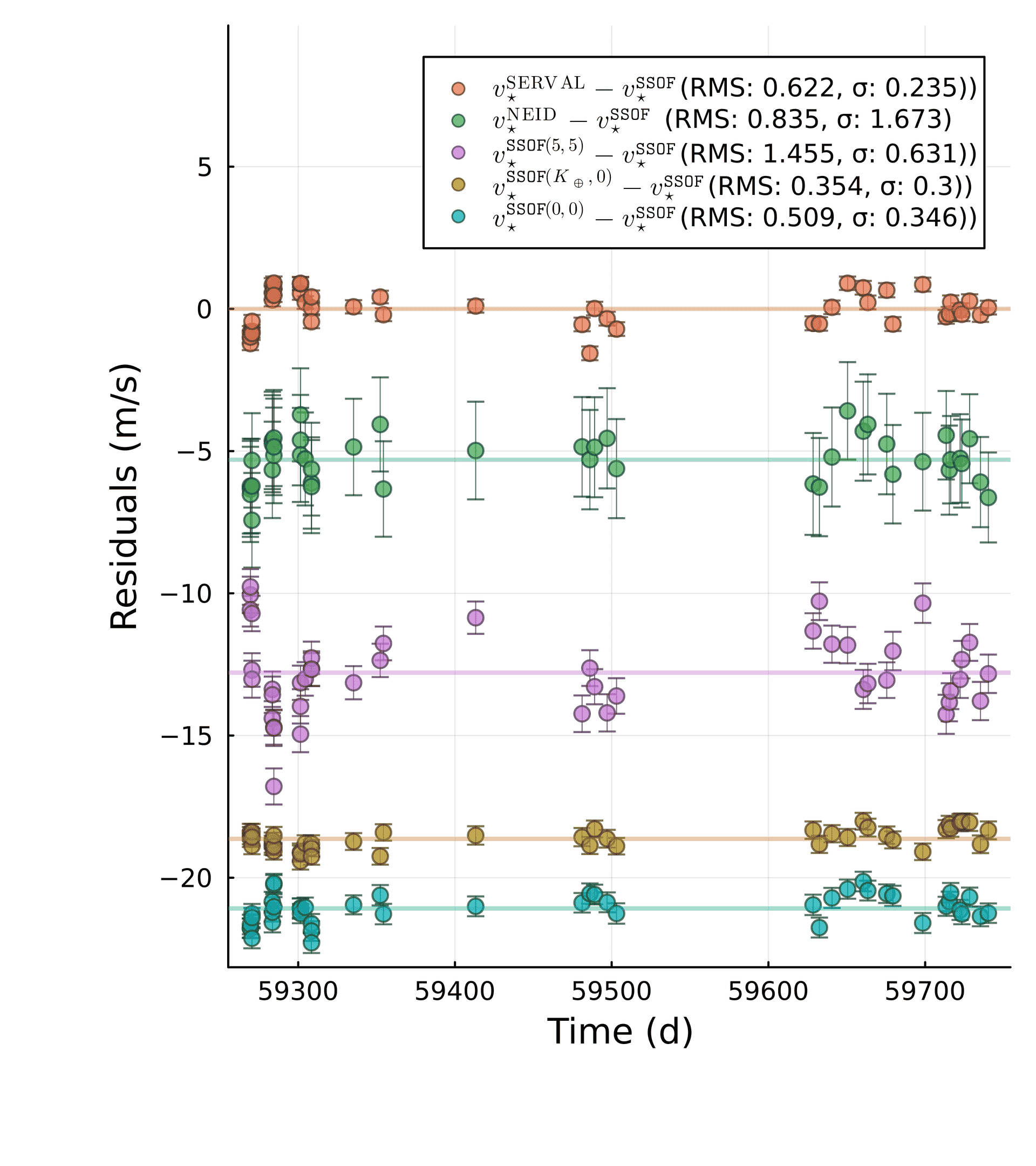}
\includegraphics[width=8.5cm]{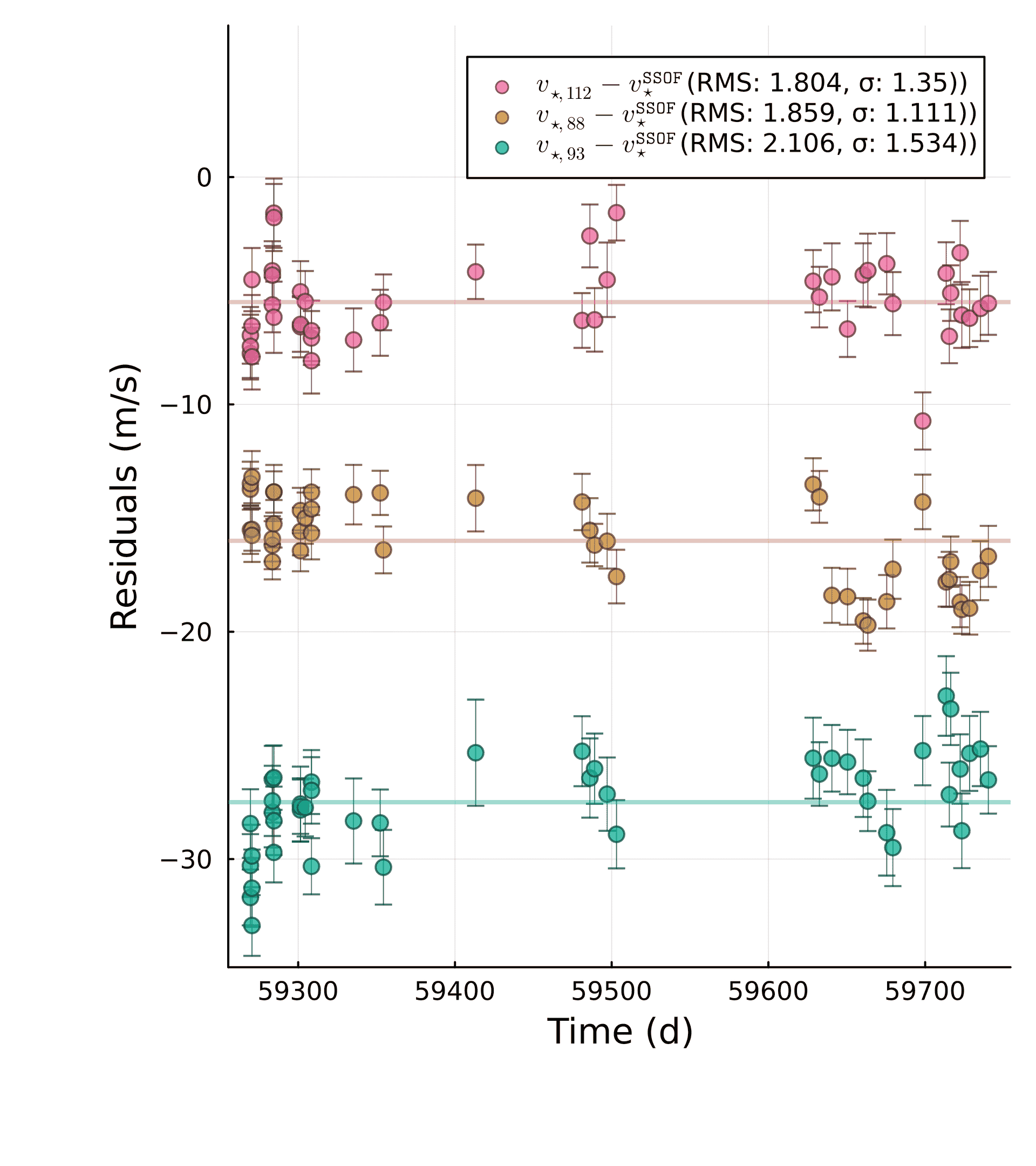}
\caption{Top: The final RV reduction of $v^{\ssofns}_\star$ (blue) for Barnard's Star combining the velocity measurements from 53 \ssof order models resulting in a 85\% tighter SMP when compared to the NEID DRP's CCF-based RVs (green) and a similar SMP and RV RMS when compared to \texttt{SERVAL}'s template-matching RVs ($v^{\textrm{SERVAL}}_\star$, orange). $v^{\ssofns(5,5)}_\star$ (purple) has a higher RMS and higher uncertainty estimates. All RV time series are shown with their corresponding 1-$\sigma$ error bars. Bottom left: The residual time series for $v^{\textrm{SERVAL}}_\star$ (orange), $v^{\textrm{NEID}}_\star$ (green), and several other versions of $v^{\ssofns}_\star$. Most of the \ssof models used in $v^{\ssofns}_\star$ did not use stellar variability so $v^{\ssofns(K_\oplus,0)}_\star$ and $v^{\ssofns(0,0)}_\star$ are practically indistinguishable from $v^{\ssofns}_\star$. Bottom right: The residual time series for the highlighted \ssof order models. The residual velocities from the ``key order" (pink) H$_2$O and O$_2$ separating order (dark orange), and H-alpha variability order (seafoam green) are all essentially consistent with white noise.}
\label{fig:barnardrv}
\end{figure*}

The final $v^{\ssofns}_{\star}$ time series is shown in Fig. \ref{fig:barnardrv}, along with residual RV time series for several different bulk RV reduction schemes and the highlighted \ssof models including: (1) $v^{\ssofns(5,5)}_\star$ which was constructed using a \ssofns(5,5) model for every order, regardless of AIC; (2) $v^{\ssofns(K_\oplus,0)}_\star$ which was constructed using the AIC-minimum \ssof models that didn't use stellar variability, and (3) $v^{\ssofns(0,0)}_\star$ which was constructed using the \ssof models with only stellar and (optionally) telluric templates.
\begin{figure*}[ht]
\centering
\includegraphics[width=18cm]{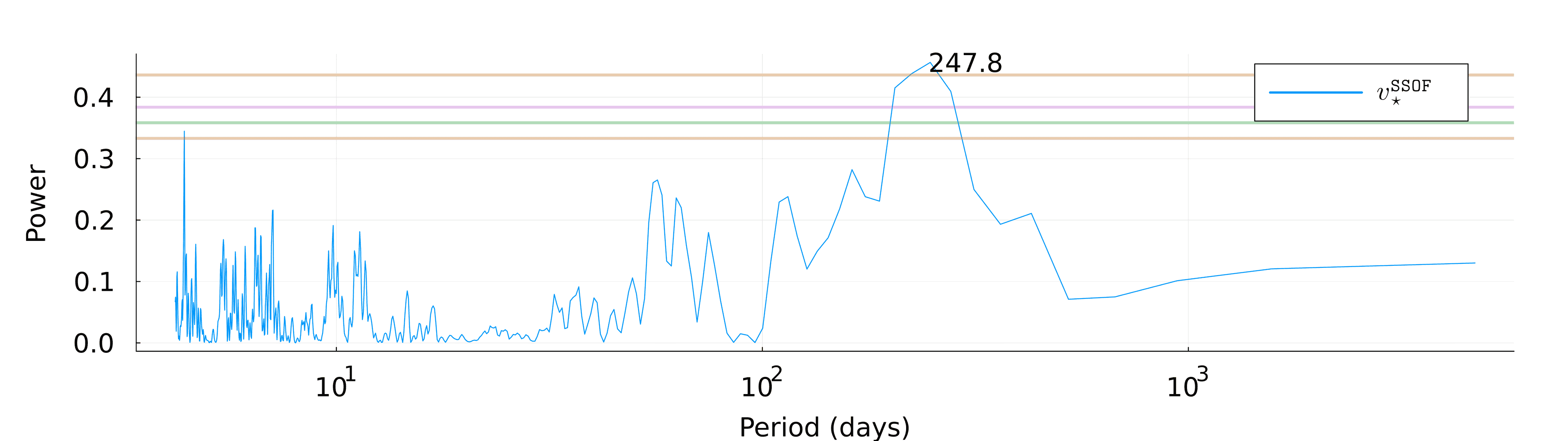}
\caption{Lomb-Scargle periodogram of $v^{\ssofns}_\star$ for Barnard's Star. The period of the maximum peak is labeled. The horizontal lines are the 95\%, 97.72\% (1-sided 2-$\sigma$), 99\%, and 99.87\% (1-sided 3-$\sigma$) false alarm probability (FAP) powers based on the distribution of maximum powers from many Lomb-Scargle periodograms of bootstrap reorderings of $v^{\ssofns}_\star$. The dominant signal remaining is at an alias of the 145 day rotation period and reminiscent of the one that has been explored by \citet{lubin2021} and \citet{artigau2022} (although these measurements are 2670-3140 days after the 1000 day observation window the original signal appeared to be confined to)}
\label{fig:barnardperiod}
\end{figure*}
$v^{\ssofns}_{\star}$ have significantly smaller error estimates than NEID's DRP (down 85\% to 25 \cms from the NEID DRP's 167 \cms) due to the DRP's default CCF mask not using a significant portion of the Doppler information available.
They also have a similar SMP and RV RMS when compared to \texttt{SERVAL}'s RVs. 
The largest signal in a Lomb-Scargle periodogram of $v^{\ssofns}_\star$ (Fig. \ref{fig:barnardperiod}) is at 248 days, a signal which is an alias of the 145 day rotation period and reminiscent of the one that has been explored by \citet{lubin2021} and \citet{artigau2022}.
There is a period of the 248-day signal at the beginning of the measured RVs where the signal does not appear, but then activity picks up in the second observing season and because it bridges observing seasons in this way, it aliases to 248 days.
$v^{\ssofns(5,5)}_\star$ has a both a higher RMS and larger error bars because fewer orders were selected to be included in the reduction. 
$v^{\ssofns(K_\oplus,0)}_\star$ and $v^{\ssofns(0,0)}_\star$ are functionally identical to $v^{\ssofns}_\star$.
The residual velocities between single \ssof orders and $v^{\ssofns}_\star$ are again largely consistent with white noise for quiescent stars.
\afterpage{
\begin{longtable*}[p]{|l|l|l|l|l|p{1.15cm}|c|c|c|p{3cm}|}
    \caption{\ssof model results for each tested NEID order for 45 observations of Barnard's Star including: echelle grating diffraction order; the observed wavelength range used by \ssof (\AA); the RMS of \ssof order RVs over time, average of RV uncertainties over time, and RMS of the RV residuals comparing each \ssof order RV to the combined \ssof RVs (\ms) at the same time; The number of feature vectors used (* indicates that the model is not the minimum-AIC model); whether or not $v_{\star,j}$ was used in the calculation of $v^{\ssofns}_{\star}$, was a ``key order" (KO), and/or was used by NEID's DRP. A number indicates the reason why the order wasn't used to calculate $v^{\ssofns}_{\star}$ (1 indicates that $\overline{\sigma_{v_\star,j}}$  was too high, 2 indicates that $\textrm{RMS}(v_{\star,j})$  was too high, and 3 indicates that $v_{\star,j}$ was inconsistent with $v^{KO}_{\star}$); any notable comments about the order (H$_2$O indicates the presence of a water-like \ssof basis vector and O$_2$ indicates the presence of a oxygen-like \ssof basis vector). \label{tab:barnard}}\\

    \hline
    \multicolumn{10}{|c|}{Beginning of Results Table for Barnard's Star (Tab. \ref{tab:barnard})}\\
    \hline
    Order & $\textrm{min}(\lambda_{D})$ & $\textrm{max}(\lambda_{D})$ & $\textrm{RMS}(v_{\star,j})$ & $\overline{\sigma_{v_\star,j}}$ & Residual RMS & $K_\oplus$ & $K_\star$ & Used & Comments \\
    \hline
    \endfirsthead

    \hline
    \multicolumn{10}{|c|}{Continuation of Results Table for Barnard's Star (Tab. \ref{tab:barnard})}\\
    \hline
    Order & $\textrm{min}(\lambda_{D})$ & $\textrm{max}(\lambda_{D})$ & $\textrm{RMS}(v_{\star,j})$ & $\overline{\sigma_{v_\star,j}}$ & Residual RMS & $K_\oplus$ & $K_\star$ & Used & Comments \\
    \hline
    \endhead

    \hline
    \endfoot

    \hline
    \multicolumn{10}{| c |}{End of Results Table for HD 3651 (Tab. \ref{tab:3651})}\\
    \hline
    \endlastfoot
    
        167 & 3665.91 & 3698.05 & 4514.59 & 4360.89 & 4514.64 & $\varnothing$ & 0 & 1, 2 & ~ \\ \hline
        166 & 3656.83 & 3727.55 & 15819.84 & 17744.5 & 15820.28 & $\varnothing$ & 0 & 1, 2, 3 & ~ \\ \hline
        165 & 3679.16 & 3750.34 & 6996.19 & 11643.72 & 6996.45 & $\varnothing$ & 0 & 1, 2 & ~ \\ \hline
        164 & 3701.76 & 3773.45 & 1341.83 & 674.65 & 1341.9 & $\varnothing$ & 0 & 1, 2 & ~ \\ \hline
        163 & 3724.37 & 3796.35 & 874.48 & 892.41 & 874.8 & $\varnothing$ & 0 & 1, 2 & ~ \\ \hline
        162 & 3747.36 & 3819.78 & 643.21 & 685.79 & 642.99 & $\varnothing$ & 0 & 1, 2 & ~ \\ \hline
        161 & 3770.61 & 3843.5 & 347.88 & 427.43 & 347.47 & $\varnothing$ & 0 & 1, 2 & ~ \\ \hline
        160 & 3794.19 & 3867.52 & 294.4 & 267.84 & 294.16 & $\varnothing$ & 0 & 1, 2 & ~ \\ \hline
        159 & 3818.07 & 3891.85 & 421.67 & 152.06 & 421.43 & $\varnothing$ & 0 & 1, 2, 3 & ~ \\ \hline
        158 & 3842.23 & 3916.5 & 179.65 & 176.23 & 179.79 & $\varnothing$ & 0 & 1, 2 & ~ \\ \hline
        157 & 3866.75 & 3941.46 & 82.66 & 61.66 & 82.91 & 1 & 0 & 1, 2 & CaII K \\ \hline
        156 & 3932.72 & 3966.7 & 126.28 & 87.24 & 126.37 & $\varnothing$ & 0 & 1, 2 & CaII K \\ \hline
        155 & 3952.68 & 3992.29 & 173.24 & 118.31 & 172.97 & $\varnothing$* & 0* & 1, 2 & CaII H \\ \hline
        154 & 3978.77 & 3993.38 & 31.71 & 26.97 & 31.72 & $\varnothing$* & 0* & 1, 2 & ~ \\ \hline
        153 & 4009.13 & 4019.57 & 93.95 & 105.9 & 94.01 & $\varnothing$ & 1 & 1, 2 & ~ \\ \hline
        152 & 4035.09 & 4044.78 & 101.55 & 87.03 & 101.46 & $\varnothing$* & 0* & 1, 2 & ~ \\ \hline
        151 & 4054.48 & 4073.09 & 0 & Inf & 1.58 & $\varnothing$* & 0* & 1 & ~ \\ \hline
        150 & 4089.94 & 4100.34 & 66.56 & 60.19 & 66.51 & $\varnothing$ & 0 & 1, 2 & ~ \\ \hline
        149 & 4109.13 & 4128.44 & 55.95 & 48.8 & 55.91 & $\varnothing$* & 0* & 1, 2 & ~ \\ \hline
        148 & 4146.98 & 4158.89 & 62.97 & 68.85 & 62.84 & $\varnothing$ & 0 & 1, 2 & ~ \\ \hline
        147 & 4165.6 & 4186.9 & 47.12 & 47.57 & 46.77 & $\varnothing$ & 0 & 1, 2 & ~ \\ \hline
        146 & 4191.87 & 4213.95 & 48.22 & 51.88 & 48.02 & $\varnothing$ & 0 & 1, 2 & ~ \\ \hline
        145 & 4235.29 & 4244.36 & 69.61 & 84.15 & 69.38 & $\varnothing$ & 0 & 1, 2 & ~ \\ \hline
        144 & 4254.53 & 4273.64 & 28.78 & 29.23 & 28.79 & $\varnothing$ & 0 & 1, 2 & ~ \\ \hline
        143 & 4276.88 & 4306.89 & 29.4 & 25.75 & 29.3 & $\varnothing$ & 0 & 1, 2 & ~ \\ \hline
        142 & 4303.15 & 4340.42 & 28 & 27.82 & 28.03 & $\varnothing$ & 0 & 1, 2 & ~ \\ \hline
        141 & 4331.57 & 4370.55 & 20.23 & 18.93 & 20.06 & $\varnothing$ & 0 & 1, 2 & ~ \\ \hline
        140 & 4361.76 & 4403.17 & 20.07 & 19.51 & 20.35 & $\varnothing$ & 0 & 1, 2 & ~ \\ \hline
        139 & 4390.33 & 4436.98 & 20.5 & 18.19 & 20.2 & $\varnothing$ & 0 & 1, 2 & ~ \\ \hline
        138 & 4421.34 & 4470.69 & 13.26 & 12.49 & 12.94 & $\varnothing$ & 0 & 1, 2 & ~ \\ \hline
        137 & 4450.35 & 4505.05 & 9.41 & 10.12 & 9.77 & $\varnothing$ & 0 & 1, 2 & ~ \\ \hline
        136 & 4481.6 & 4539.53 & 9.31 & 8.67 & 9.22 & $\varnothing$ & 0 & 1, 2 & ~ \\ \hline
        135 & 4512.61 & 4574.23 & 6.65 & 6.92 & 6.35 & $\varnothing$ & 0 & \ssofns, NEID & RVs start to be useful \\ \hline
        134 & 4545.45 & 4609.34 & 4.68 & 5.19 & 4.57 & $\varnothing$ & 0 & \ssof (KO), NEID & ~ \\ \hline
        133 & 4577.8 & 4644.09 & 4.37 & 4.1 & 4.35 & $\varnothing$ & 0 & \ssof (KO), NEID & ~ \\ \hline
        132 & 4611.49 & 4679.09 & 3.44 & 3.56 & 3.34 & $\varnothing$ & 0 & \ssof (KO), NEID & ~ \\ \hline
        131 & 4645.83 & 4714.81 & 5.02 & 4.08 & 5.22 & $\varnothing$ & 0 & \ssof (KO), NEID & ~ \\ \hline
        130 & 4681.72 & 4752.43 & 4.79 & 4.74 & 4.27 & $\varnothing$ & 0 & \ssof (KO), NEID & ~ \\ \hline
        129 & 4716.83 & 4788.36 & 5.09 & 4.01 & 4.69 & $\varnothing$ & 0 & \ssof (KO), NEID & ~ \\ \hline
        128 & 4752.79 & 4826.87 & 3.73 & 2.87 & 3.32 & 0 & 0 & \ssof (KO), NEID & First telluric features \\ \hline
        127 & 4790.97 & 4865.52 & 2.61 & 2.71 & 2.67 & 0 & 0 & \ssof (KO), NEID & ~ \\ \hline
        126 & 4828.09 & 4907.03 & 3.72 & 2.99 & 3.04 & 0* & 0* & \ssof (KO), NEID & ~ \\ \hline
        125 & 4866.42 & 4949.71 & 3.71 & 3.38 & 3.57 & $\varnothing$ & 0 & \ssof (KO), NEID & ~ \\ \hline
        124 & 4903.34 & 4986.91 & 3.11 & 2.57 & 2.74 & 0 & 0 & \ssof (KO), NEID & ~ \\ \hline
        123 & 4941.73 & 5026.8 & 2.48 & 1.81 & 1.96 & 0 & 0 & \ssof (KO), NEID & ~ \\ \hline
        122 & 4982.62 & 5070.36 & 3.03 & 1.96 & 2.35 & 0 & 0 & \ssof (KO), NEID & ~ \\ \hline
        121 & 5023.64 & 5112.2 & 2.7 & 2.17 & 2.05 & $\varnothing$ & 0 & \ssof (KO), NEID & ~ \\ \hline
        120 & 5065.29 & 5156.01 & 3.62 & 2.33 & 3.22 & $\varnothing$ & 0 & \ssof (KO), NEID & ~ \\ \hline
        119 & 5107.87 & 5199.72 & 2.39 & 2.13 & 2 & $\varnothing$ & 0 & \ssof (KO), NEID & ~ \\ \hline
        118 & 5148.9 & 5243.79 & 18.2 & 1.61 & 18.08 & $\varnothing$* & 0* & NEID, 2, 3 & ~ \\ \hline
        117 & 5192.74 & 5288.6 & 2.92 & 1.61 & 2.65 & $\varnothing$ & 0 & \ssofns, NEID & ~ \\ \hline
        116 & 5233.76 & 5334.19 & 3.01 & 2.11 & 2.58 & $\varnothing$ & 0 & \ssof (KO), NEID & ~ \\ \hline
        115 & 5279.1 & 5380.57 & 3.19 & 2.39 & 2.28 & $\varnothing$ & 0 & \ssof (KO), NEID & ~ \\ \hline
        114 & 5326.81 & 5427.77 & 3.19 & 2.75 & 2.87 & $\varnothing$ & 0 & \ssof (KO), NEID & ~ \\ \hline
        113 & 5372.55 & 5475.8 & 3.46 & 2.37 & 2.95 & $\varnothing$ & 0 & \ssof (KO), NEID & ~ \\ \hline
        112 & 5420.52 & 5524.69 & 2.57 & 1.35 & 1.8 & $\varnothing$ & 0 & \ssof (KO), NEID & ~ \\ \hline
        111 & 5469.36 & 5574.46 & 2.81 & 1.93 & 2.21 & $\varnothing$ & 0 & \ssof (KO), NEID & ~ \\ \hline
        110 & 5519.09 & 5625.14 & 3.06 & 2.08 & 2.63 & $\varnothing$ & 0 & \ssof (KO), NEID & ~ \\ \hline
        109 & 5569.73 & 5675.91 & 2.57 & 1.67 & 1.95 & $\varnothing$ & 0 & \ssof (KO), NEID & ~ \\ \hline
        108 & 5621.31 & 5729.31 & 2.88 & 2.13 & 2.47 & $\varnothing$ & 0 & \ssof (KO), NEID & ~ \\ \hline
        107 & 5673.85 & 5782.85 & 2.77 & 2.37 & 2.33 & $\varnothing$ & 1 & \ssof (KO), NEID & ~ \\ \hline
        106 & 5727.38 & 5837.4 & 3.02 & 2.05 & 2.15 & $\varnothing$* & 0* & \ssof (KO), NEID & ~ \\ \hline
        105 & 5781.93 & 5892.99 & 2.93 & 2.44 & 2.29 & $\varnothing$* & 0* & \ssof (KO), NEID & ~ \\ \hline
        104 & 5837.53 & 5949.66 & 4.48 & 2.38 & 4.27 & 0* & 0* & \ssof (KO), NEID & ~ \\ \hline
        103 & 5895.76 & 6007.42 & 4.89 & 2.24 & 4.53 & 1 & 0 & \ssof (KO) & H$_2$O \\ \hline
        102 & 5952.01 & 6066.31 & 2.89 & 2.01 & 2.63 & 0 & 0 & \ssof (KO), NEID & ~ \\ \hline
        101 & 6010.94 & 6126.37 & 3.24 & 2.31 & 3.11 & $\varnothing$* & 0* & \ssof (KO), NEID & ~ \\ \hline
        100 & 6071.06 & 6187.63 & 3.85 & 2.46 & 3.29 & $\varnothing$* & 0* & \ssof (KO), NEID & ~ \\ \hline
        99 & 6132.39 & 6250.13 & 2.6 & 1.77 & 1.91 & $\varnothing$ & 2 & \ssof (KO), NEID & ~ \\ \hline
        98 & 6194.97 & 6313.91 & 4.11 & 1.68 & 3.84 & 0* & 0* & NEID, 3 & ~ \\ \hline
        97 & 6258.84 & 6379 & 5.27 & 1.46 & 5.21 & 0* & 0* & NEID, 3 & ~ \\ \hline
        96 & 6324.04 & 6445.44 & 2.06 & 1.45 & 1.33 & 1 & 1 & \ssofns, NEID & ~ \\ \hline
        95 & 6390.62 & 6513.29 & 2.91 & 1.33 & 2.33 & 3 & 0 & \ssofns, NEID & H$_2$O \\ \hline
        94 & 6458.61 & 6582.58 & 2.33 & 1.58 & 2.06 & 3 & 0 & \ssofns & H-alpha, H$_2$O \\ \hline
        93 & 6528.06 & 6653.35 & 2.51 & 1.53 & 2.11 & 2 & 2 & \ssofns, NEID & H-alpha (variance detected), H$_2$O \\ \hline
        92 & 6599.03 & 6725.67 & 1.9 & 0.98 & 1.39 & $\varnothing$* & 0* & \ssofns, NEID & ~ \\ \hline
        91 & 6671.55 & 6799.58 & 2.04 & 1.11 & 1.44 & $\varnothing$ & 5 & \ssofns, NEID & ~ \\ \hline
        90 & 6745.69 & 6875.13 & 2.42 & 1.17 & 1.97 & 2 & 1 & \ssofns, NEID & O$_2$ \\ \hline
        89 & 6821.49 & 6952.37 & 4.17 & 1 & 3.82 & 0* & 0* & NEID, 3 & ~ \\ \hline
        88 & 6899.01 & 7031.37 & 2.03 & 1.11 & 1.86 & 2 & 1 & \ssofns & H$_2$O, O$_2$ \\ \hline
        87 & 6978.32 & 7112.19 & 2.59 & 0.95 & 2.34 & 5* & 0* & 3 & H$_2$O \\ \hline
        86 & 7059.47 & 7194.89 & 2.66 & 0.65 & 2.17 & 5* & 0* & 3 & H$_2$O, O$_2$? \\ \hline
        85 & 7142.53 & 7279.53 & 3.08 & 0.73 & 2.56 & 5* & 0* & 3 & H$_2$O \\ \hline
        84 & 7227.56 & 7366.19 & 2.35 & 1.06 & 1.53 & 5* & 0* & \ssofns & H$_2$O \\ \hline
        83 & 7314.65 & 7454.94 & 3.01 & 1.86 & 2.49 & 1 & 1 & \ssofns & H$_2$O \\ \hline
        82 & 7403.86 & 7545.85 & 2.34 & 2.05 & 2 & 0 & 1 & \ssofns & ~ \\ \hline
        81 & 7495.27 & 7639.4 & 17.12 & 23.26 & 16.85 & 5* & 0* & 1, 2, 3 & ~ \\ \hline
        80 & 7588.97 & 7734.89 & 4.92 & 1.02 & 4.58 & 5* & 0* & 3 & O$_2$ \\ \hline
        79 & 7685.04 & 7832.8 & 3.18 & 0.81 & 2.83 & 3* & 0* & 3 & O$_2$ \\ \hline
        78 & 7783.58 & 7933.21 & 2 & 1.05 & 1.29 & 2 & 0 & \ssofns & H$_2$O \\ \hline
        77 & 7884.67 & 8036.24 & 2.24 & 1.68 & 1.54 & 2* & 0* & \ssofns & H$_2$O \\ \hline
        76 & 7988.42 & 8141.98 & 4.02 & 3.03 & 3.51 & 1 & 2 & \ssofns & H$_2$O \\ \hline
        75 & 8094.95 & 8250.53 & 8.33 & 3.24 & 7.87 & 5 & 0 & 2, 3 & H$_2$O \\ \hline
        74 & 8204.34 & 8362.02 & 4.73 & 2.43 & 4.28 & 2 & 1 & \ssofns & H$_2$O \\ \hline
        73 & 8316.74 & 8476.57 & 4.16 & 1.65 & 3.76 & 1 & 1 & 3 & H$_2$O \\ \hline
        72 & 8432.26 & 8594.29 & 2.66 & 2.18 & 1.97 & 2* & 0* & \ssofns & Ca IR triplet 1 \& 2 (variance detected), H$_2$O \\ \hline
        71 & 8551.04 & 8715.34 & 5.76 & 2.42 & 5.68 & 0 & 1 & 3 & Ca IR triplet 3 (variance detected) \\ \hline
        70 & 8673.2 & 8839.84 & 17.97 & 1.97 & 17.87 & 1* & 0* & 2, 3 & H$_2$O \\ \hline
        69 & 8798.91 & 8967.95 & 3.64 & 1.83 & 3.07 & 2* & 0* & \ssofns & RVs stop being useful, H$_2$O \\ \hline
        68 & 8928.32 & 9099.82 & 10.95 & 2.74 & 10.73 & 5* & 0* & 2, 3 & H$_2$O \\ \hline
        67 & 9061.59 & 9235.64 & 24.2 & 4.46 & 24.66 & 5* & 0* & 2, 3 & H$_2$O \\ \hline
        66 & 9198.89 & 9375.57 & 0 & Inf & 1.58 & 0* & 0* & 1 & ~ \\ \hline
        65 & 9340.42 & 9519.8 & 0 & Inf & 1.58 & 0* & 0* & 1 & ~ \\ \hline
        64 & 9486.38 & 9668.54 & 165.82 & 1.85 & 165.82 & 3* & 0* & 2, 3 & H$_2$O \\ \hline
        63 & 9636.97 & 9822.01 & 71.98 & 9.97 & 72.02 & 5* & 0* & 1, 2, 3 & H$_2$O \\ \hline
        62 & 9792.43 & 9980.42 & 13.57 & 5.57 & 13.46 & 1* & 0* & 2, 3 & H$_2$O \\ \hline
        61 & 9952.97 & 10143.94 & 8.63 & 6.14 & 8.92 & 1 & 1 & 2 & H$_2$O \\ \hline
        60 & 10118.94 & 10313.31 & 11.63 & 10.84 & 11.56 & 0 & 0 & 1, 2 & ~ \\ \hline
        59 & 10290.31 & 10487.71 & 34.59 & 25.26 & 34.78 & $\varnothing$ & 0 & 1, 2 & ~ \\ \hline
        58 & 10467.81 & 10653.98 & 70.98 & 73.48 & 70.99 & $\varnothing$ & 0 & 1, 2 & ~ \\ \hline
        57 & 10726.9 & 10754.69 & 616.8 & 418.96 & 616.98 & $\varnothing$ & 0 & 1, 2 & ~ \\ \hline
        56 & 10842.54 & 11049.97 & 1233.15 & 327.39 & 1233.75 & $\varnothing$ & 0 & 1, 2, 3 & ~ \\ \hline
\end{longtable*}
\clearpage 
}





\section{Discussion and Conclusions}\label{sec:discuss}

\subsection{Summary}
Extremely precise radial velocity measurements are an important tool in the efforts for exoplanet discovery and characterization.
Many EPRV spectrographs are now online and have been providing high-resolution and high signal-to-noise spectra for this express purpose.
Maximizing the scientific return from these impressive instruments requires 
advanced algorithms and efficient software.
In this work, we have presented an open-source, instrument-agnostic pipeline, \ssofjlns, to address many of the challenges shared by these instruments in going from measured 1D spectra to precise RV measurements.

By running \ssof on NEID spectra for several stars, we have demonstrated \ssofns's effectiveness on multiple spectral types, with or without the presence of a planet, and with various levels of known stellar variability.
Using only high-resolution spectra with significant barycentric changes, but notably without the input of synthetic spectra or line lists, \ssof can automatically recover EPRV measurements with state-of-the-art precision and accuracy, and produce data-driven models for the de-convolved, time-varying telluric transmission and stellar spectrum along the way.
In all cases, the RVs from combining the AIC-minimum \ssof models reduced the RV RMS and uncertainties compared to NEID's CCF-based DRP.

\ssof demonstrates that data-driven approaches at the spectral level are computational feasible, and can reliably recover known atmospheric and astrophysical signals in EPRV spectra. Leveraging the empirical models derived by \ssof could be a springboard for more detailed studies in micro-tellurics and stellar activity, both of which are currently driving the systematic precision floor in EPRV studies. \ssof is an open-source and freely available tool for the astronomy community and we hope that it can be used to help get the most from the impressive data collected by EPRV instruments.


\subsection{Directions for Future Research}\label{ssec:future}

While \ssof is an excellent tool for analyzing the spectra of individual stars with significant amounts of data taken with a single spectrograph, it has limitations that can be mitigated with additional development.

\subsubsection{Model improvements}\label{sssec:improvements}

\emph{Regularization} --- \ssof has shown some promise for measuring spectral activity indicators, but its feature vectors span the entire spectral order, leading to the inclusion of other spectral features of less interest.  
This could be improved by enabling the user to have feature vectors which can only model the desired, subsections of the given spectra or by fine tuning the regularization to be stronger in parts of the spectra where we do not expect relevant features.
For example, regularization could account for locations of known spectral lines \citep{czekala2015}, or even which lines are likely to be sensitive to stellar activity \citep{wise2022}. 
\ssof already allows users to mask out complicated regions of stellar or telluric variability but the focus of the current work was to try to model all parts of the measured spectra.
More detailed explorations of selective masking are outside the scope of the current work.

\emph{Improving telluric models} --- 
\ssof is currently built to model one star at a time and has to relearn the telluric features for each star.
In principle, the telluric transmission spectra should share the same features and should not have to be relearned from scratch.
In a strictly data-driven spirit, one way to improve \ssofns's telluric model would be to combine information from analyzing several stars to build an empirically-motivated, high-confidence telluric model that characterizes the range of atmospheric variability for the observing conditions at individual observatories.
This could be initially approximated by combining the separately-fit telluric models from some amount of initial \ssof analyses or performed more rigorously by fitting a shared telluric model across several EPRV data sets, each with their own stellar model.
In a science-driven spirit, \ssof could also leverage the vast amount of knowledge of telluric line strengths and locations from line-by-line radiative transfer models \citep[e.g. \texttt{TelFit}, \texttt{TAPAS}, \texttt{Molecfit}, \texttt{TERRASPEC},][]{gullikson2014,clough2005,bertaux2014,smette2015,bender2012} to ensure that \ssof knows where to look for spectral features or to have pre-trained feature vectors for individual molecular species. 
Using a consensus, pre-learned telluric model would also enable \ssof to model stars with small amounts of barycentric velocity samples.
This includes the sun (by far the largest source of EPRV data) and would allow us to probe the spectral imprints of stellar variability on much smaller timescales (for solar-like stars).

\emph{Jointly modeling multiple datasets to improve stellar models} --- 
Modeling stars with small amounts of observations is possible by transferring models learned from previous \ssof analyses or jointly modeling similar stars.
Stars with low amounts of data could have their model initialized with \ssof models from similar-type stars and then be fine-tuned (as long as their absolute model wavelengths are matched correctly).
\ssof could also be expanded to jointly model datasets for different stars, of similar stellar type, to learn from all stars simultaneously.
Each star would likely need its own stellar template, but they could share some amount of feature vectors if one expected that their variability should be of a shared nature.
Thus, each star's variability could then be better learned than if each one had to be characterized on its own.

\emph{Jointly modeling multiple datasets from different instruments} --- 
\ssof should be able to used a shared stellar model for EPRV observations of the same star from several spectrographs, as long as they overlap in wavelength.
This would only require a slight mathematical adjustment to the overall framework, minimizing the sum of several $\mathcal{L}_{R}(\beta_M, \beta_R)$, each with their own $Y_D$ and $T_I$ from each instrument instead of just the one.
The stellar model parameters ($\beta_\star$) could be shared, but the telluric model parameters ($\beta_\oplus$) and regularization values ($\beta_R$) would likely have to be different for each instrument.

\emph{LSF model} --- 
\ssof is built to include the effects of an arbitrary LSF, including one which varies as a function of detector location and time.
The only requirement is that it can be precomputed and does not change during model inference.
In its current state, \ssof cannot currently learn the LSF shape and can only use LSF information if the LSF has been previously characterized.
While it is in principle possible to perform this joint fitting, we anticipate that it will significantly increase inference times and the amount of model degeneracies and have left it for future work.

\emph{Evaluating spectral models at higher resolution} --- 
We currently evaluate the telluric and stellar models at the resolution of the instrument before multiplying them together. 
While the majority of stellar lines are much broader than the instrument pixels (and are therefore still well sampled at this resolution), the same cannot be said for many telluric lines.
As an example, take a case where there are two deep, narrow spectral lines within a instrument pixel, one telluric and one stellar.
In our current model, they would both be averaged out over the instrument pixel before being multiplied together. 
This works fine if the lines are separated but, if they overlap during a given observation, the overall flux in the pixel should be higher when the lines are taking away flux from the same part of the spectrum.
Our current model is unable to account for this. 
We could rewrite \ssof to be more robust to these effects by replacing equations \ref{eq:basic}-\ref{eq:LSF} with 
\begin{equation}
	Y_{M,n} = T_{I,n} \cdot (Y'_{\oplus,n} \circ (T_{\star,n} \cdot Y'_{\star,n})) + \epsilon
\end{equation}
where $T_\star$ brings the high-resolution stellar model to the observer frame from where it can be multiplied with the (still high resolution) telluric model and $T_{I}$ then convolves the total (still high resolution) model with the pixel response (and LSF if available).


\emph{Incorporate knowledge of instrumental systematics} --- 
\ssof relies on instrumental pipelines to perform the important step of transforming the 2D spectra contained in the CCD detector readouts measured my EPRV instruments into many 1D spectral orders. \ssof could also benefit from additional information related to detector properties and known systematics, such as charge transfer efficiency amplitude and directionality, locations of interstitial boundaries between CCD detectors, an informed model of the pixel response, and information related to the detector fringing amplitude and lengthscale. Measured wavelengths that cross over a boundary in readout directions could be flagged and removed. This would prevent \ssof from having to model a type of complicated, non-linear effect that is outside its model specifications. \ssof currently assumes that the detector pixels have a uniform top-hat response function, fully cover the detector, and are centered at the reported wavelengths. This could be improved by integrating the spectrum over each data pixel’s wavelength boundaries with a calibrated pixel response function. The linear interpolation used by the VIL method to perform the RV shift and resampling could also be performed more correctly by instead performing an equivalent (if slower) integration. \ssof telluric feature vectors are regularized to look for spectral lines. If the spectral order to be modeled has a known, significant fringing component, an additional fringing feature vector in the observer frame could be added with a regularization term that instead encourages features of the corresponding lengthscale. These scores could then provide insight into the detector and illumination stability.

\emph{Simultaneous modeling in spectral and temporal domains} --- 
\ssof does not currently use any temporal information when trying to infer the model templates, feature vectors, scores, or RVs.
In principle, we know that the spectral variations and RVs are continuous processes whose measurements should have temporal correlations.
Instead of post-processing the RVs and indicators with \texttt{GLOM} as we did in \secref{ssec:3651}, we could treat the \ssof scores and RVs as a Bayesian model with a GP likelihood during the model fitting.
We could either use a physically motivated GP model like \texttt{GLOM} or a simpler, fast-inference GP model like the one used for the template and feature vector regularization. 

\emph{Complicated orders} --- 
There are some spectral orders that are consistently ignored even though they are within the wavelength ranges where orders have high SNR and are being consistently used. 
Namely orders 117-118 (5145-5289\AA), 81-80 (7495-7735\AA), and 75 (8095-8251\AA).
Orders 117-118 have stellar feature vectors that appear to be driven by non-uniform Doppler shifts across the order.
This could be indicative of an inaccurate wavelength solution for these orders, causing a changing apparent shift for each line.
These two orders draw their wavelength calibration from the NEID DRP's laser frequency comb which didn't have consistently good SNR in these orders over the time period of our observations, which could have affected the wavelength solutions.
Orders 81-80 are dominated by saturated O$_2$ lines, and the easiest path for improvement is likely to mask out some portion of the lines $>$7595\AA.
Similarly while there is variable H$_2$O lines throughout orders 74-76, the amount in order 75 seems to be overwhelming \ssofns, which could likely be helped by masking some of the H$_2$O lines between 8100-8400\AA.

\subsubsection{Opportunities for Science with \ssof}\label{sssec:science}



 
\emph{Application to additional instruments} --- 
\ssof has been designed to be an instrument-agnostic data reduction pipeline but it requires some data homogenization to get convenient inputs.
We have already demonstrated that this is possible for NEID (and EXPRES data from the EXPRES Stellar Signals Project \citep{zhao2022}, not shown in this work), but there remain many high-resolution EPRV spectrographs for which the specific intricacies and foibles for data ingestion into \ssof have not yet been addressed \citep[such as HARPS, ESPRESSO, CARMENES, HPF, KPF, Maroon-X;][]{crass2021}.
Combining extensions to tools such as \texttt{EchelleInstruments.jl} (which ingests and preprocesses data from various instruments into uniform data structures) with \ssof would facilitate multi-instrument analyses and comparisons.


\emph{Residual stellar variability signals} --- 
Ideally, the \ssof framework would result in intrinsic stellar variability being completely modeled by feature vectors, with no imprint on the measured radial velocities. 
In practice, we found that \ssof could reduce RV ``noise'' presumably due to stellar variability, but did not eliminate such signals for the datasets considered.
Future work could explore whether improving the signal-to-noise, spectral resolution and/or number of observations leads to further reductions in stellar variability being mistaken for radial velocities.   

\emph{Generating synthetic datasets for evaluating other algorithms} ---  
\ssof provides a powerful way to generate realistic synthetic data sets for evaluating \ssofns, other EPRV pipelines, or any other stellar variability modeling tools.  
One could combine the \ssof telluric model and \ssof stellar model from any well-observed star (or an independent set of synthetic stellar spectra) to create simulated spectra with arbitrary RV signals and spectral noise realizations.
This could be particularly useful for creating high-quality spectral-level datasets for a data challenge (\`a la \citet{zhao2022}) to evaluate the performance of \ssof and other spectra-level analysis tools. 


\subsection{Comparison to other EPRV pipelines}  
There exist many other open-source data reduction pipelines which can measure precise radial velocities, each with their own benefits.
Other tools for modeling EPRV spectra trade off some ability to match the observed data with increased theoretical motivation.
For example, by assuming a shared or specific line profile shape and/or relying on synthetic spectra to acquire line parameters \citep[e.g. ][]{lienhard2022,gully-santiago2022,czekala2015,serval}.
Recently, other data-driven methods which measure RVs in a line-by-line framework have also shown significant promise \citep{artigau2022,cook2022}.

Of particular note, \wobble \citep{bedell} is the intellectual predecessor to \ssof having pioneered the joint fitting of high-SNR stellar templates and time-variable telluric lines.
Indeed several of \ssofns's features were mentioned as possibilities in \citet{bedell}, including assuming Gaussian noise in linear flux rather than logarithmic
flux, accounting for LSF effects after combining the telluric and stellar components, and modeling stellar variability in a way that avoids degeneracy with RV effects.
We have avoided this degeneracy by forcing our stellar feature vectors to be orthogonal to a Doppler feature vector and with our careful model initialization.
We have additionally improved the model regularization with fast GP-likelihood terms and model selection with an information-criterion based model search.

\ssof demonstrates that data-driven approaches at the spectral level are computational feasible, and can reliably recover known atmospheric and astrophysical signals in EPRV spectra. Leveraging the empirical models derived by \ssof could be a springboard for more detailed studies in micro-tellurics and stellar activity, both of which are currently driving the systematic precision floor in EPRV studies. \ssof is an open-source and freely available tool for the astronomy community and we hope that it can be used to help get the most from the impressive data collected by EPRV instruments. 
\\

{\em Acknowledgments: }
We thank an anonymous referee for their helpful suggestions and patience.
C.G. and E.B.F. acknowledge the support of the Pennsylvania State University, the Penn State Eberly College of Science and Department of Astronomy \& Astrophysics, the Center for Exoplanets and Habitable Worlds and the Center for Astrostatistics.
This research was partially supported by Heising-Simons Foundation Grant \#2019-1177.
This research or portions of this research were conducted with Advanced CyberInfrastructure computational resources provided by ICDS at The Pennsylvania State University (http://icds.psu.edu), including the CyberLAMP cluster supported by NSF grant MRI-1626251.
G.S. acknowledges support provided by NASA through the NASA Hubble Fellowship grant HST-HF2-51519.001-A awarded by the Space Telescope Science Institute, which is operated by the Association of Universities for Research in Astronomy, Inc., for NASA, under contract NAS5-26555.
The citations in this paper have made use of NASA's Astrophysics Data System Bibliographic Services.
This paper contains data taken with the NEID instrument, which was funded by the NASA-NSF Exoplanet Observational Research (NN-EXPLORE) partnership and built by Pennsylvania State University. NEID is installed on the WIYN telescope, which is operated by the NSF's National Optical-Infrared Astronomy Research Laboratory, and the
NEID archive is operated by the NASA Exoplanet Science Institute at the California Institute of Technology. Part of this  work was performed for the Jet Propulsion Laboratory, California Institute of Technology, sponsored by the United States Government under the Prime Contract 80NM0018D0004 between Caltech and NASA. We thank the NEID Queue Observers and WIYN Observing Associates for their skillful execution of the NEID observations.

\software{\julia \citep{Bezanson2017}, \optimjl \citep{Mogensen2018}, \nablajl \citep{nabla}, \texttt{RvSpectML.jl} \citep{rvspectml}, \texttt{MatrixEquations.jl} \citep{matrixequations}, \texttt{ExpectationMaximizationPCA.jl} \citep{empcajl}}

\bibliography{main}

\clearpage
\appendix
\restartappendixnumbering

\section{NEID Line Spread Function}\label{app:lsf}

We modeled NEID's line-spread function (LSF) as the convolution of a top hat filter with a Gaussian.
\begin{equation}
	\mathrm{LSF}(\delta,\alpha,\sigma) \propto \text{erf}\left(\dfrac{\alpha + \delta}{\sqrt{2} \sigma}\right) + \text{erf}\left(\dfrac{\alpha - \delta}{\sqrt{2} \sigma}\right)
	\label{eq:box}
\end{equation}
where $\delta$ is the separation from the center of the LSF, $\alpha$ is the half-width of the top hat filter and $\sigma$ is the Gaussian standard deviation. This analytic form resulted in a more accurate representation of the NEID LSF, as opposed to a typical Gaussian (often used in other instruments). We fit $\alpha$ and $\sigma$ as a function of detector location using a series of 1D spectra from the NEID laser frequency comb (LFC) source, which produces unresolved lines in the spectrometer.
We reliably measured $\alpha$ and $\sigma$ from echelle orders 121-61, 
which the NEID LFC covers with reasonable flux (spanning roughly 5200-10000 \AA).

\section{Gaussian Process Regularization}\label{app:gpreg}

One of the regularization terms we use for our spectral templates and feature vectors is a GP log-likelihood \citep{gps}
\begin{equation}
 \ell(\beta_{GP}|\textbf{y}, \boldsymbol\xi)=-\dfrac{1}{2} \left(N \ \text{log}(2\pi)+\text{log}(|\textbf{K}+\sigma^2_k\textbf{I}|)+\textbf{y}^T (\textbf{K}+\sigma^2_k\textbf{I})^{-1} \textbf{y} \right)
 \label{eq:gplikelihood}
\end{equation}
where $\textbf{y}$ is the vector of parameters, $\boldsymbol\xi$ are the uniformly-spaced model log wavelengths, $\textbf{K}$ is the covariance matrix which is constructed by evaluating a kernel function for each pairwise couple of inputs ($\textbf{K}_{i,j}=k(\beta_{GP},|\xi_i - \xi_j|)$), $\sigma_k^2$ is the measurement noise (there isn't really any since $\textbf{y}$ is a model component but we will take to be a very small value, 1e-12, to help with numerical stability) and $\textbf{I}$ is the identity matrix.
This term encourages nearby points in $\textbf{x}$ to be correlated and suppresses overall $\textbf{x}$ amplitude.
Using the standard $\mathcal{O}(n^3)$ algorithms for GP regression to calculate the GP likelihoods used in \eqref{eq:reg} would quickly become intractable for our problem sizes (typically $\sim 10^4$ model parameters). 
Instead, we turn to an approach that re-frames GP regression models as the output of a linear time invariant (LTI) stochastic differential equation (SDE), which can be solved exactly with classical $\mathcal{O}(n)$ Kalman filtering theory.
While \tgpjl already exists in Julia to perform fast GP inference using this methodology, it is built to take efficient gradients with respect to $\beta_{GP}$ in $\ell(\beta_{GP}|\textbf{y}, \boldsymbol\xi)$ but not $\textbf{y}$.
Using automatic differentiation on \tgpjlns, while $\mathcal{O}(n)$, was dominating our run times at our problem size.
Thus we re-implemented the necessary portions of \tgpjl and expanded it to take the gradients we desire with extremely high performance. 

The main focuses of this section are to briefly motivate how some GPs can be expressed as output of LTISDEs, then show how (without fully justifying the interim steps) one can calculate the likelihoods and gradients that we need to use GPs for the regularization purposes described in \secref{ssec:reg}.
Large portions of the following section are reproduced (with some additions) from \citet{sarkkashort} and \citet{sarkka2019applied}.

GPs are fully specified by their mean and covariance functions.
For example, a zero-mean GP prior for the values of a function $f$ on a collection of inputs sampled at $\boldsymbol\xi$ (e.g. often time but in our case log-wavelengths) is commonly denoted as
\begin{equation}
	f(\xi) \sim \mathcal{GP}(0, k(\beta_{GP}, \boldsymbol\xi))
\label{eq:gp}
\end{equation}
where $k$ is a covariance function with hyperparameters $\beta_{GP}$. 
This GP would correspond to a joint-prior on the function values of $p(f(\xi_1), f(\xi_2), f(\xi_3), ..., f(\xi_{D,n}))) \sim N(0,\textbf{K})$.

One commonly used kernel function is the Mat\'ern $^5/_2$ kernel, which produces twice mean-square differentiable GPs whose correlations exponentially decay with input separation.
\begin{equation}k_{M}(\lambda,\tau) = \sigma_M^2 
    \dfrac{2^{1-\nu}}{\Gamma(\nu)} 
    (\lambda \tau)^\nu
    K_\nu(\lambda \tau) = 
    \sigma_M^2 \left(1 + \lambda \tau + \dfrac{(\lambda \tau)^2}{3}\right) \ e^{-\tau}
\label{eq:m52}
\end{equation}
where $\nu=5/2$, $\tau = |\xi - \xi'|$, $\Gamma$ is the gamma function, $K_\nu$ is the modified Bessel-function of the second kind, $\lambda = \sqrt{2\nu} / l$, and $l$ is the length-scale of local variations.
In this case, the spectral density exists and the covariance function and the spectral density are Fourier duals of each other (i.e. $S_{M}(\omega) = \mathcal{F}(k_{M}(\xi))$ via the Wiener-Khinchin theorem \citep{gps}).
From the spectral density equation provided in \citet{sarkkashort} (after equation 8), we get
\begin{equation}
    S_{M}(\omega) = 
    \dfrac{\sigma_M^2 2\pi^{1/2} \Gamma(\nu+1/2)}{\Gamma(\nu)} \lambda^{2\nu} (\lambda^2+\omega^2)^{-(\nu+1/2)} = 
    \dfrac{\sigma_M^2 16 \lambda^5}{3} (\lambda^2+\omega^2)^{-3} = 
    q (\lambda^2+\omega^2)^{-3} = 
    (\lambda+i \omega)^{-3} q (\lambda+i \omega)^{-3}
    \label{eq:sm52}
\end{equation}
where $q=\dfrac{\sigma_M^2 16 \lambda^5}{3}$ is the constant (i.e. doesn't change with $\omega$) portion of $S_{M}$.
Now consider the the linear LTISDE of the following form
\begin{equation}
	\dfrac{d^m f(\xi)}{d\xi^m}+a_{m-1} \dfrac{d^{m-1} f(\xi)}{d\xi^{m-1}} + ... + a_{1} \dfrac{d f(\xi)}{d\xi} + a_{0} f(\xi) = w(\xi)
\label{eq:sde}
\end{equation}
where $\{a_0, ... , a_{m-1}\}$ are constants and $w(\xi)$ is a white-noise process with constant spectral density $S_w(\omega) = q$ (which will end up being the same $q$ from \eqref{eq:sm52}).
To aid in the calculation of $f(\xi)$'s spectral density, \eqref{eq:sde} can be rewritten into the following Markov process form
\begin{equation}
	\dfrac{d \textbf{x}(\xi)}{d\xi} = \textbf{F}\textbf{x}(\xi) + \textbf{L}w(\xi)
\label{eq:markov}
\end{equation}
where $\textbf{x}(\xi)$ is the state vector (i.e. $\textbf{x}(\xi) = (f(\xi), \dfrac{d f(\xi)}{d\xi}, ... , \dfrac{d^{m-1} f(\xi)}{d\xi^{m-1}})^T$) and $\textbf{F}$ and $\textbf{L}$ are defined as
\begin{equation}
\textbf{F} = 
\begin{pmatrix}
0 & 1 &   &   \\
  & \ddots & \ddots & \\
  &   & 0 & 1 \\
-a_0 & \ldots & -a_{m-2} & -a_{m-1} \\
\end{pmatrix}
, \ \textbf{L} = 
 \begin{pmatrix}
0 \\
\vdots \\
0 \\
1 \\
\end{pmatrix}
\label{eq:FL}
\end{equation}
Taking the Fourier transform of \eqref{eq:markov} (relying on the fact that $\mathcal{F}(\dfrac{dx(\xi)}{d\xi}) = i \omega \mathcal{F}(\textbf{x}(\xi)) = i \omega \textbf{X}(\omega)$) leads to $i \omega \textbf{X}(\omega) = \textbf{F}\textbf{X}(\omega) + \textbf{L}W$ where $W = \mathcal{F}(w)$.
With some rearranging, $\textbf{X}(\omega) = -(\textbf{F} - i \omega \textbf{I})^{-1}\textbf{L}W$.
$f$ can be by extracted from $\textbf{x}$ by multiplying $\textbf{x}$ by $\textbf{H}$ where $\textbf{H}^T = (1\ 0 \ \hdots \ 0)^T$.
The power spectrum of $f$ is as follows.
\begin{equation}
	S_f(\omega) = |\textbf{H}\textbf{X}(\omega)|^2 = \textbf{H}\textbf{X}(\omega)\overline{\textbf{H}\textbf{X}(\omega)} = \textbf{H}(\textbf{F} - i \omega \textbf{I})^{-1}\textbf{L}q\textbf{L}^T(\textbf{F} + i \omega \textbf{I})^{-T}\textbf{H}^T
\label{eq:smarkov}
\end{equation}
where $q \equiv W\overline{W}$ is the constant spectral density of $w(\xi)$.
$S_f$ will have the same spectral density as the Mat\'ern $^5/_2$ GP, $S_M$, (and thus we will have succeeded in our goal of finding a LTISDE whose outputs have the correct properties) if the $q$ values are the same, and
\begin{equation}
	(\lambda+i \omega)^{-3} = \dfrac{1}{\lambda^3 + 3 i \lambda^2 \omega - 3 \lambda \omega^2 - i \omega^3} = \dfrac{1}{a_0 + i a_1 \omega - a_2 \omega^2 - i \omega^3} = -\textbf{H}(\textbf{F} - i \omega \textbf{I})^{-1}\textbf{L}
\end{equation}
Thus the $a_i$ values for our desired LTISDE are binomial coefficients ($a_i=\left(\nu+1/2 \atop i\right)\lambda^{\nu+1/2-i}$) and \textbf{F} is the following $3\times3$ matrix
\begin{equation}
\textbf{F} = 
\begin{pmatrix}
0 & 1 & 0 \\
0 & 0 & 1 \\
-\lambda^3 & -3\lambda^2 & -3\lambda \\
\end{pmatrix}
\end{equation}

Now that we've shown that we can find a LTISDE with the right properties, we will move extremely quickly through the numerical details of how to perform the inference without much further justification. 
We direct the motivated reader to \citet{sarkkashort} and \citet{sarkka2019applied} for further reading.

In a stationary state, the covariance function of $f(\xi)$ is the inverse Fourier transform of its spectral density which can be calculated as
\begin{equation}
k(\lambda,\tau) = \textbf{H}(\textbf{P}_\infty\textrm{exp}(\textbf{F} \tau)) \textbf{H}^T
\end{equation}
where $\textbf{P}_\infty$ is the stationary covariance of $\textbf{x}(\xi)$ and can be obtained as the solution of the following matrix Riccati equation.
\begin{equation}
    \dfrac{d\textbf{P}}{d\xi} = 0 = \textbf{FP}_\infty + \textbf{P}_\infty \textbf{F}^T + \textbf{L}q\textbf{L}^T
\end{equation}
which we solved with \texttt{lyapc} function from \texttt{MatrixEquations.jl}.
The continuous time LTI model \eqref{eq:markov} can be transformed into discrete time model of the following form:
\begin{equation}
\textbf{x}(\xi_{k+1}) = \textbf{A}_k\textbf{x}(\xi_{k}) + \textbf{q}_k, \ 
\textbf{q}_k \sim \textrm{N}(0,\Sigma_k)
\end{equation}
where $\textbf{A}_k$ is the state transition matrix and $\Sigma_k$ is the process noise which can be calculated as
\begin{equation}
\textbf{A}_k = \textrm{exp}(\textbf{F}\tau), \ 
\Sigma_k = \textbf{P}_\infty - \textbf{A}_k \textbf{P}_\infty \textbf{A}_k^T
\end{equation}
In our case, $\textbf{A}_k$ and $\Sigma_k$ are constant since we use a single $\tau$ (uniformly spaced model log wavelengths).
Now that we know the state transition matrix and process noise, we can perform our Kalman filtering.
The Kalman filter is a recursive estimator where only the estimated state from the previous step and the current measurement are needed to compute the estimate for the current state (hence its inherent $\mathcal{O}(n)$ scaling). 
Kalman filters are usually conceptualized into two phases.
First is the state prediction where one gets a prediction for the mean and covariance of the current state (at $\xi_k$) based on the previous state.
\begin{gather}
\overline{\textbf{m}_k} = \textbf{A}_k \textbf{m}_{k-1} \\
\overline{\textbf{P}_k} = \textbf{A}_k \textbf{P}_{k-1} \textbf{A}_k^T + \Sigma_k
\end{gather}
where $\overline{\textbf{m}_k}$ is the predicted (a priori) state estimate, and $\overline{\textbf{P}_k}$ is the predicted (a priori) covariance estimate.
Second is the update step where one updates the posterior of the current state (at $\xi_k$) based on the current measurement.
\begin{gather}
v_k = y_k - \textbf{H}\overline{\textbf{m}_k} , \\
S_k = \textbf{H} \overline{\textbf{P}_k} \textbf{H}^T +\sigma^2_k, \\
\textbf{K}_k = (\overline{\textbf{P}_k}\textbf{H}^T) / S_k, \\
\textbf{m}_k = \overline{\textbf{m}_k} + \textbf{K}_k v_k, \\
\textbf{P}_k = \overline{\textbf{P}_k} - \textbf{K}_kS_k \textbf{K}_k^T, \\
\ell_k = -(\text{log}(S_k) + v_k^2/S_k+\text{log}(2\pi))/2 \label{eq:lk}
\end{gather}
where $v_k$ is the innovation or the difference between the predicted state and the measurement, $S_k$ is the variance of innovation, $\sigma^2_k$ is the same ``measurement noise" from \ref{eq:gplikelihood}, $\textbf{K}_k$ is the Kalman gain that minimizes the residual error, $\textbf{m}_k$ is the updated (a posteriori) state estimate, $\textbf{P}_k$ is the updated (a posteriori) covariance estimate, and $\ell_k$ is the log-likelihood of $y_k$ given $\overline{\textbf{m}_k}$.
We can get $\ell(\beta_{GP}|\textbf{y}, \boldsymbol\xi)$ by adding together each $\ell_k$
\begin{equation}
\ell(\beta_{GP}|\textbf{y}, \boldsymbol\xi) = \sum^J_{k=1} \ell_k
\end{equation}
To perform efficient inference, we wish to have gradients of $\ell(\beta_{GP}|\textbf{y}, \boldsymbol\xi)$ with respect to $\textbf{y}$.
Taking the partial derivative of $\ell_k$ with respect to $y_i$ gives the following since $S_k$ does not depend on any $y_i$.
\begin{equation}
\dfrac{\partial\ell_k}{\partial y_i} = -\dfrac{\partial v_k}{\partial y_i}\dfrac{v_k}{S_k}
\end{equation}
By going through the recursive steps of the Kalman filter inference, it can be shown that
\begin{equation}
    \dfrac{\partial v_k}{\partial y_i}= 
\begin{cases}
    -\textbf{HA}_k\dfrac{\partial \textbf{m}_{k-1}}{\partial y_i},& k > i\\
    1,              & k = i\\
    0              & \text{otherwise}
\end{cases}, \ 
    \dfrac{\partial \textbf{m}_k}{\partial y_i}= 
\begin{cases}
    (1-\textbf{K}_k\textbf{H})\textbf{A}_k\dfrac{\partial \textbf{m}_{k-1}}{\partial y_i},& k > i\\
    \textbf{K}_k,              & k = i\\
    0              & \text{otherwise}
\end{cases}
\end{equation}
Note that $\dfrac{\partial v_k}{\partial y_i}$ ultimately does not depend on any of the $y_i$ values themselves.
Thus these values do not change over the course of inference and can be precomputed, turning the exact evaluation of the gradient of $\ell(\beta_{GP}|\textbf{y}, \boldsymbol\xi)$ into a matrix multiplication of the precomputed gradient coefficients times $\textbf{v} \oslash \textbf{S}$.
This can be further sped up (if you are willing to sacrifice a minuscule amount of accuracy) by replacing the coefficient matrix with a sparse version, noting that these gradient coefficients vanish rapidly as $k$ gets further from $i$ (i.e. when your separation gets larger than several $l$).
Using the precomputed, sparse coefficient matrix to calculate the gradients is $\mathcal{O}(n)$ and gives a speedup of $\sim$ 150x when compared to using automatic differentiation on \tgpjlns.

\pagebreak

\begin{deluxetable*}{cp{10cm}c}
    \tabletypesize{\scriptsize}
    \tablecaption{Notation used in this paper\label{tab:notation}}
    \tablehead{
        \colhead{Symbol} & \colhead{Meaning} & \colhead{Size}
    }
    \startdata
    \hline
    \multicolumn{3}{c}{Matrices}\\
    \hline
    $Y_D$ & Observed flux values (i.e. data) & $P \times N$\\
    $Y_M$ & \ssof joint model output fluxes & $P \times N$\\
    $Y_\oplus$ & Telluric transmission spectra output fluxes interpolated onto $\xi_D$ & $P \times N$\\
    $Y_\star$ & Stellar model output fluxes interpolated onto $\xi_D$ & $J_\star \times N$\\
    $Y'_\oplus$ & Telluric transmission spectra output fluxes & $J_\oplus \times N$\\
    $Y'_\star$ & Stellar model output fluxes & $J_\oplus \times N$ \\
    $\xi_D$ & Native log-wavelength coordinates of the observed fluxes at each observation & $P \times N$ \\
    $\lambda_D$ & Native wavelength coordinates of the observed fluxes at each observation & $P \times N$ \\
    $\sigma_D$ & White noise uncertainties for $Y_D$ & $P \times N$\\
    $T_I$ & Transformation matrix used to mimic the effects of an instrumental LSF & $P \times P$ (sparse) \\
    $T_\oplus$ & Transformation matrix that interpolates $Y'_\oplus$ from $\xi_\oplus$ to $\xi_D$ & $P \times J_\oplus$ (sparse) \\
    $T_\star$ & Transformation matrix that interpolates $Y'_\star$ from $\xi_\star$ to $\xi_D$ & $P \times J_\star$ (sparse) \\
    $W_\oplus$ & Telluric feature vectors & $J_\oplus \times K_\oplus$\\
    $W_\star$ & Stellar feature vectors & $J_\star \times K_\star$\\
    $W_{RV}$ & Doppler feature vector & $J_\star \times 1$\\
    $S_\oplus$ & Telluric feature scores & $K_\oplus \times N$\\
    $S_\star$ & Stellar feature scores & $K_\star \times N$\\
    $S_{RV}$ & Doppler feature scores $\propto v_\star$ & $1 \times N$\\
    \hline
    \multicolumn{3}{c}{Vectors}\\
    \hline
    $\xi_\oplus$ & Native, evenly-spaced log-wavelength coordinates of the \ssof telluric transmission spectrum & $J_\oplus$ \\
    $\xi_\star$ & Native, evenly-spaced log-wavelength coordinates of the \ssof stellar spectrum & $J_\star$ \\
    $\lambda_\star$ & Native wavelength coordinates of the \ssof stellar spectrum, $\textrm{exp.}(\xi_\star)$ & $J_\star$ \\
    $\mu_\oplus$ & Telluric transmission spectra template & $J_\oplus$\\
    $\mu_\star$ & Stellar spectra template & $J_\star$\\
    $\beta_M$ & Set of \ssof model parameters, $\{\beta_\oplus,\beta_\star\}$  \\
    $\beta_\oplus$ & Set of telluric model parameters, typically $\{\mu_\oplus,M_\oplus,S_\oplus\}$\\
    $\beta_\star$ & Set of stellar model parameters, typically $\{\mu_\star,M_\star,S_\star,v_\star\}$ \\
    $\beta_R$ & Set of regularization coefficients used in $\mathcal{L}_R$, $\{a_1, a_2, ..., a_{11}, a_{12}\}$ & 12\\
    $z$ & Total Doppler shift estimated by \ssofns, $z_\star + z_D$ & $N$\\
    $z_D$ & Barycentric correction Doppler shift & $N$\\
    $z_\star$ & Differential Doppler shift estimated by \ssof & $N$\\
    $v_\star$ & Approximate differential radial velocity estimated by \ssofns, calculated as $c \ z_\star$ & $N$\\
    $v_\theta$ & Radial velocities produced by Keplerian model parameters, $\theta$ & $N$\\
    $\sigma_{z_\star}$ & Estimated white noise uncertainties for $z_\star$ & $N$\\
    $\sigma_{v_\star}$ & Estimated white noise uncertainties for $v_\star$, calculated as $c \ \sigma_{z_\star}$ & $N$\\
    $\theta$ & Set of Keplerian model parameters, $\{P, T_p, e, \omega, K\}$ & 5 \\
    \hline
    \multicolumn{3}{c}{Scalars}\\
    \hline
    $P$ & Number of pixels in the spectral order being analyzed by \ssof \\
    $N$ & Number of observations\\
    $K_\oplus$ & Number of telluric feature vectors  \\
    $K_\star$ & Number of stellar feature vectors  \\
    $\mathcal{L}$ & Modified negative log-likelihood loss function used by \ssof\\
    $\mathcal{L_R}$ & $\mathcal{L}$ including regularzation \\
    $a$ & Regularization coefficient \\
    $\ell_{\textrm{LSF}}$ & Log-likelihood of a GP using a Mat\'ern $^5/_2$ kernel whose lengthscale is set to mimic the wavelength covariance of a LSF\\
    $\ell_{\texttt{SOAP}}$ & Log-likelihood of a GP using a Mat\'ern $^5/_2$ kernel whose lengthscale is set to mimic the covariance structure found solar spectra \\
    $l_{GP}$ & Lengthscale of the Mat\'ern $^5/_2$ kernel which characterizes the latent GP used by \texttt{GLOM} in the analysis of $v_\star^{\ssofns}$ for HD 3651 \\
    $c$ & Speed of light, 299792458 \ms \\
    \hline
    \multicolumn{3}{c}{Operators}\\
    \hline
    $\circ$& Hadamard product, the element-wise product of two arrays of the same size\\
    $\oslash$& Hadamard division, the element-wise division of two arrays of the same size\\
    \enddata
\end{deluxetable*}

\end{document}